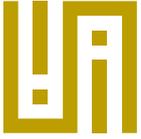

**ADELARD**

# SAFETY CASE TEMPLATES FOR AUTONOMOUS SYSTEMS

24 Waterside
44–48 Wharf Road
London
N1 7UX

T +44 20 7832 5850
F +44 20 7832 5870
E office@adelard.com
W www.adelard.com

**Authors**

Robin Bloomfield
Gareth Fletcher
Luke Hinde
Philippa Ryan
Heidy Khlaaf

**Produced for**

Dstl, under contract number 1000140535

# SAFETY CASE TEMPLATES FOR AUTONOMOUS SYSTEMS

## Summary

This report documents safety assurance argument templates to support the deployment and operation of autonomous systems that include machine learning (ML) components. The document presents example safety argument templates covering

- the development of safety requirements

- hazard analysis

- a safety monitor architecture for an autonomous system including at least one ML element

- a component with ML, a sensor in our example

- the adaptation and change of the system over time

It also presents generic templates for "argument defeaters" and evidence confidence that can be used to strengthen, review, and adapt the templates as necessary.

An interim version of this report was made publicly available, in order to get feedback on the approach and on the templates. Feedback is incorporated in this final version. This work is being sponsored by Dstl under the R-cloud framework.

# SAFETY CASE TEMPLATES FOR AUTONOMOUS SYSTEMS

## Document control

**Reference:** D/1294/87004/1
**Status:** FINAL
**Verified:** Elitza Karadotcheva
**Approved:** Robin Bloomfield

| Version | Review no./Issued | Date |
|---------|-------------------|------|
| v1.0 | R/4893/87004/5 | 7 August, 2020 |
| v2.0 | R/4951/87004/6 | 2 November, 2020 |
| v3.0 | R/4970/87004/7 | 30 November, 2020 |
| v4.0 | R/5043/87004/9 | 26 February, 2021 |

## Distribution

Alec Banks, Dstl
Adam Jeffery, Dstl
Publicly available at https://arxiv.org/abs/2102.02625

## QA statement

This work was carried out under Adelard's Quality Management System, which is certified to BS EN ISO 9001 (BSI registration FS 29600).

# SAFETY CASE TEMPLATES FOR AUTONOMOUS SYSTEMS

## Contents

## Figures

## Tables

# 1   Introduction

The aim of this work is to provide safety assurance argument templates to support the deployment and operation of autonomous systems which include machine learning (ML) components. This report presents safety argument templates covering

- the development of safety requirements for an autonomous system with ML
- hazard analysis
- a safety monitor architecture for an autonomous system
- an ML based sensor
- a template for adaptation of the system over time

It also presents generic templates for "argument defeaters" and evidence confidence (see Section 3.1), both of which can be used to strengthen, review, and adapt the other autonomous system templates as necessary. This means the many known safety and verification challenges in ML and autonomy can be expressed as part of the case.

The templates were developed as part of a set covering a range of different safety assurance issues for autonomous systems, with the assumption these would be assured using UK regulations and legislation, including some defence requirements. The templates we present here have been developed to provide a basis for discussion and feedback, with the emphasis on peer review and refinement of key issues.

Furthermore, we have presented elements of these for review at workshops of key stakeholders and experts in this area. We have held two such defeater workshops with attendees from MOD, Dstl and with external experts from the USA and Germany working on autonomous systems.

There are a number of approaches to "templates". At one extreme, they provide a structure that just needs to be parameterised with the application variables. All the thinking has been done and it is a matter of filling in details (as in a legal template for a will say). The other extreme is one in which the template does some of the standard things and frees the user to do the more thoughtful heavy lifting. A template document in Microsoft Word might be an example of this.

We take the view that in this area we need to provide generic structures as scaffolding for the user so they can develop and communicate their understanding. To enable this there needs to be sufficient discussion of the issues and explanations of how the template can be adapted. Therefore, in this report we support the argument structure with a discussion of the issues involved and consider the reasoning and identification of relevant and proportionate evidence.

This work is being sponsored by Dstl under the R-cloud framework.

# 2   How to use this document

This document is laid out as shown in Figure 1.

Section 2 presents a guide of how to use the contents of the report. In Section 3, we present our approach to developing the assurance templates in more detail, providing scope and assumptions on users and target systems. In Section 4, we provide a very high level set of claims, which shape the overall argument and Figure 9 shows how the templates can be used together. Section 5 presents the first template, looking at requirements. In Section 6 we cover the hazard analysis template. Section 7 presents the third template for the monitor architecture. Section 8 and Section 9 focus on the sensor template and performance measures for a sensor based component with ML. Section 10 contains the template for adaption and future changes. Section 11 provides a discussion, and some conclusions and potential next steps for the work. In the appendices, we have provided additional background material and supporting analyses.

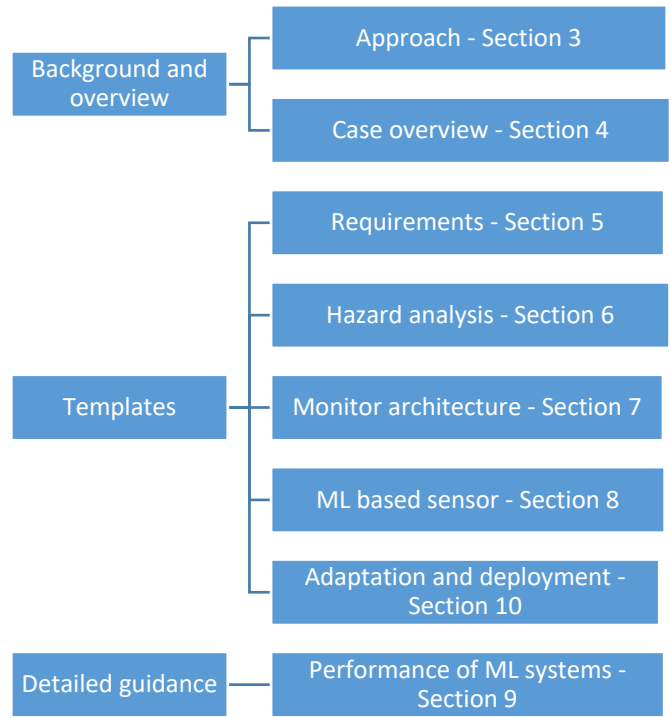

Figure 1: Overview of document structure

## 3 Approach

In this section of the report, we introduce the methodology we have used to describe and develop the templates, as well as our criteria for their acceptability by Dstl and potential users, with assumed use cases and stakeholders.

### 3.1 Supporting methodology

We are developing templates and a supporting methodology for autonomous systems. In doing this we place particular focus on the use of evidence, and within each template, we suggest types of evidence that could be used to support the case and the challenges to ensure they are robust. Further background to the methodology can be found in the following paper [52] on Assurance 2.0.

There are a number of key concepts to our approach to the templates:

- The templates are presented using an extension of the Claims Argument Evidence (CAE) notation with blocks [10][21]. Each template is supported by detailed walkthroughs of the presented reasoning, describing potential evidence, and specific issues for ML or autonomous elements of the system.
- We focus on the reasoning: with supporting narrative (and models) that are important to capture the reasoning. A case is far more than the graphical presentation.

- We use mini-patterns (see Section 3.1.3) for confidence building, property-based claims and evidence trustworthiness.
- The concept of defeaters is used to capture the essential challenges to the claims and their mitigation and help build a robust case. In informal logic [46], the concept of defeaters is used to articulate reasons why a claim might not be supported.
- As defeaters get addressed, the case structure will evolve with additional claims and evidence.

The following section provides more detail on this approach.

### 3.1.1  CAE blocks

The key elements of the Claims, Argument, Evidence (CAE) approach are:

- *Claims*, which are assertions about a property of the system or some subsystem. Claims that are asserted as true without justification become assumptions and claims supporting an argument are called sub-claims.
- *Arguments* link the evidence of the claim; the reasoning rules need to justify the claim from the evidence.
- *Evidence* that is factual and used as the basis of the justification of the claim.

In order to support the use of CAE, a graphical notation is used to describe the interrelationship of the claims, arguments and evidence. In practice, the desired top claims we wish to make such as "the system is adequately safe" are too vague or are not directly supported or refuted by evidence. It is therefore necessary to develop them into sub-claims until the final nodes of the assessment can be directly supported (or refuted) with evidence.

The basic concepts of CAE are supported by the ISO/IEC 15026-2:2011 international standard and industry guidance [10]. The framework additionally consists of CAE blocks that provide a restrictive set of common argument fragments and a mechanism for separating inductive and deductive aspects of the argumentation. These were identified by empirical analysis of actual safety cases [21]. The Blocks are

- Decomposition: Partition some aspect of the claim, or "divide and conquer".
- Substitution: Transform a claim about an object into a claim about an equivalent object.
- Evidence incorporation: Evidence supports the claim, with emphasis on direct support.
- Concretion: Some aspect of the claim is given a more precise definition.
- Calculation or proof: Some value of the claim can be computed or proved.

Figure 2 illustrates CAE block use.

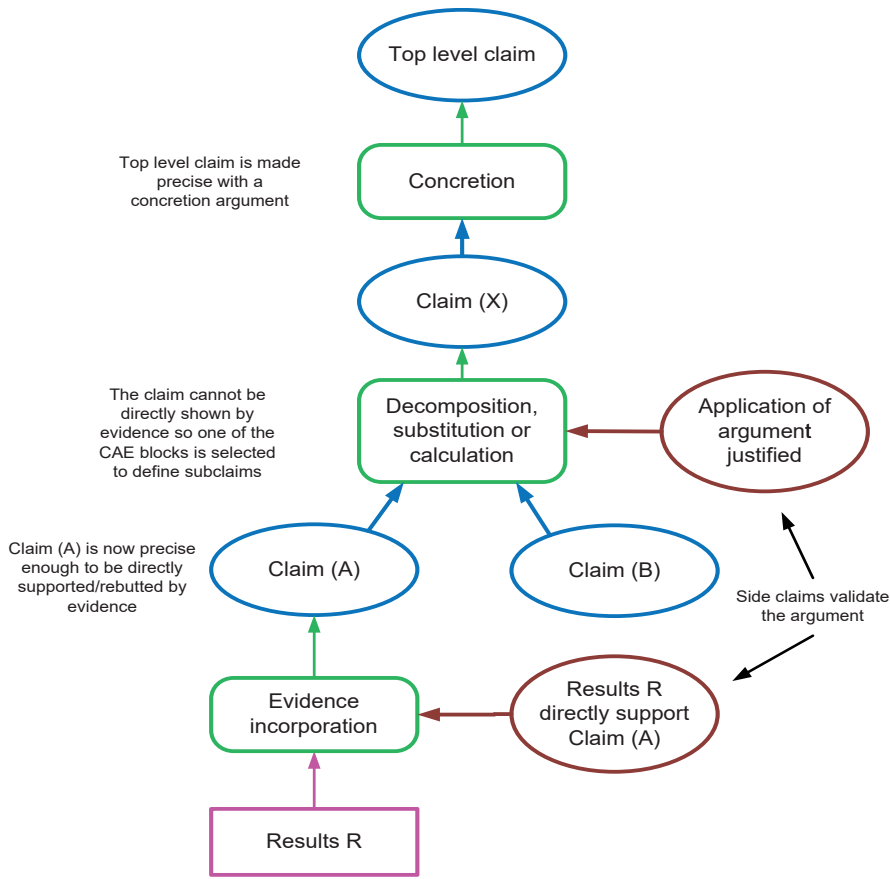

**Figure 2: An example of CAE block use**

One important aspect of the CAE blocks is the use of side claims to justify how sub claims or evidence support the block's top claim. In outlining the assurance case, we often omit the side claim for conciseness: in a real application they would need detailing and justifying.

The framework also defines connection rules to restrict the topology of CAE graphical structures.

Technical background to the CAE framework and guidance material is available on the *https://claimsargumentsevidence.org* website.

### 3.1.2  Property based approaches and requirements

Many safety critical certification approaches rely on showing detailed conformance to standards. However, where there are novel technologies or applications there is increasingly a trend to propose outcome, property and behaviour focused, assurance approaches. These propose high level objectives but avoid describing the exact means of demonstrating that they are met. The work in space and aviation (e.g. for UAVS for NASA [23]), on property based methods for FAA [24], in autonomous standards exemplified by UL4600 [25], in the UK by PAS 1881 [22] and in the variety of "Frameworks" from industry, all envisage an outcome based approach and structured safety or assurance cases as does earlier work in the nuclear industry [26][27].

In developing the case we will be concerned with a variety of properties of subsystems (e.g. sensor reliability, guard performance) and in deploying the templates we can use the generic fragment in Figure 3 to make the link between specific requirements and the properties of interest.

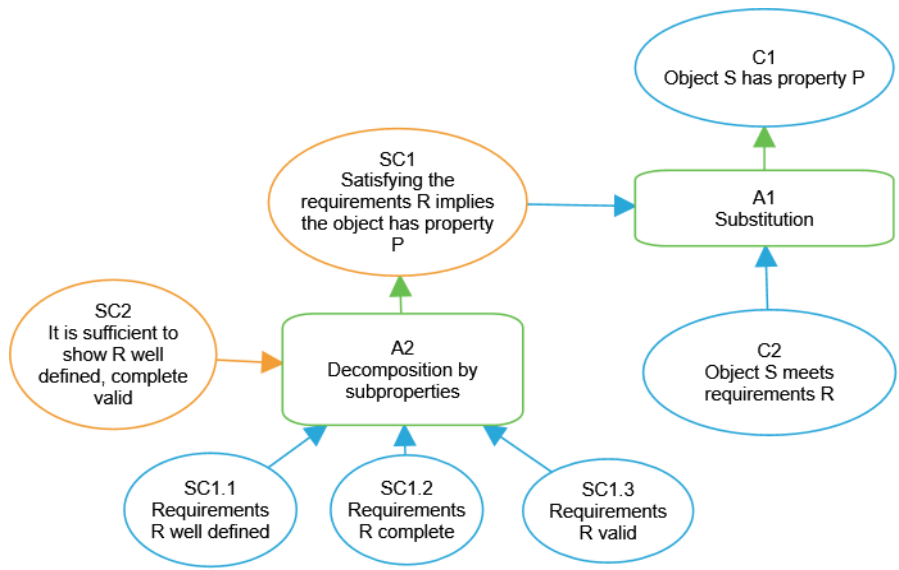

**Figure 3: Linking properties to requirements**

The claim C1 that an object (e.g. a system, subsystem or component) has a certain property P is replaced with a claim C2 that it meets a set of requirements R. The case would then develop under C2 demonstrating why the requirements have been satisfied. The requirements could be specific and local to the discussion or they could be subset of overall system requirements. To justify the equivalence of these claims (C1 and C2) we have a side claim SC1[1]. This side claim is justified by decomposition over the necessary attributes that the requirements need for this equivalence to hold: the key one being the notion of "validity" where we will need to show that requirements are well-defined, complete and valid, and hence they capture the properties of interest. Note that Section 5 expands the requirements pattern.

### 3.1.3 Defeaters and confidence building patterns

The concept of defeaters is used to capture the essential challenges to the case and their mitigation and help make that case more robust. In informal logic [46], the concept of defeaters is used to articulate reasons why a claim might not be supported. [47] distinguishes between two kinds of defeaters: *rebutting defeaters*, which are reasons for believing the negation of the conclusion, and *undercutting defeaters*, which provide a reason for doubting that claim. Rebutting defeaters can be addressed with negated sub-claims.

The identification and mitigation of defeaters are foundational to assurance argumentation, but are often addressed implicitly, as all safety engineers question whether the arguments they put forward are valid. To identify relevant defeaters, we need to address both deficiencies in reasoning within the assurance argumentation (e.g. a known fallacy, such as reasoning from the specific to the general, or the wrong instantiation of an applicable rule) and problems due to incorrect reasoning about behaviours and dependencies. We address some of the former by our more rigorous approach to assurance case construction (e.g. use of predefined CAE Blocks (Section 3.1.1)). The latter we address through our defeaters approach.

---

[1] The identifiers used for nodes in the CAE patterns are used as unique codes to aid review and reference, but do not have specific meanings attached.

The sources of doubt can be thought of as the potential weaknesses in evidence and claims about behaviour, standards and vulnerabilities. Example sources of doubt might be:

- the case is incomplete (e.g., certain hazards have not been treated)
- personal bias has affected the case
- the coverage of testing was insufficient

Another way of identifying sources of doubt is to develop a counter case that aims to refute the claim under consideration. Depending on the details of the system, claim, evidence and postulated doubt, different techniques might be deployed to build confidence. Indeed, there are a number of approaches to reducing cognitive and organisational biases, but these are outside the scope of this study.

A pattern of reasoning that we have found useful is to explicitly show undercutting defeaters, which can be deployed by showing them in the CAE structure. For example, Figure 4 illustrates an example where the assessor has identified sources of doubt – undercutting defeaters – in their judgement.

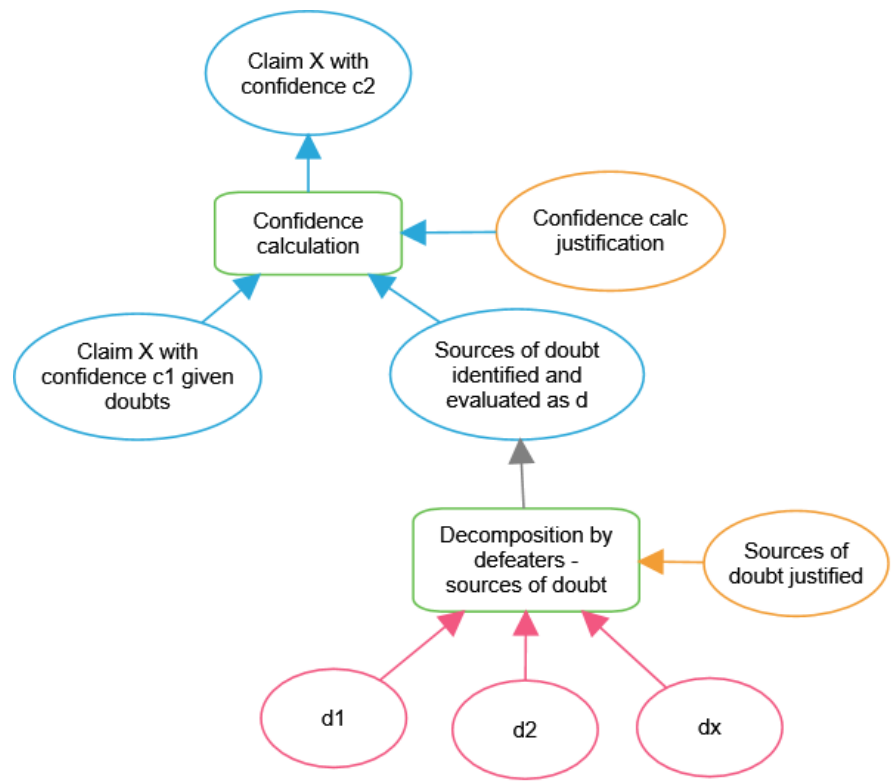

**Figure 4: Confidence building or defeater pattern**

In Figure 4, Claim X has confidence explicitly associated with it, for example it might include aleatory doubt from a statistical technique. Other doubts such as those from assumptions being wrong are captured by identifying defeaters and evaluating these to give an additional doubt d in X. The "confidence calculation[2]" analyses these to support a claim with a different confidence. The side claim has to justify why the "calculus of confidence" used is valid and how it supports the top claim.

The "calculus of confidence" is very much a research topic: in many situations it will be a qualitative approach where we mitigate the defeaters so that we judge them as no longer significant sources of doubt. In that case "confidence c2" is effectively equal to "confidence c1" and it might be expressed qualitatively as "adequate". More sophisticated approaches might be feasible and appropriate when arguing about specific properties such as reliability.

As defeaters get addressed the case structure will evolve with additional claims and evidence. For example, d1 might lead to new claims in the left leg of Figure 4, d2 might be addressed directly with new evidence being incorporated underneath it and dx might be judged not applicable at all. How the evolution of the case and defeaters are managed is an ongoing topic and a key part of how tool support is provided for the application of the templates.

As part of the project, we are running a series of workshops to help identify defeaters, and to explore and develop the methodology. We have held two such workshops to date:

- the first focused specifically on the requirements for autonomous systems in June 2020
- the second was around the sensor, guard and monitor architecture being used in autonomous systems in July 2020

The workshops are reported in [20] and the outputs from them have been used to inform the templates presented in this report.

---

[2] We use "calculation" rather than say "evaluation" to denote that it is a CAE Calculation Block.

### 3.1.4   Confidence in evidence and direct evidence

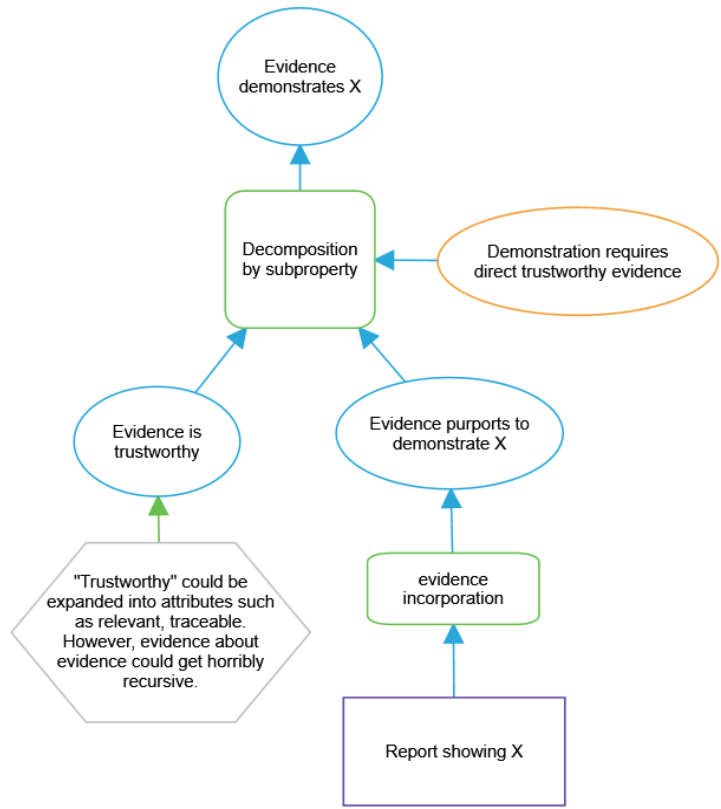

**Figure 5: Evidence incorporation expanded**

We need to ensure that we have confidence in the trustworthiness and relevance of the evidence; not only do we need evidence that purports to demonstrate a claim, but also to trust that evidence.

The evidence incorporation pattern is shown in Figure 5, in which we include claims about the evidence trustworthiness. This could be justified by an expansion of "trustworthy" into attributes such as consistent, traceable and reliable, and these could be demonstrated by a combination of good process, competent people, configuration management tools and adequate security.

The trustworthiness of evidence may be a general issue covering many parts of the case. If the arguments and evidence for this are common across the case, it may be appropriate to separate the evidence trustworthiness claim into its own separate top-level claim. Another complication is that evidence regarding the evidence needs to be trusted, causing the argument to become recursive.

In applying the templates, the claim that the evidence connects to should be directly demonstrated from the evidence. By this, we mean that the connected claim should be an agreed fact and that any inference should be elaborated in the case under consideration or in a subcase.

### 3.1.5   Notation summary

The extension of the use of CAE blocks and associated narrative can capture required properties, challenge, doubts and rebuttal and to illustrate how confidence can be considered as an integral part of the justification. A summary of the additions to the CAE blocks notation is provided in Figure 6. The symbols are labelled apart from the decorator on the top claim that indicates it continues in a structure above.

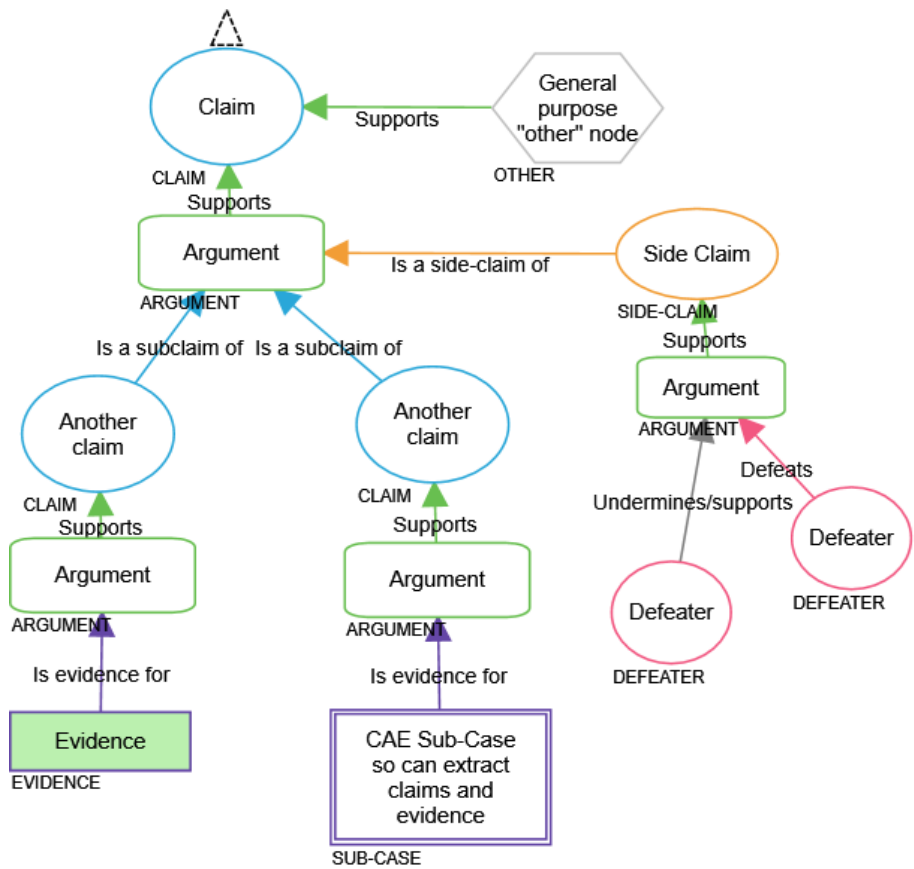

**Figure 6: Summary of graphical notation**

We may wish to show the structure of the defeaters, e.g. decomposing them into sub-defeaters. The need for notations and special CAE Blocks for defeaters is a research topic which is evolving as we investigate their use. Related to this is the notion of defeaters being defeated themselves and the best way of showing this graphically (e.g. by additional decoration, types of arrow).

## 3.2   Criteria for acceptability

While developing the templates we have considered a number of different criteria to judge their acceptability.

- Usability – do the templates guide without constraining as well as not stopping the user from thinking beyond the bare template?
- Understanding and focus – does it provide guidance on key challenges, supporting an engineer or decision maker?

- Communication to stakeholders – does the template support presentation of key risks and issues including to a non-technical audience?
- Coverage of technical issues
  o does the template make a convincing argument for deployment?
  o does the template make a convincing argument for continuing use?
  o does it cover known issues, areas of uncertainty, and counter arguments?
- Feasibility – can we actually produce evidence to support claims and is it feasible to build such a system?
- Scope – is the scope of the overall safety case template clear and do the templates help in identifying the scope for a specific system?

Our validation exercises against these criteria are ongoing, but feedback to date shows that the templates were well received across a wide audience. Our next step is to instantiate the patterns directly for a specific autonomous vehicle as further validation and for potential refinement.

## 3.3    Stakeholders

The templates are developed for the following stakeholders directly:

- engineers and developers of autonomous systems with safety requirements, specifically assumed to be in the defence sector although the majority of the guidance is applicable to any autonomous safety system
- regulators or duty holders who need to review the residual risks and consider accepting the safety case

There will be a number of other groups impacted by the system, such as the general public, users, operators and maintainers. This may include end-users of the autonomous system as a service, e.g. delivery. We assume these would have less direct involvement in either review or development of the safety case but should be considered when instantiating it for a particular deployment or as part of a capability.

## 3.4    System use cases

The templates were developed with the following system use cases in mind:

- autonomous transport vehicle carrying equipment over reasonably well managed terrain (roads, grass, not marsh or minefields)
- autonomous surveillance drone

We assume that these systems have passive defences only, such as armour plating or self-deactivation. In other words, there is limited tension between operational needs and safety.

A system overview is presented in Figure 7. We have assumed there is likely to be more than one autonomous vehicle deployed in the operating environment, all of which require demands from the system support infrastructure, such as repair, maintenance, upgrades and fuelling to operate; these services are assumed to be performed by SQEP technicians and engineers.

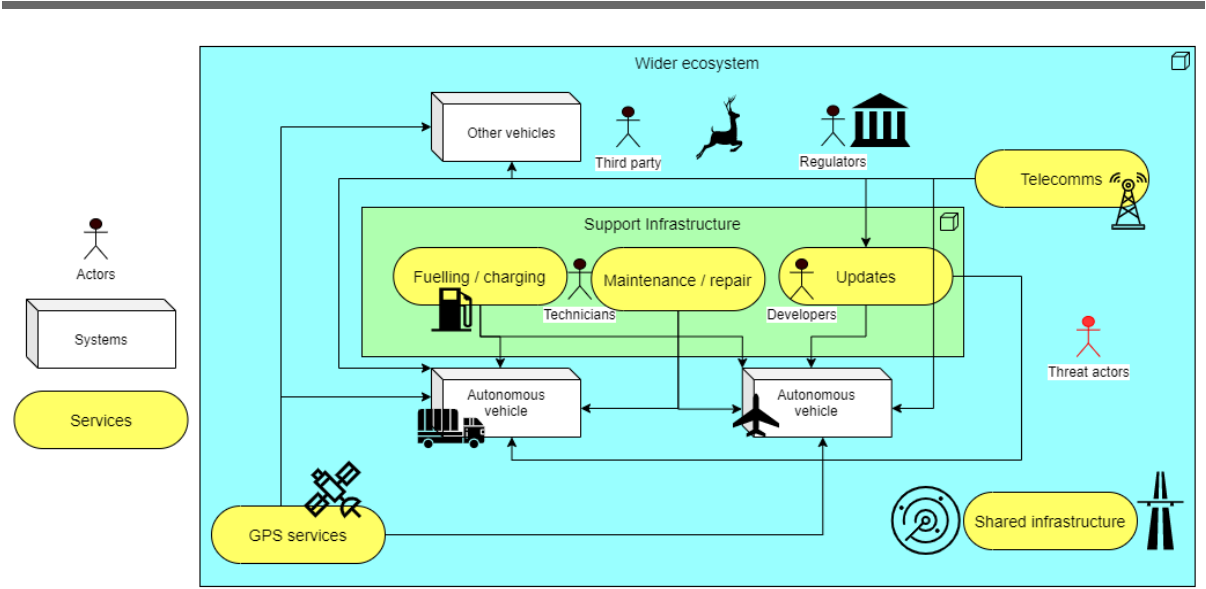

**Figure 7: Autonomous vehicle system overview**

The autonomous vehicles will be likely to require software updates as more data is recorded and to fix vulnerabilities. The autonomous vehicle will have to share the operating space with personnel safely, so human factors and interactions between operators will need to be considered. If the end-user or operator isn't specifically trained, this should be considered in the safety case and hazard analysis.

We consider changes to the wider ecosystem that impact the autonomous vehicles; they are likely to be operating in domains where conventional vehicles are present and potentially members of the public or animals. All can be unpredictable particularly if the situation is not familiar. Non-friendly actors may also be present in the operating space of the autonomous vehicle, and the vehicle should be reasonably resilient against potential attacks or exploits against them, but additional analysis may be needed to determine the threats in a specific scenario and whether it is acceptable to deploy the system.

The vehicles may rely on some shared static infrastructure not controlled by the system owners (such as roads, runways and air space) and shared services, such as telecommunications, digital services, data and information (such as global positioning and maps). The ecosystem could be a congested environment where all vehicles need their own demands met to operate safely and are competing for a finite set of resources.

At a lower level we have tried to consider a broad range of ML element types embedded in the system; this includes ML used in the following scenarios where there is

1. a formally defined input space, e.g. probabilistic graph, decision tree, but to speed up processing or to infer from a sparse input data ML is used

2. a poorly constrained input space, e.g. object recognition in live image streams, for which ML is used as no detailed specification for object identification is possible, and also for speed

3. a changing input space, e.g. the behaviour of other systems is expected to change over time, and hence the ML may need to adapt

Our templates are not limited to these use cases or assumptions, but they help provide illustrative examples to their application, and also help to ensure a broad range of issues are considered. The only restriction is that we do not address the broader issues of ethics and legality that are raised by autonomous war fighting systems. The rather modest use cases also reflect the current challenge of justifying the safety of ML based systems, a topic we explore in detail in this report.

# 4 Overview of safety case

The top level structure of the case is shown in Figure 8. Our top level claim (C1[3]) is that the system is acceptably safe. Additional information for this claim should be provided as for any safety case, describing the system in more detail as well as the target environment.

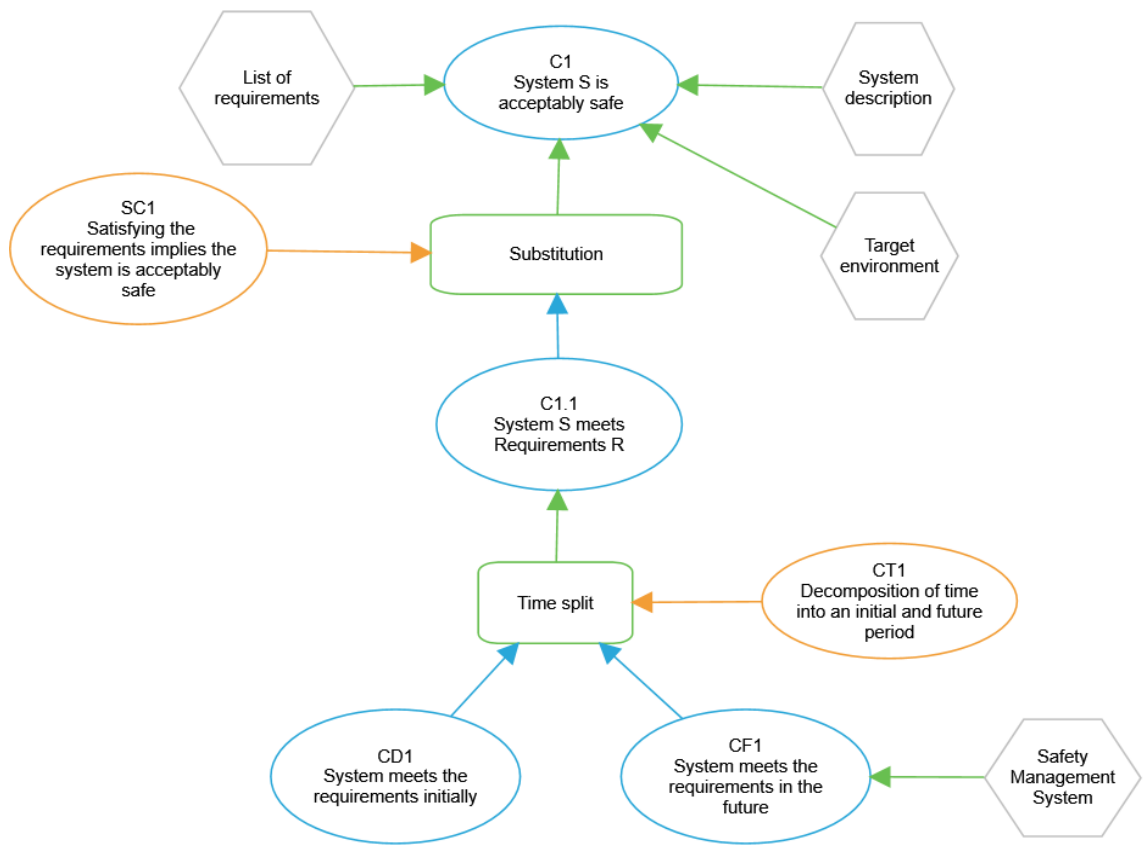

**Figure 8: Top level argument**

Our definition of "acceptably safe", is that all safety related requirements have been adequately met such that they can be signed off by the appropriate duty holders and accepted by safety responsible personnel, in accordance with U.K. defence policy (e.g. Def Stan 00-056 [3]). In other domains, there may be different criteria, such as regulator acceptance or compliance criteria. Typically, this includes a demonstration that the system risk is tolerable and reduced As Low As Reasonably Practicable (ALARP). The justification is that the requirements capture all necessary safety behaviour (SR1) and this is expanded as a template in Section 5. This is a more detailed expansion of the template shown in Figure 3, where we argue that if the properties of interest (system safety) are captured adequately in the requirements then showing we have met them will ensure safety. This argument is further split with a time decomposition that the requirements are met initially (CD1) (i.e. at deployment) and also in the future (CF1).

---

[3] We have used IDs, such as "CR1", to aid referencing to nodes within CAE structures being discussed in the main body of text. No particular node ID format is specified in our text or required when using the patterns.

This structure can be viewed as broadly analogous to the split in a three part safety case frequently used in the UK defence domain (e.g. Land systems [4]). The Part 1 argument contains requirements for the system (SR1), the Part 2 argument demonstrates that the system is safe by design (CD1) and the Part 3 argument (CF1), i.e. it continues to meet its requirements in the future and in-service by being monitored, installed, maintained and operated effectively. There are challenges to all three elements of this case for an autonomous system. The requirements may be difficult to present in a way which can be demonstrably met, the design argument may be difficult to support with sufficient confidence, and the operation argument may have novel elements relating to how mitigations are managed if there is no human operator involved. Further, the design of the system may have a more rapid update cycle than usually expected, creating a challenge to the process of sign off and review. In addition, as the systems are novel there might be different stages of deployment from trials through phases of operation with a variety of limitations.

An overview of the templates' scope is shown in Figure 9.

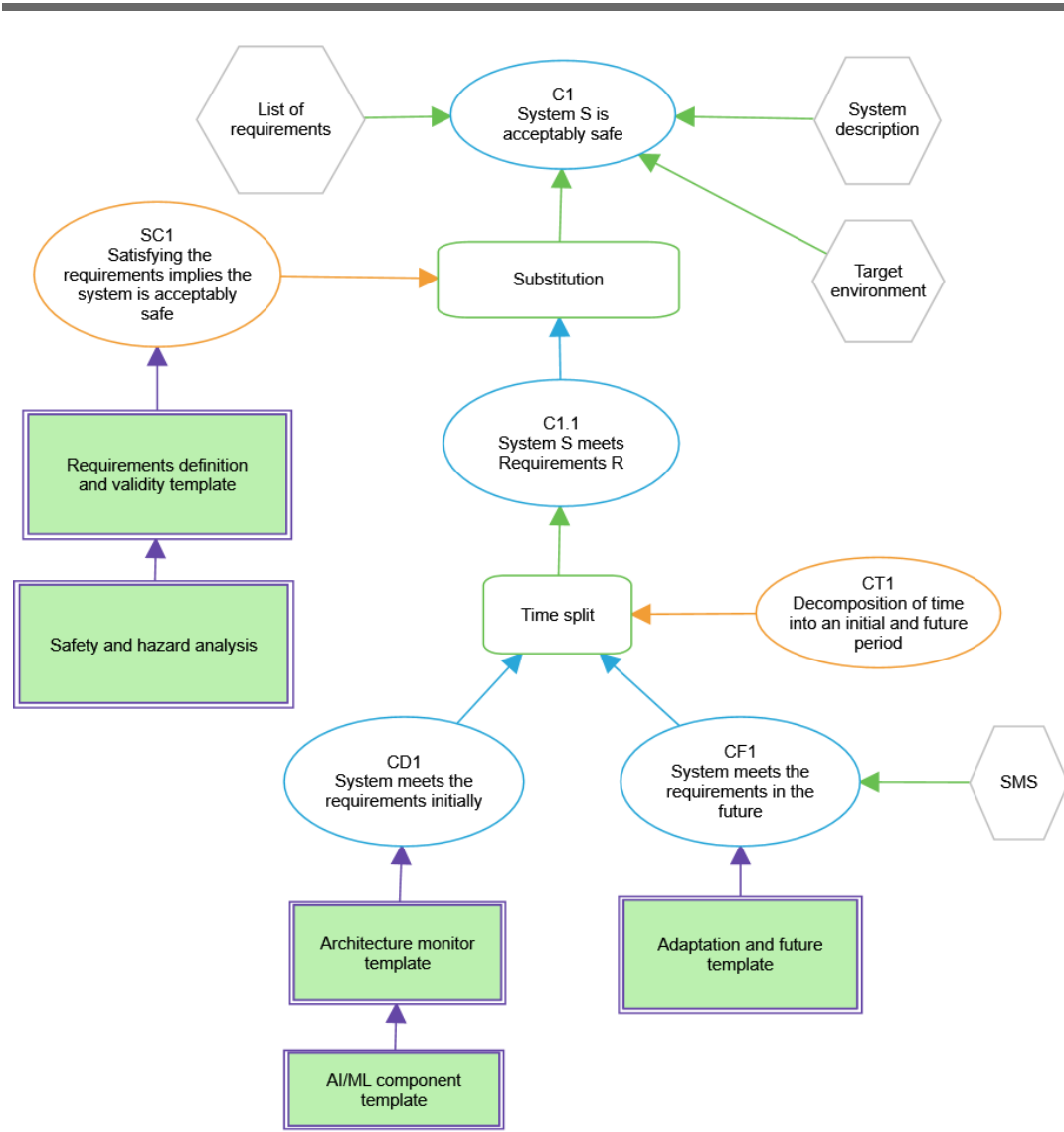

**Figure 9: Overview of template areas**

We identified a number of specific topics for argument templates. Figure 9 indicates those parts of the case that the templates support, specifically:

1. Requirements and policies for the system including a review of risk and tolerability criteria. This is provided in Section 5.

2. System safety and hazard analysis assurance, and how these have dealt with epistemic uncertainties, including the changing nature of hazard initiations and hazard controls. This is addressed in Section 6.

3. Assurance of ML and AI based algorithms including training, update and delivery including impact on planning, development, testing, and static analysis evidence. We have developed a monitor/ML template (Section 7), a sensor template (Section 8) and discuss ML issues in the requirements template (Section 5).

4. Learning and adaptation of the system - this template is presented in Section 10.

A number of significant issues are not fully addressed in this collection of templates:

- Architecture and integration of the role of the human/AI/software aspects, including task analysis and associated human factors evidence. This will be addressed in future work.
- Assurance of ML platforms and implementations including impact on planning, development, testing, and static analysis evidence. This is an important issue as shown by the analysis of disengagement we summarise in Section 9.2 but the overall topic has not been addressed.
- Ensuring a principled design and deployment approach. There are a number of different principles which are becoming well established for machine learning and becoming increasingly high profile, including freedom from bias, fairness of outcome, data protection, transparency and ethical decision making (such as decisions made in the event of an accident). We argue that a system should be safe and principled in design. Whilst our top claim is "System S is acceptably safe" it could just as easily be "System S is acceptably safe and principled", acknowledging the importance of meeting principles throughout the lifecycle.

For a complex system in a complex environment to be judged safe, there are a plethora of issues that need to be addressed and the templates provide a structure for identifying, prioritising and understanding these. For autonomous system we propose the following key questions are:

- do we understand and have we described what it is we are developing and assuring?
- do we understand how big a challenge this is and how are we dealing with this?

The evidence of the effectiveness of ML based systems is used to shape our judgment of feasibility, to define what should be collected during the project and to form the basis for the safety case. Evidence of effectiveness is challenging to define and find. We have reviewed the current state of the art in metrics and where possible evidence, and this is summarised in Section 9 with Appendix B and Appendix C providing supporting analyses. This has resulted in some initial guidance on what level of performance might be achieved in industry.

# 5 Requirements and policies template

## 5.1 Objectives of template and context

The aim of this template is to support the user in developing an assurance case about requirements for an autonomous system. The template itself is generic, but has been explored at two levels of abstraction here as follows:

- demonstrating that the system safety requirements are well-defined, complete and valid for the autonomous system being developed, as well as being feasible
- demonstrating that a ML component's requirements are well-defined, complete and valid for that component, including traceability back to system requirements

In practice the template can be developed at multiple levels, such as from high level system requirements (e.g. of the form "The system shall limit the risk of injury or death for non-involved persons" see Appendix A.1), down to subsystem requirements to detailed implementation specifications. In a complete assurance case, traditional (i.e. non-ML) software and hardware components will also have requirements branches, but this is not discussed explicitly here as no new requirements challenges are anticipated. We note though that the traditional components may need to compensate for shortfalls in assurance and performance in the ML components, and so may have more complex sets of requirements to meet than in traditional systems.

Superficially, the claims and strategies presented in this template are not very different to a similar template for a non-autonomous system. The purpose of the requirements case needs to be considered, however. Whilst the system safety requirements should not dictate a design solution, they will provide the baseline against which a solution will be judged for effectiveness. An additional acceptability criterion for this template is to ensure that this forward use of the requirements is supported, i.e. that the requirements are feasible.

## 5.2 Structure and reasoning

### 5.2.1 Top level of requirements template

The top level requirements' claim is that satisfying the requirements will show that the system is acceptably safe. This relies on the safety related requirements being sufficient to capture all the necessary safety assurance properties, which is the top claim in this template (SC1). A different property could be similarly demonstrated, e.g. security, but this would involve a different set of lower level claims.

We argue that this is sufficient if the requirements can be shown to be well-defined, complete and valid. Well-defined requirements are those which are expressed in a way that meets established quality criteria, and can be considered feasible. A complete set of requirements will address all applicable safety issues (note that this should be true at all phases of the lifecycle, including deployment, operation, maintenance and even disposal). A valid set of safety requirements covers the functional and non-functional behaviour needed to ensure that the system risk is reduced to an acceptable level. These three branches are now described in turn.

Whilst some of these attributes may overlap, we argue that the conjunction is needed to cover all the issues required. For example, a set of requirements can be valid but incomplete. Requirements may be well-defined, but not valid and vice versa.

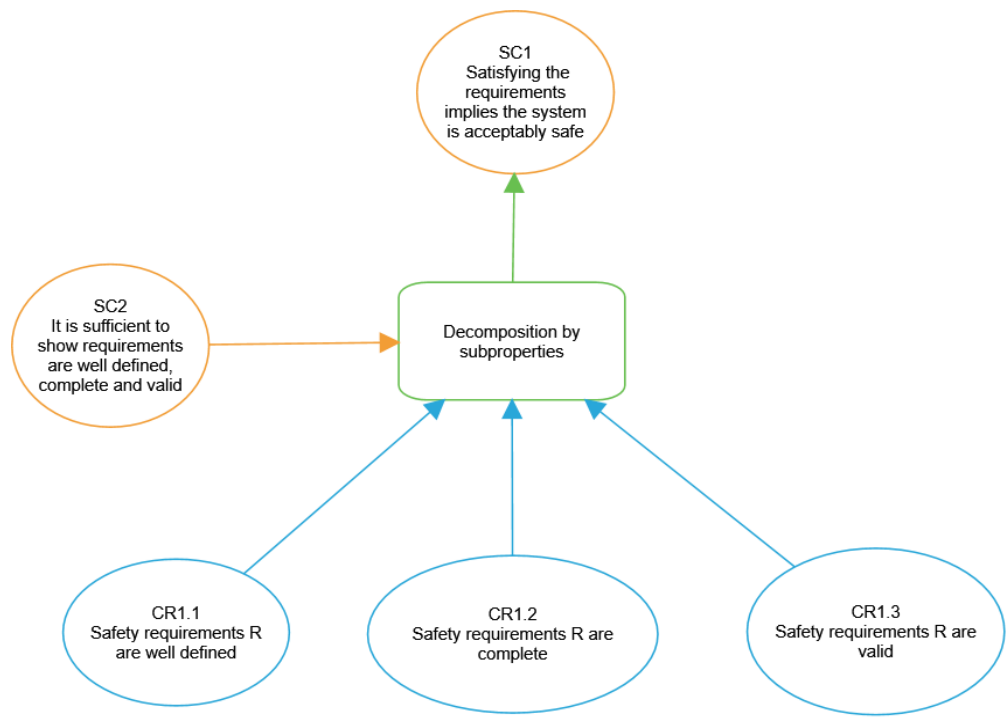

**Figure 10: Requirements top level template**

### 5.2.2   System level requirements are well-defined

The claim CR1, that the requirements are well-defined, requires supporting evidence (ER1) to demonstrate this property. This means they meet desirable quality characteristics both individually and as a set (over a subsystem or system). If the requirements are well-defined, it will be easier to demonstrate that they are met by the subsequent design.

To support CR1, a review of the requirements (or multiple reviews) will be required. This should consider adherence to best practice for requirements expression [5], and be conducted by Suitably Qualified and Experienced Personnel (SQEP).

We now consider what these quality characteristics are, and how they are influenced by the use of autonomy and ML.

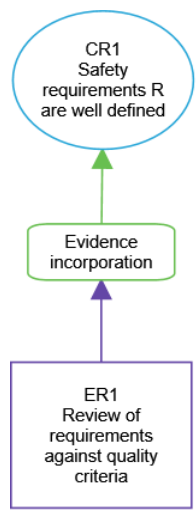

**Figure 11: Requirements well-defined**

In [5], the authors provide a number of types of requirement, such as event driven, state driven, and for managing unwanted behaviour. They give a number of typical problems such as untestability, inappropriate design detail, wordiness, duplication, omission, complexity, vagueness and ambiguity.

Desirable characteristics (as opposed to undesirable) are that the requirements are

- prioritised

- unambiguous – see below

- verifiable

- complete - see Section 5.2.3

- consistent - i.e. do not contradict one another – see Section 5.3.1 for an ML specific discussion

- modifiable

- traceable – see below

- feasible - see Section 5.2.4 and below

- necessary

- accurate

- unitary - addressing one issue at a time

- up to date - see Section 10 and below

We consider the following to have specific issues for AS with ML, however this is not an exhaustive list. The later section on defeaters describes challenges to the case, including some of these issues.

**Traceable**

Traceability goes in two directions. First, the requirement should meet all or part of a business need as stated by stakeholders and it should be authoritatively documented. For a defence system this is often provided by a Concept of Use (CONUSE) document describing a need for a capability. Secondly, the requirement should be traceable to a part or parts of the implemented system design and to individual test cases. In the case of ML, it may be very difficult to trace a particular requirement except at a high level of abstraction – such as to an entire Neural Network (NN) or broad set of training data. Alternatively, where there are, for example, multiple decision trees or an ensemble of NN, a requirement (or several) may trace to all of them.

**Verifiable**

We define this to mean that the implementation of the requirement can be established formally through typical verification methods such as inspection, static analysis, dynamic testing or developing methods such as simulation.

Assuming that the requirements themselves are otherwise well-defined, i.e. unambiguous, simple, precise, it may still be impossible to demonstrate they are met with 100% guarantee for ML and or autonomous systems. In other words, constructing *testable* requirements is problematic. Therefore we would recommend a threshold level of acceptability is defined as clearly as possible, or any shortfalls are required to be fully justified. As noted in [12] this can be difficult for non-experts in ML as it may need interpretation of statistical confidence measures into risk criteria. Hence, expertise in both risk management and ML would be needed. Where there is uncertainty this must be clearly expressed and managed and a level of tolerability must be understood. In Section 9 we provide more discussion on confidence measures that are available for ML.

**Feasible**

For requirements to be feasible, it must be possible for them to be implemented adequately. Specifically, they must both provide the functionality required and be acceptably safe. It may be possible to compromise on original requirements, and provide partial functionality. Feasibility analysis at the start of a project should be part of any requirements review, however it may be more difficult to understand feasibility of a system which makes use of unproven and/or hard to assure technology such as ML.

**Unambiguous**

For tasks such as object recognition or planning, it may be very difficult to describe the input space in formal terms as it is so large. In this situation, presenting unambiguous or precise requirements will be extremely difficult with current methodology. At the system level the requirement may be described in informal terms such as "detection and classification of objects in air space" or "always yield to traffic on the left".

**Up to date**

This temporal aspect of requirements is to ensure they capture the actual required behaviour at a given time. The types of change, internal and external to the system, need to be considered.

At the system level we can consider issues both where the initial requirements were not correct, for example where there was some ambiguity or a misunderstanding of the deployment use cases. This can be refined and captured in the requirements set. It may also be the case that the external environment has changed, for example if there are more autonomous vehicles being deployed at once. Hence there are different types of uncertainty in the requirements and we can review and update them considering these.

We also need to consider where the ML has effectively altered the system requirements or their allocation, based on its solutions to a problem and its performance. For example, it may perform better or worse than anticipated, in different aspects. A means to trace the impact of this to the system level is needed, particularly if there is also a change in hazardous behaviour. With ML that changes rapidly over time it may be very difficult to continuously document system requirements that capture the current situation so tolerances and a behavioural envelope over an interface might be defined. It may also mean there are requirements for continuous monitoring and logging.

Better understanding of ML performance may also impact on other attributes of the requirements, such as feasibility.

### 5.2.2.1   Types of evidence

Suggested types of evidence to support this claim include

- minutes of a requirements review by SQEP with knowledge of the system, how it will be deployed and expertise in testing/verifying ML and autonomous systems
- presentation of the requirements using formal notation/restricted language sets where possible
- process for periodic review or updates of requirements as part of an open systems approach
- input from safety analyses reviewing system safety

### 5.2.3   Completeness of system safety requirements

In this section of the argument (see Figure 12), we consider the completeness of the requirements set. The requirements may be both well-defined and valid but not cover all the required behaviour or performance. Our argument is that consideration of legislation, policy and system specific safety issues is both necessary and sufficient to support completeness. We consider both safety specific policy and legislation, as well as other legislation or principles which may influence the safety of the system. We note that there is a growing body of literature and legislation relating to principles of autonomous systems design (such as for data protection, data fairness, accountability and transparency [9][11]), which may either complement or conflict with safety requirements for the system.

System specific safety requirements are considered complete if the safety analysis used to derive them covers all the identified hazardous situations adequately. (This further assumes that the hazardous situations have been identified in a rigorous manner and are as complete as can be expected; we address this in Section 6.4.) In terms of evidence of coverage, the safety analysis including a traceability matrix mapping the requirements to each known hazard could be used to support this claim. Completeness is a difficult property to claim with absolute assurance except in trivial cases, therefore there will need to be backing evidence of SQEP, coverage of previously encountered hazards, and a justification that no further requirements are needed to address each hazard.

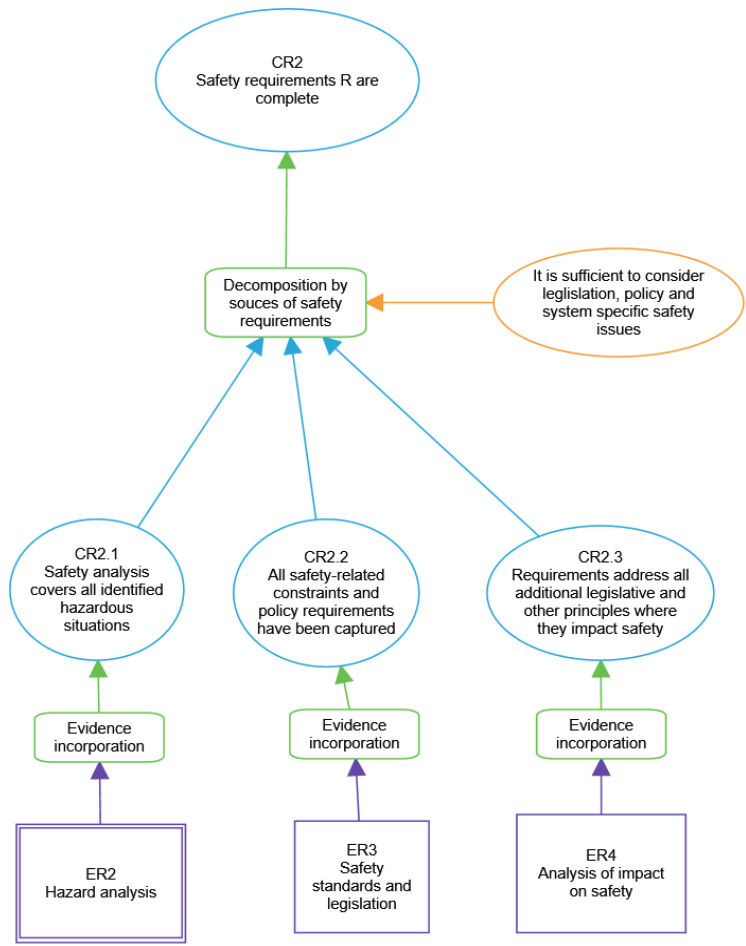

**Figure 12: Requirements complete**

Part of using the template will be to identify where justifying the evidence incorporation identifies the need for explicit argument structure. For example, the need to demonstrate that the dynamic nature of the environment has been considered, that the adaptation of other actors in the overall ecosystem to the autonomous system is understood and that assumptions of behaviour that are made in the other systems safety cases are not violated. This indicates the need for a wider systems of systems safety case.

Policy, standards and legislation cover existing legal or contractual obligations which should also be considered for completeness. We have specified two separate claims – CR2.2 covers safety specific legislation, whereas CR2.3 considers other legislation and principles where they influence safety.

Safety legislation may be wide-ranging, including high level policy documents governing the safety process, and extant standards for systems and software. Additional guidance may also be drawn upon where possible, such as established good practice. A legislation register can be used to capture and agree on the required documents to be followed. For each of these, a compliance matrix can be used to demonstrate adherence and justify any none or partial compliances.

Challenges for the policy and standards claim (CR2.2) is demonstrating their applicability to an autonomous system with ML, as well as demonstrating that this list of documents is sufficient. When demonstrating compliance, we will further need to consider whether the compliance requirements are *appropriate* for these types of systems. The process for this should be considered business as usual, for example in the UK defence sector it is common to tailor a standard for its applicability to a particular system or subsystem prior to its production of a compliance matrix. However, this assumes that the standard broadly covers all the necessary issues for the type of system. A real issue for autonomous and ML-based systems is that there are a limited number of standards available, and these are generally unproven in practice. In [7], the authors provide a detailed analysis of issues with the ISO 26262 automotive standard when applied to AI and ML. A community standard for autonomous systems is being developed in [8] by an industry wide body, but this is as yet unproven in practice.

In other words, even if all available applicable guidance may be found and complied with to an acceptable level, there are still likely to be confidence issues over whether good practice is adequately considered. Further, even if we are complete with respect to the identified standards, we may not be able to argue completeness in the sense that the types of policy and standards expected to be sufficient are covered.

There may be wider policies and issues of concern, particularly with autonomous systems and ML and these are addressed in CR1.2.3. These may not be directly about safety, instead covering ethical policies, security, environment and data protection. However, they may influence some of the decisions made and design solutions. Therefore we have included analysis of the impact on safety as part of the evidence, but there may need to be higher level claims in a broader assurance case about adherence to principles.

For example, ethical principles [32][34] of fairness (that the outcome of ML is fair to all affected groups), no bias (that the training data is not, for example, skewed towards a particular ethnicity or gender) and explainability are relevant. Other issues may include environmental requirements which restrict the design space. There may be confidentiality restrictions on the storage of training or operational data. In the ideal situation these will complement or even improve safety, but at least it should be shown that they can be incorporated without negatively impacting on both operational effectiveness and safety of the system. At the system level these requirements will be very high level, and the detail may be developed at the ML level.

### 5.2.3.1   Types of evidence

Suggested types of evidence to support these claims include:

- ER2
    - hazard analysis including hazards identified, and demonstrating coverage that each are considered (see Section 6.4)
    - SQEP used for the analysis
- ER3
    - high level standards such as Def-Stan 00-055 [6]
    - low level guidance documents such as [8]
- ER4
    - impact analysis of safety with principled use of autonomy
    - guidance on principles of principled use of autonomy, such as freedom from bias, fairness and transparency [9]

### 5.2.4 Validity of system safety requirements

This part of the template addresses whether the potential requirements are valid (CR3). We define valid to mean that the requirements identified will reduce the risk of deploying and operating the system to be both tolerable and ALARP (CR3.1). This means that levels of acceptability will need to be specified, such as risk to both involved and non-involved persons. In this situation, we assume that the deployers and maintainers of the autonomous system would be considered involved as they directly interact with the system, and can be trained in interacting with it, even if they are not operating it directly. They would be exposed to different risks to non-involved persons, which would include the public and other staff.

There may be issues of validity if assumptions about the behaviour of non-involved persons are incorrect. For example, assuming pedestrians always use the dedicated crossings. Therefore, these assumptions should be captured and reviewed. Changes in behaviour in response to an autonomous system should also be considered, as well as any possible emergent issues (such as interactions between autonomous systems and strain on infrastructure).

There is a potential issue in that societal acceptance of risk may be different than for a traditional system, potentially with a lower tolerance. This would have a knock-on effect of needing the system to meet stringent levels of assurance which may difficult to achieve.

We decompose this argument over tolerability and ALARP. The high level claims for determining if risks are tolerable and ALARP are no different to those required at present. In other words, if there is a control of any kind (operational or design) which can be used to reduce the risk and is practicable it should be incorporated. Additionally, the risk should be at least considered tolerable.

We anticipate that the tolerability claim (CR3.1.1) will be partially supported using the same type of evidence as at present, e.g. using the risk matrix, classifying the likelihood and severity before and after controls are identified, and with accident classifications dependent on severity of outcome and numbers affected. However, it may be difficult to translate likelihood of failure of systems incorporating ML (e.g. confidence measures and performance over a trial data set) into the probability of failure that is considered. This is in addition to the already difficult problem of traditional software reliability.

An additional source of evidence of tolerability may be simulation of the system, which is also often used for systems including ML to help identify requirements, for example both Tesla and Waymo have performed millions of hours of simulation and logging of autonomous vehicles to determine safe behaviours, as well as verifying their systems [14]. This could be used to indicate that risks were reduced using the simulation method to examine safety, as well as to determine tolerability of potential outcomes.

Demonstrating that the risks are ALARP or reduced to SFAIRP means both identifying potential controls (CR3.1.2.1.1) and then assessing which of these are considered practicable (CR3.1.2.1.2). Details of all potential controls that have been identified are part of the hazard analysis (see Section 6). However, those which are practicable become part of the requirements hence are discussed here.

Controls may be considered in design (ideal), by operating procedures and limitations, or at least have measures to limit the severity of impact after an event. Use of techniques such as bow-ties may be useful to support this claim. One architectural design control we are considering is the use of a monitor architecture (see Section 7). Other design controls may include development of the ML to known good practice or use of ensembles.

Operational controls may be difficult if there is not a human directly in the loop (e.g. to take remote piloting of an autonomous vehicle).

After the potentially practicable controls and alternatives have been identified, these should then be reduced to a set which is considered practicable. We have identified that this will require a trade-off analysis. There are a number of considerations for this trade-off. The traditional consideration is one based on cost. This is not explored here, however we note that the cost of some design or verification techniques, such as millions of miles of real-world testing (to reduce our uncertainty in risk), would not be practicable for both cost and time reasons.

There are also issues of requirements clashes, emergent requirements, and adaptive behaviour to consider. All of these will impact on the completeness of the requirements.

### 5.2.4.1   Types of evidence

Suggested types of evidence to support these claims include:

- ER5
  - o  risk assessment
  - o  simulation of the system
  - o  information from projects with similar systems or deployments if available
- ER6
  - o  see hazard analysis template, particularly Section 6.4 on outputs
- ER7
  - o  trade-off and cost benefit analysis of practicability

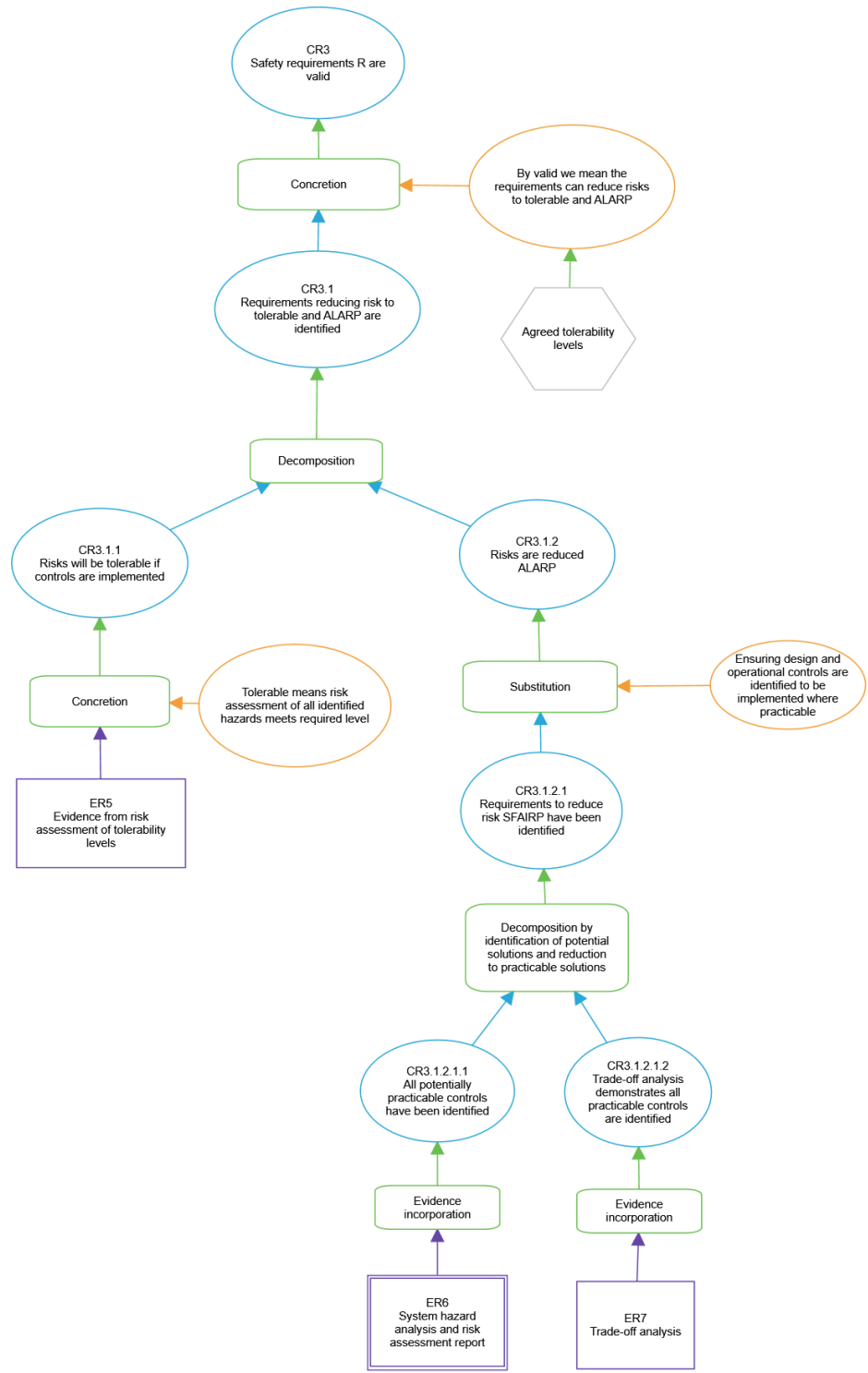

**Figure 13: Validity of safety requirements**

## 5.3  Considering the requirements templates for an ML component

In this section we provide an example of using the three branches of the requirements template specifically for an ML component. To ensure that the discussion has sufficient breadth, we consider three types of machine learning as well the nature of their requirements. The functional requirements are considered to be captured in the training data used for developing and training the ML. This is a similar concept to, for example, a MATLAB model being used to auto-generate code. Hence much of the discussion is around training data and ML performance parameters as these are used to generate the final ML components. We also consider traceability from that data back to system level requirements, as well as into the final ML component.

The three broad ML categories, and their structured requirements, are

- Supervised learning – this typically uses labelled data sets for training, verification and testing. For classification tasks, performance can be measured based on ratios of true/false positive/negative values (see Appendix C) as well as using self-generated confidence measures (see Appendix B).
- Unsupervised learning – this typically uses unlabelled data sets for training, allowing the ML to determine clusters and identify patterns and commonality. Performance can be measured using ground truth comparisons or various statistical measures of clustering.
- Reinforcement learning – this typically is given fixed goals, penalties and rewards based on potential actions. Performance can be measured against those rewards as part of the training process and throughout deployment.

A mixture of these methods can be used in practice. For example, You Only Look Once (YOLO) uses clustering methods to determine typical object bounding boxes and supervised learning for classification [15]. Sometimes classification uses partially labelled data, for example a set of images is labelled as containing cars but their location is not marked.

As well as the training data used for certain ML tasks, we consider aspects of its construction and validation to be part of the requirements argument. This is in keeping with relevant literature such as the Safety of Autonomous Systems Working Group (SASWG) guidance for assurance of ML [8], as well as work elsewhere on the challenges of requirements engineering for ML in [12][13].

The architecture or construction requirements may not be explicit, but there may be a number of issues to consider including numbers and types of layers in a CNN, tuning of hyper-parameters and learning constraints, numbers of decisions trees and trade-offs against processing power, timing bounds and memory usage.

Finally, there may also be requirements on transparency and explainability. The ability to support this will depend on the type of ML being used but may lead to logging, live data analysis and similar types of functionality being required.

### 5.3.1  ML requirements are well-defined

As discussed in the previous section, the training data and learning parameters used for ML can be seen as the source of the majority, if not all, of the low-level functional requirements for an ML component. There are also non-functional requirements, such as performance requirements, which are considered at the end of this section.

### 5.3.1.1   ML training data and learning parameters are well-defined

The first leg of the requirements template (CR1.1) can be interpreted at the next level down with minimal change, although we consider an additional quality criterion of being *representative*. Since data sets or constraints may not be able to capture every situation, they must provide representative samples with appropriate distribution instead.

Well-defined training data sets should meet typical quality criteria for formality and being unambiguous. Interpreting this, it would mean that where the data is labelled it should be labelled correctly, with all relevant items located or positioned or earmarked correctly. For example, a car should have a bounding box which surrounds it, not missing part of it or with a very wide border. Probabilities, formulae and weights should be attached to the correct items in reinforcement criteria. For unlabelled data sets, the types of items being considered should be represented in appropriate quantities. Lack of ambiguity might also apply in the sense that it should actually be possible to determine an outcome from the data, an extreme example might be a blank image marked as containing an item of interest, but this could be more subtle depending on the data type and whether ambiguous data might be normal. The qualities of training data sets are discussed further in the sensor template in Section 8.2.2.1.

The quantity and distribution may not itself be representative of the final environment distribution; for example, if we are looking for unusual events to be detected, then training and test data may represent them in higher quantities. However, there can be issues of privacy of data, data storage (time restrictions) and so on which impact on usable data. This can be a complex issue as discussed in [12] where the authors interview data scientists who share experiences of difficulties storing and gathering appropriate data. The argument will need to justify any trade-offs required and determine safety impact.

In other aspects, training data should be representative of the final environment, e.g. weather, background noise, gender distributions. Sets of actions should be representative of the options likely to be available to the ML.

Consistency would ensure that examples of similar objects were presented in the same way, for example bounding boxes of similar scale to objects. It would ensure penalties or rewards were applied consistently in similar situations so that the ML can reach an appropriate optimum/solution.

Traceability would be of interest based on where data was sourced, but it would also apply to traceability to system requirements. Particularly where the ML can contribute to, or mitigate against, a hazardous situation. This may mean that training and testing data is partitioned and labelled in a way that is clear for traceability (such as a group of edge cases relating to a particular hazardous cause), but this labelling is not used for the actual training of the ML. However, whilst this would demonstrate that edge cases are represented and even tested for, it may not provide confidence that those cases are completely covered as it is likely to be impossible to trace to a specific data response or lines of code in the ML.

The types of constraint placed on re-enforcement learning should also be well-defined. Goals should be consistent with one another, and not conflict. Rewards and penalties should be proportional to the impact of taking the correct or the wrong action. Constraints may be needed to prevent so-called reward hacking.

Feasibility is an important attribute; can the ML reasonably reach a solution given the data sets or constraints provided?

In terms of verifiability, there is work on development of formal verification for ML, which we detail in [35]; this provides measures of "pointwise robustness"[4] to determine the ML's ability to cope with subtle changes in the input space. Our conclusions were that this provided limited additional confidence in the ML and the risks of the perturbations were in practice very small for scenarios such as autonomous vehicles.

All the data should be up to date, for example if new objects have been introduced into the environment, this should be represented in the data and re-training performed. In this situation, an additional requirement is needed to measure performance improvements or at least equivalence. The impact of this change could have far reaching effects on other verification, e.g. system level simulation and field data.

### 5.3.1.2  Other ML requirements are well-defined

Performance requirements, such as expected numbers of false positives or confidence measures, should be clearly stated with both appropriate measures for the type of ML being used, and thresholds clearly stated. Other performance requirements, such as timing or memory usage, can be defined using existing means and no particular challenge is anticipated in expressing them.

Requirements on explainability may include logging, or sending live data and trend information to a remote observer. Again, we would expect the ability to express these requirements as unchallenging, but the ability to meet them is more challenging.

### 5.3.1.3  Types of evidence

- data sets, justified simulated environments and/or criteria for reinforcement learning
- performance requirements and measures including confidence measures and timing constraints
- reviews of data sets and profiling including
    - o  distribution, quantities, quality, security, how representative it is of the environment
- reviews of rewards, penalties and performance measures including
    - o  consistency, appropriateness of penalties/rewards, consideration of reward hacking, appropriateness of potential outputs
- reviews of successful training criteria
- comparisons with ML architectures and types used for similar tasks
- evidence of SQEP for these specialised tasks

### 5.3.2  Completeness of ML requirements

The completeness argument considers both hazardous situations, and policy and legislative constraints.

If the malfunctioning or performance of the ML component can contribute to a hazardous event, all ways in which it can contribute need to be understood, as well as how and if those contributions can be detected. A Hazops or similar process could be used for this analysis (see hazard analysis template in Section 6). This may need to be represented or mitigated in the data sets or learning constraints. This could impact, for example, the range of objects or situations represented in a data set or the rewards or penalties for certain actions.

---

[4] Pointwise robustness gives an indication of how well a classifier continues to perform with small perturbations on parts of the input.

As previously noted, there may be requirements on performance explainability, logging and other functions to support detection of issues or mitigation of other component issues.

Providing all of these would support the completeness claim CR2.1. A checklist could be used to identify typical sets of requirements, but this would need expert review to ensure all aspects of the list were relevant or whether the list was sufficient.

The evidence supporting CR2.2 and CR2.3 should be largely unaffected, as the legislative items will flow down to this level. However, we anticipate that there may be planned partial or non-compliances which would be justified or waivered under CR2.3. For example, a waiver on security or privacy of personal data used for training to ensure maximum safety. There may be other non-compliances presented elsewhere in the assurance case under a satisfaction of requirements claim.

### 5.3.2.1  Types of evidence

- ER2
  - o  traceability of each contribution and where it is represented in data, constraints or ML functions
  - o  SQEP used for the analysis
- ER3
    high level standards such as Def-Stan 00-055 [6]
  - o  low level guidance documents such as [8]
- ER4
  - o  principles analysis e.g. distributions or encoded actions
  - o  waivers and non-compliance justifications

### 5.3.3  Validity of ML requirements

The final element of the requirements argument is on validity and was broken down over tolerability and ALARP at the system level on the basis that this means safety is acceptable. At the next level down this will be ensured by CR2.1 that all the hazardous contributions and mitigations are considered.

Tolerability requirements would need to be expressed as performance requirements. As discussed previously these would need to have meaning for the ML component (as something that could be objectively measured), as well as having meaning in terms of risk assessment. We have provided detailed discussion on different ML performance measures in Section 9.

Supporting an ALARP claim would involve considering all design measures that could be taken to improve the safety performance of the ML. As it is software based, this would be down to either design improvements or quality improvements. Design improvements might include ensembles, additional NN layers or self-monitoring. Quality improvements might include static analysis of ML libraries, use of trusted ML components or increased testing. An understanding of the risk reduction provided by these methods is needed, as well as how practicable they are to implement and verify. This may involve cost (e.g. too costly to verify ensemble) and performance indicators (e.g. too slow to get a response, too much uncertainty introduced with many components). It might also involve effort and benefit, e.g. too long to verify with expensive real-world simulations, with limited additional confidence.

In addition, there may be scope for the actual top-level claim to be adjusted by restricting the operation and claims about the system. For a novel system it might be appropriate to assess whether the early operation of the system with limitations could be used to significantly increase confidence in the tolerability claim and that this period of increased risk was acceptable. This would probably require some quantitative modelling of the interplay of confidence and claim. Relevant background is in the following papers [36][37][38].

### 5.3.3.1   Types of evidence

- ER5
  - o   performance measures
  - o   analysis showing interpretation of performance in terms of risk
- ER6
  - o   brainstorming of potential design controls
  - o   use of good practice for ML design
  - o   SQEP with ML and autonomy experience considering the controls
- ER7
  - o   trade-off analysis results showing improvements gained and cost of implementing controls

## 5.4   Defeaters

As part of the project we have held two workshops at which a number of different experts, both internal to Adelard and a range of industrial experts, performed some brainstorming exercises to identify defeaters to the arguments. The full write up is in [20]. This section discusses defeaters for the requirements.

### 5.4.1   Requirements and policy defeaters

The defeaters are summarised in Table 1 and in each case we have considered which parts of the pattern are affected, as well as whether the defeater impacts on the system and/or ML levels. Possible mitigations are described in each situation. These would be used to provide additional evidence to counter the defeaters and thus shore up the safety case (see section 5.4.2).

The defeaters sometimes impact upon more than one part (e.g. both well-defined and valid). As there is some overlap between the evidence supporting the different parts; this is not considered to indicate that there is an issue with the structure of the current argument. However, the defeaters should be considered in the context of every leg they could impact, as their influence on a related claim may be different. For example, a defeater was identified that it could be difficult to derive test cases from the requirements. This would impact the claim that the requirements are well-formed, as one of the typical quality attributes is that requirements should be testable. However, it may also impact the validity of the argument if the requirement does not adequately capture the risk situation and hence our understanding of tolerability may be flawed.

A number of the defeaters concern issues across multiple levels of requirements, hence multiple sections of the argument, in particular considering traceability and allocation of requirements. These would also need to be considered at every level they could impact. For example, one defeater was that it may be difficult to ensure consistency between different levels of requirements, i.e. translating from one level to another. This may be an issue with poorly defined system level requirements, but would also impact on completeness at the ML level if we cannot be sure all requirements have been adequately decomposed.

No defeaters were identified which were not currently represented somewhere in the overall case, however some could impact upon other areas as well as requirements; this further provides some assurance in the overall structure.

| Description | Part of pattern | Level | Possible mitigations |
|---|---|---|---|
| Traceability of requirements may be difficult for ML components. | Well-formed. | System and ML. | Compensate for lack of detailed traceability with more testing, including impact analysis test after changes.<br><br>Use of agile development process revising requirements. |
| Allocation of requirements to ML or traditional components may be difficult. | Well-formed. | System. | Use of monitor architectures, with conservative estimates of performance.<br><br>Use of agile development process revising requirements. |
| Difficult to derive test cases from the requirements when using simulation for testing. | Well-formed, valid.<br>V&V case. | System and ML. | Review of test cases.<br><br>Additional test cases generated.<br><br>Many levels of testing (module, subsystem, simulations, field testing).<br><br>Use of agile development process revising requirements. |
| Requirements on human review of machine decision, where operator is skewed by machine decision. | Valid.<br>Human factors case. | System. | Robust training of operators.<br><br>Additional checking mechanisms.<br><br>Clear HMI indicators. |
| Requirements may not capture issues in testing such as sampling, quantisation. | Well-formed. | System. | Ensure review of requirements by SQEP.<br><br>Ensure review of testing issues against requirements to justify differences. |
| Ensuring consistency between different levels of requirements, e.g. translating from one level to another. | Well-formed, complete. | System and ML. | Relates to traceability.<br><br>Detailed reviews of requirements, with broad range of SQEP to avoid mistranslation. |
| General existing issues with requirements engineering amplified by use of ML. | All requirements areas. | System and ML. | Use of existing quality criteria.<br><br>Consider use of expanded quality criteria. |
| Accident investigation should be supported. | Complete, valid. | System and ML. | Ensure appropriate and feasible level of logging and explainability is included as a requirement. |

| Description | Part of pattern | Level | Possible mitigations |
|---|---|---|---|
| Maintenance of requirements long-term. | All. | System and ML. | Use of agile development process revising requirements. See discussions on future assurance in Section 10. |
| Contractual issues. | Well-formed. | System. | Issues may be around scope of supply of components and relate to allocation of requirements being feasible. |
| Limited expertise in requirements engineers. | Well-formed, complete, valid. | System. | Use machine learning experts to assist requirements engineering process at system level. |
| Certification and acceptance. | Valid, well-formed. Hazard analysis template. | System. | It may be unclear how to accept risk or to certify the system. This would be related to requirements which define risk criteria. To mitigate would need requirements which are clear to a potentially non-technical audience. |
| Quantitative requirements may be wrong/mis-matched. | Well-formed, valid. Hazard analysis template. | System, ML. | Use/specify conservative measures of performance. Use ML experts to help translation of performance measures to risk criteria. |
| Unable to identify established standards. | Complete. | System, ML. | Use of broad range of SQEP. Use available good practice even if not well established with justifications for deviations. Ensure robust risk and safety analysis. Deploy in limited capacity to gain confidence. |
| Impossible to identify all hazardous events. | Complete. | System. | Use of simulation models to identify potential accident sequences [39]. |
| Inability to identify practicable design controls for ML. | Valid. | System, ML. | Implement guard/monitor architectures. Implement more operator controls where still safe to do so. Consider not deploying system if risk cannot be demonstrably reduced to tolerable and ALARP. Deploy in limited capacity to gain confidence. |

| Description | Part of pattern | Level | Possible mitigations |
|---|---|---|---|
| Inability to identify practicable operator controls for ML. | Valid. | System. | Consider not deploying system if risk cannot be demonstrably reduced to tolerable and ALARP.<br><br>Deploy in limited capacity to gain confidence. |
| Requirements miss long term issues such as sensor calibration. | Valid, complete. | System, ML. | Consider deployment case and use Hazops to identify potential long term issues. |

**Table 1: Requirements template defeaters and possible solutions**

### 5.4.2   Incorporating defeaters in the case

The following diagrams provide an example of how defeaters could be incorporated into the requirements case[5], and subsequently discharged.

The first stage of incorporation is to present the defeaters in the appropriate part of the pattern. We have used the basic defeaters pattern (Section 3.1.3), with decomposition over sources of doubt. This means a justification is required as to the sources of doubt. In the case of the requirements pattern, we have the categories of likely defeaters from the workshop, which can be used during a requirements review to generate issues. Any additional defeaters that are identified should also be incorporated and addressed.

An example is shown in Figure 14, with some of the defeaters from Table 1 presented as the red ovals, with "defeats" links. Our example justification is that a SQEP review identified the sources of doubt. These defeaters undermine our confidence that the requirements are well-defined.

There is potentially another layer to this. If we are not confident in the review (for example the panel SQEP) we may also lack confidence in the sources of doubt identified. Consideration would be needed to avoid adding unnecessary and disproportionate complexity to addressing defeaters.

---

[5] We use the term "case" as an umbrella term incorporating both fully formed arguments with evidence, as well as a partially developed set of claims and evidence.

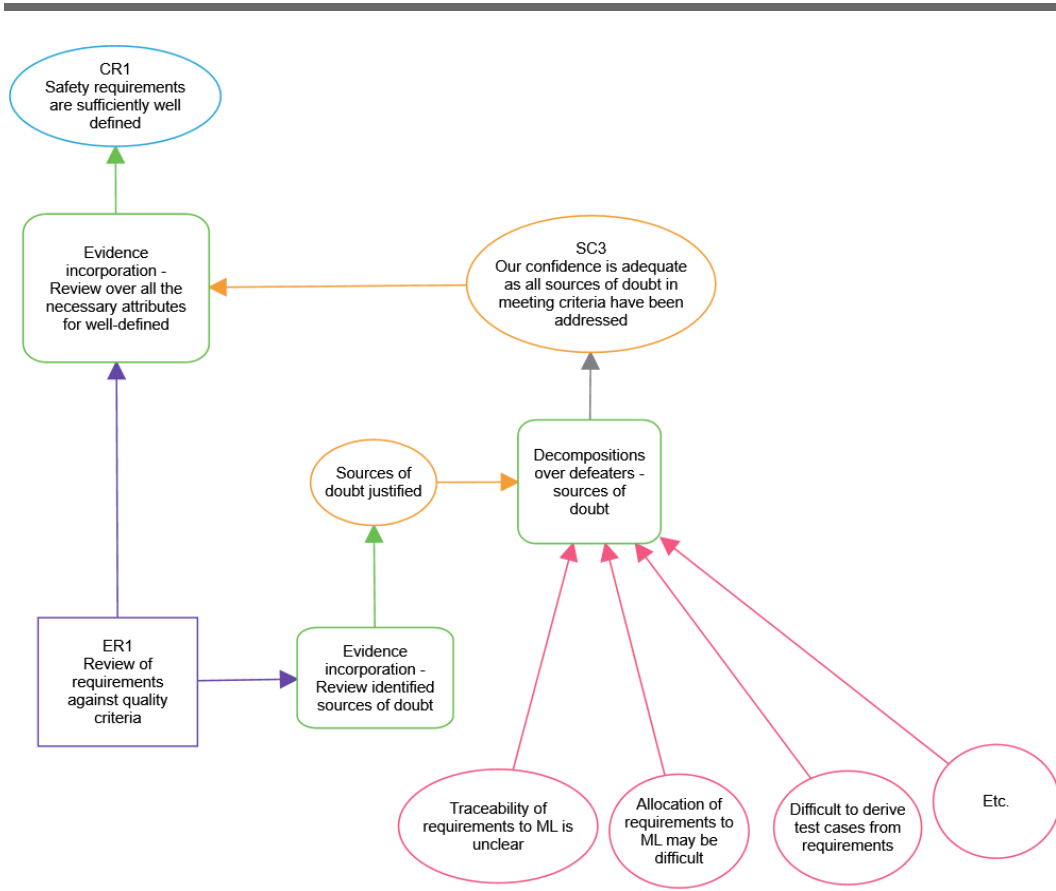

**Figure 14: Initial incorporation of defeaters in the requirements are well-defined template**

The aim will then be to convert all the defeaters into normal claims, by supporting them with adequate evidence as shown in Figure 15. We have provided some examples from the earlier defeaters table, for example using testing to bolster our belief that requirements have been addressed where we may not be able to get traceability. Each of the defeaters in the example is considered below.

### Traceability of requirements to ML is unclear

The main issue identified with traceability is that although functional requirements can be traced to, for example, training data or learning parameters and even to test cases, they cannot typically be traced to the actual lines of code or elements of ML which specifically address a particular issue. This means that although single examples may be shown to be covered, it is not possible to show that in the general case all incidents of a particular requirement are met. Additional evidence to reduce the significance of this defeater may include additional testing (i.e. more test cases/simulations are traced to the specific requirements and shown to be supported) and improved requirements, potentially expressing more clearly the cases, which can and can't be detected. In the latter case, changed requirements may have wider consequences to be considered e.g. reduced overall efficacy or limitations on use, but with greater assurance in that which can be achieved.

### Allocation of requirements to ML may be difficult

One issue with allocation of requirements is that it may be hard to sub-divide requirements over components. Alternatively, a requirement placed on an ML component may have better or worse performance for certain aspects of its behaviour. Where behaviour cannot be guaranteed to the level originally required with ML, it may be possible to add safeguards and monitors to manage certain aspects of that behaviour. Whilst this may not manage every situation, it may improve the behaviour enough so that we can deploy the overall system. Alternatively, it may be that since the performance is different to that anticipated originally, we can change the requirements to be more realistic and achievable. Again, there may be wider consequences to consider about overall efficacy of the system.

**Difficult to derive test cases from requirements**

It may be difficult to derive test cases where requirements are too broadly expressed or ambiguous. Again, we may use a combination of additional testing and improved requirements to reduce uncertainty. For example, the testing itself may uncover issues with the requirements if it is impossible to express test cases well (either what is being tested or the expected results), leading to improved or more specific descriptions of requirements.

It should be noted that in each case we may not be completely removing the defeater but reducing its impact on the case such that other claims are sufficiently met, or even changing the case that is being made by altering what can be evidenced. This needs careful consideration and management. It is not unusual in current practice to list waivers or non-compliances to requirements with a justification as to why they do not impact on overall safety. Changing what the system can do, and potential limitations, does need to be expressed clearly and succinctly to risk owners (e.g. duty holder) and passed on to end users e.g. through training.

Additional claims might be needed which further address the confidence we have now in the defeaters. However, for brevity and readability we have not added these to the structure and they may be better presented in supporting narrative in general.

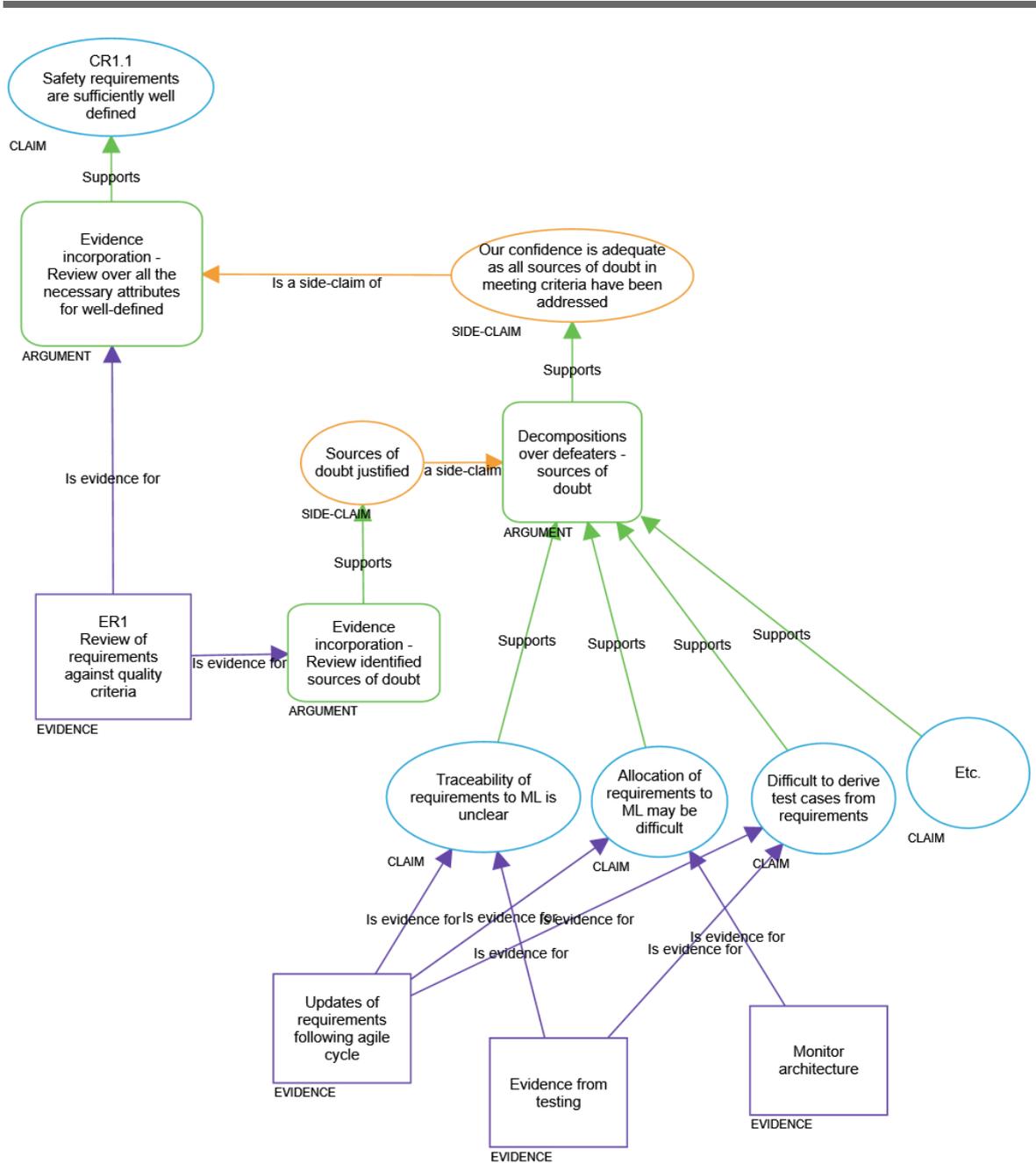

Figure 15: Rebuttal of defeaters in requirements are well-defined template

## 6    Hazard analysis template

In this section we propose a template for hazard analysis for an autonomous system. Our definition of hazard analysis encompasses identification of accidents, hazards, causes and the analysis of risk and identification of potential mitigations. This information is typically contained in a hazard log for a system. The template covers each of these areas.

## 6.1 Objectives of template and context

The main objective of the hazard analysis template is to provide the safety analysis scope for the whole of the system. It directly supports the requirements templates in both the completeness and validity branches.

## 6.2 Structure and reasoning

We argue that the hazard analysis can be considered complete and valid if appropriate hazard analysis methods are followed by SQEP. We structure our argument around the inputs and the outputs to the analysis, each in turn. This high level structure of the template is generic across technologies: the challenge for AI/ML is in showing how we might meet these claims. As for the previous template, we concentrate the supporting discussion on elements that may be particularly challenging or different for an autonomous system, including when containing ML.

The guidance in the UK Defence domain describes Preliminary Hazard Identification and Analysis [16] as input into the initial system requirements and to help scope the project, with ongoing iterative Hazard Identification and Analysis [17] through life. Our template can be used for both.

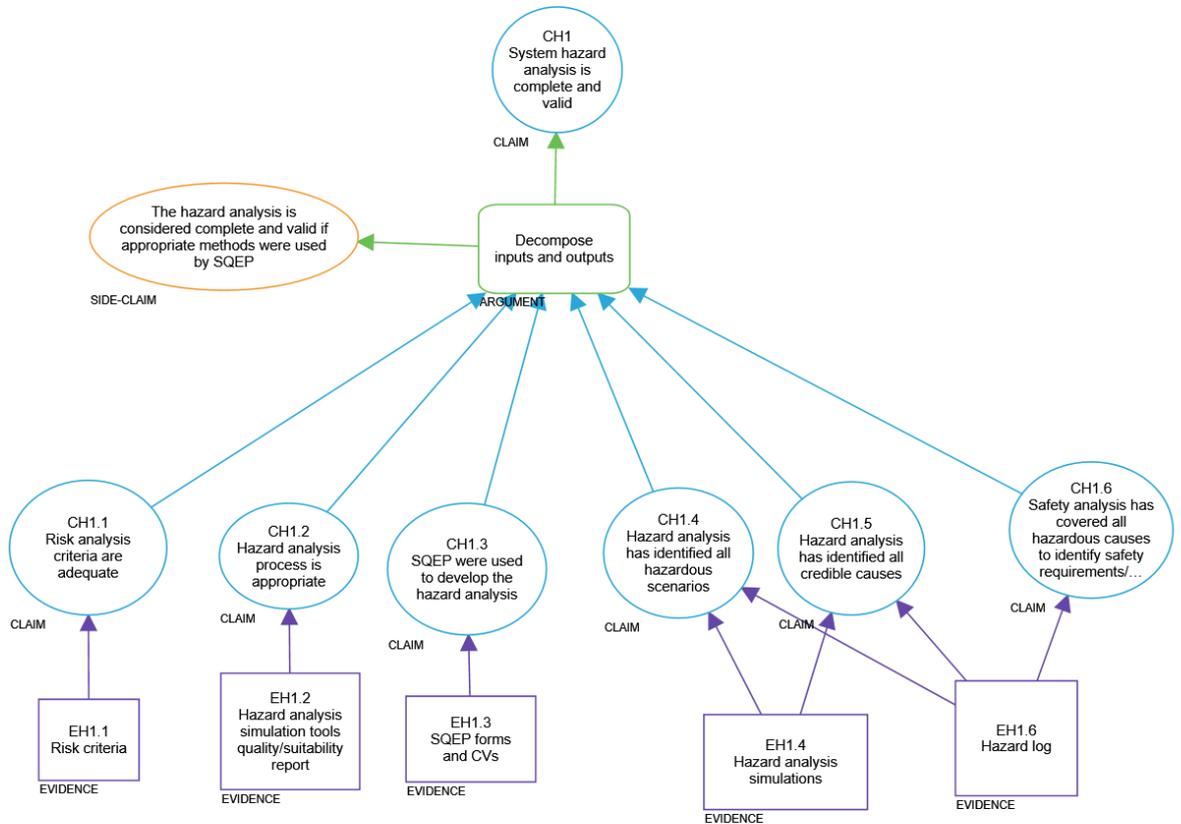

**Figure 16: Hazard analysis template**

## 6.3   Inputs to hazard analysis

The three claims (CH1.1-CH1.3) consider the inputs to the hazard analysis process.

### 6.3.1   Risk analysis criteria are adequate

CH1.1 considers whether the risk analysis criteria are adequate. These will be typically presented as a matrix combining likelihood (either quantitative or qualitative) and severity of outcome. Based on this matrix it can be determined whether the risk of hazards and accidents can be considered tolerable. A risk assessment before additional mitigations and controls can be compared with one after mitigations and controls are applied to assess risk reduction (see the validity argument in Section 5.2.4).

In terms of likelihood, figures may need to be converted to performance metrics for ML. We might normally consider for example, acceptable numbers of failures per operating hours (e.g. *remote* might relate to a failure with likelihood of $10^{-5}$ over one year or frequent might be expressed qualitatively as "would be likely to be experienced continually"). One issue to contemplate is what might be considered a dangerous failure, for example only a mis-classification of a certain type, or only a subset of planning decisions might lead to a hazard. Autonomous systems may be deployed in novel situations where there may be trade-offs that impact the risk criteria e.g. temporary increases in risk but a longer-term benefit.

In terms of severity in defence, typical categories to consider include

- major injury/one fatality, catastrophic multiple deaths, or marginal minor injury to involved personnel
- impact on general public or non-involved persons compared with involved persons (users)

Involved and non-involved would need consideration here for autonomous systems as there are no direct operators, so the "involved" boundary may not be clear. Damage to equipment without an operator may also need to be considered as an accident outcome, as may environmental damage.

Overall, the level of tolerability may be different for an autonomous system, in particular it may be reduced based on public perception or due to limited initial deployment. Different risk outcomes may need different levels of sign off which should be considered. The uncertainty associated with autonomous system performance and their impact on the wider system should be considered (e.g. how others might change behaviours when they know the system is autonomous). As noted in the introduction, we limit these templates to systems that do not have any autonomous war fighting capability.

#### 6.3.1.1   Types of evidence

Typical evidence would be approved risk matrices, covering different categories of affected personnel and depending on deployment situations. Additional evidence may come from wider debate and research into tolerability of autonomous system failures.

### 6.3.2   Hazard analysis process is appropriate

CH1.2 considers the appropriateness of the hazard analysis process. The processes used for hazard analysis can be diverse. They include experience brought forward from similar systems/projects, checklists of issues, guideword based walkthroughs and so on.

Systems containing ML often rely on simulation tools as part of their verification process and to examine the consequences of failures. Simulations may be desktop-based, or system level lab controlled experiments. As a result, we assume they are likely to be used to complement the hazard analysis process, potentially uncovering undetected causes of hazards and even new hazards and accidents. If they are being used then both the tools themselves (software and hardware) should be justified as fit for purpose. This may be difficult if they are off the shelf tools, not designed for safety critical use. Additionally, any configuration, whether physical or software based, should also be demonstrated to be valid for the systems use. Where there is a source of doubt, SQEP review of the results can be used to consider applicability (see 6.3.3).

An additional consideration is the change in the nature of an autonomous system. Hazards can be broadly categorised into inherent/physical and functional. Physical hazards cover problems such as interactions with hazardous materials, slips in the rain and impact of lightning. An autonomous system with no operator will have less continuous interactions with end users, but may still have interactions with maintenance crews or at specified times during a mission. Therefore, the potential number of situations in which some physical hazards can occur may be smaller. Functional hazards on the other hand will likely have less potential for operational controls, making it harder to reduce risk levels and the interactive and possible non-deterministic nature of the autonomous systems makes assessment of behaviour hard. The hazard analysis process should take this into consideration, particularly where existing methods such as checklists or experience from previous systems is used as they may need adaptation.

### 6.3.2.1   Types of evidence

Types of evidence include:

- verified checklists
- provenance of existing methods such as Hazops, STPA
- simulation tool justification
- simulation setup review

## 6.3.3   SQEP

For completeness we note that this analysis should be carried out by SQEP.

### 6.3.3.1   Types of evidence

Evidence that staff are SQEP to perform the hazard analysis should be evidenced by background and expertise, qualifications and can be bolstered by training in specific areas. A broad range of expertise is typically required, including end users, safety experts, and in this case autonomous systems and ML experts are required. This includes expertise in specific types of ML that are being used.

## 6.4   Outputs of hazard analysis

Claims CH1.4-CH1.6 cover the completeness and validity of the outputs of the process. Generally for qualitative analysis, it is hard to demonstrate completeness, nevertheless it is important to demonstrate that all reasonable means to achieve complete coverage of issues have been used.

### 6.4.1   Hazard analysis identifies all relevant scenarios

This claim covers completeness of the identified hazards and accident scenarios, i.e. the end situations of concern. As discussed previously, accidents will be categorised based on severity and impact to numbers of people and can be expressed in a generic way. Therefore, completeness and validity of these accident categories may be relatively straightforward to justify.

Some categories of hazards will also be generic, including physical issues such as fire or exposure to hazardous materials. However, functional hazards will be more system and even mission specific such as having an incomplete sensor picture, or incorrect classification of a waypoint. There may be overlap to consider, for example problems with cold surfaces would be considered a physical hazard, but cold may lead to problems with computer performance and hence impact on functionality.

Unexpected hazardous situations may be identified by simulation based tools or even post deployment (see Section 10). In which case, review of the results would be needed to appropriately categorise and classify them.

In the autonomous vehicle world, scenarios are captured in terms of operational design domains (ODD) that describe "the operating conditions under which a given driving automation system or feature thereof is specifically designed to function" [19] and a specification of the intended functionality. ODDs are often categorised by frequency and criticality and become the basis for simulation and hazard analysis. The assessment could be done in two parts: assessing the validity and completeness of the ODD (as part of the inputs claims), and assessing the hazard analysis of the intended functionality within the ODD.

Overall, assurance of completeness and validity would be based on continuous review by appropriate persons, at the right time. For example, during design and at fixed points during deployment as part of ongoing safety management system activities.

### 6.4.1.1  Types of evidence

Types of evidence to support this claim would include

- minutes from review meetings
- hazard analysis simulation results
- incident and accident reports during operation
- ODD analysis

## 6.4.2  Hazard analysis identifies all credible causes

This claim covers completeness and validity of all the credible causes of the hazards and hazardous scenarios. A cause encompasses many items. It could be an internal failure of a system (such as a mechanical failure) or a logical problem (such as a software error). Alternatively, it could be an external issue, such as weather or loss or malfunction of wider infrastructure (see Figure 7). This claim is caveated as "all credible causes" as it may be the case that a conservative or limited set of causes would be adequate for the analysis, ignoring some very unusual events which might be theoretically possible but unlikely in real-life. If so, this would need to be justified.

There are many existing techniques for capturing and performing causal analysis, such as fault trees, STPA, FRAM and Hazops. These would be applicable to an autonomous system, and to ML. More dynamic methods of hazard analysis might be needed because of the way in which autonomous systems can interact with the environment and also change the environment. For ML, detection of fault categories would need expertise in the type of ML being used, particularly when considering the likelihood of particular faults and failures.

Again, we may use simulation tools to identify potential causes of failures, particularly with ML, and this would need manual review of the results to consider why particular problems occurred (e.g. overtraining, bias, unexpected input data, bugs, failures within expected tolerances (the performance can only be 90% reliable), slow response).

The causal analysis should include an assessment of likelihood of a failure, impact and severity. This may be difficult for ML.

### 6.4.2.1   Types of evidence

Types of evidence to support this claim would include

- minutes from review meetings
- hazard analysis simulation results
- incident and accident reports during operation
- cause/consequence analysis

### 6.4.3   Hazard analysis identifies all potential controls

Claim CH1.6 considers whether all potential controls have been identified. Our definition of control includes either removal of a cause (ideal), or design or operational mitigations to detect and manage a cause. As noted previously, operational mitigations may be harder to implement for autonomous systems without an operator. Remote monitoring may be used, but would likely need trained personnel and may not be effective in all situations. It is helpful when listing/brainstorming controls to consider their likely effectiveness or feasibility in implementing them as this makes the ALARP argument simpler.

One basic check for completeness is to demonstrate that every identified cause has been reviewed to consider potential controls. However, there may be further controls to reduce risk which may not be related directly to a cause, such as using more reliable components. Additionally, more controls may be identified at a later stage of the system lifecycle and added. It may also be the case that a control is shown to be ineffective and may need to be removed or replaced.

In the requirements template, we discuss the output of this as supporting the validity branch of the requirements. The ALARP argument should demonstrate that all reasonably practicable controls are implemented, which requires an assessment of whether they are practicable (in terms of technical effort, benefit and cost) as well as simply listing the possible controls (the latter is the focus of this claim, the former is discussed in 5.3.3). One problem with controls associated with ML is that it can be very expensive, time consuming or technically difficult to incorporate them, meaning only a small subset could be used. Additionally, some controls may conflict with other controls. This should be taken into consideration in the requirements validity branch, but will impact on risk reduction and only implemented controls will ultimately be stored in the hazard log.

### 6.4.3.1   Types of evidence

Types of evidence to support this claim would include

- minutes from review meetings
- hazard analysis simulation results demonstrating effectiveness
- feedback from in service performance
- lists of potential controls and review of feasibility and practicality

## 6.5   Hazard analysis defeaters

The following table identifies some potential defeaters and their rebuttals for this template.

| Description | Part of pattern | Possible mitigations |
|---|---|---|
| No feasible controls can be identified. | Safety analysis to identify controls (CH1.6). | Consider fundamental redesign of system. Consider whether the system is considered tolerable. Deploy with limitations on use. |
| Hazards may vary considering deployment. | Hazardous scenarios (CH1.4, CH1.5). | Review hazard analysis before deployment in a new environment. Produce strict definitions of appropriate deployment environment to review before each mission. Deploy with limitations on use in new environment. |
| Tolerability levels lower than for traditional systems or hard to assess. | Risk criteria (CH1.1). | Review by SQEP. Consider wider implications of tolerability. Deploy with limitations on use. |
| Hard to identify involved personnel. | Risk criteria (CH1.1). | Present clear boundaries of impacted/involved personnel and agree with SQEP and duty holders. Limit interactions with personnel where possible. |
| Simulation tools are unreliable: both due to lack of fidelity or implementation issues. | Multiple claims. | Review of the simulation tool outputs for validity, benchmarking. Improve overall quality of simulation tools, or conduct review into their likely failures. |
| SQEP are hard to find. | SQEP (CH1.3). | Training of staff. Expand team used for hazard analysis. |
| Causes difficult to identify. | Causal analysis (CH1.5). | Use simulations to expand results or gain confidence in existing results. Use ML SQEP. |
| Likelihood of ML failures hard to quantify. | Causal analysis (CH1.5). | Use ML SQEP. Assume worst case scenarios if necessary as input to overall analysis. Gain confidence/experience with the system with limited initial deployment. |

| Description | Part of pattern | Possible mitigations |
|---|---|---|
| Completeness of analysis is difficult to justify. | Multiple claims. | Use broad range of SQEP during all stages of analysis.<br><br>Use simulations to expand results or gain confidence in existing results. |
| Controls introduce new failures. | Multiple claims. | Review all controls for secondary failures. |

Table 2: Defeaters for the hazard analysis template

# 7 Monitor template with AI/ML component

## 7.1 Introduction

Given that it is difficult to assure ML based components, especially perception systems, approaches are needed to reduce the assurance burden and allow their use. A common approach in engineering complex system architectures is to limit parts of the system that need to be highly trusted: safety and security protections are provided in a simpler system or safety monitor that detects when a system is close to being in an unsafe or insecure condition and acts accordingly.

The concept is common across different application domains: nuclear power plants have a variety of protection and limitation systems and in the finance sector most trading venues have a number of processes in place aimed at avoiding extreme price movements [83]. In ecologically oriented work, viability domains describe a similar concept to protection envelopes [82] and in security there are intrusion detection systems which infer potentially dangerous behaviour from the complex system state and knowledge of threats. The safety monitor architecture is proposed as a standardised approach in the air domain [28], in autonomous vehicles [30] and generally for cyber physical systems [29].

Note that the monitor architecture could be deployed at different levels within an autonomous system:

- at a system architecture level where diverse sensing and planning information is being used to provide control of the autonomous system
- at a component level such as a sensor where the monitor's role is to protect the wider system from problems from the sensors

This section considers the use of safety monitors with an AI/ML component. In Section 7.2 we first describe the objectives of the template and the safety monitor concept before describing the different aspects of the template itself in Section 7.3.

Issues of evidence and the extent to which claims could be substantiated are discussed in some detail in Section 9 for systems with AI/ML based sensors. In Section 7.2.1.2, we note the limitations of the sensor monitor approach.

## 7.2 Objectives of template and context

The template aims to show how the safe deployment of a subsystem component (a sensor in our example) which has AI/ML technology might be feasible and justified using a safety monitor architecture or at least that the overall level of safety is improved. In this architecture, the functions provided by the subsystem component are checked by guards in the monitor for anomalies that may indicate a failure state or that it is operating outside its intended environment.

## 7.2.1 Safety monitors and AI/ML component

The safety monitor architecture limits parts of the system that need to be highly trusted. It is common across different disciplines (such as aircraft, railway systems, nuclear power plants, etc.) and is proposed as a standardised approach in the air domain [28], and more generally for cyber physical systems [29]. It is a common control mechanism, as identified in our requirements template (see Section 5.2.4). In this section, we illustrate how we might reason about a safety monitor within an assurance case.

### 7.2.1.1 The need for safety monitors

Safety monitors can vary in sophistication from comparison between diverse sensors (e.g., comparison of LIDAR measured distance with that from a stereo camera) to a monitor implementing a complex set of equations and constraints (e.g., see Responsibility-Sensitive Safety (RSS) [30]). This architectural approach often seeks to reduce the trust needed in ML components by monitoring both the state of the environment and the vehicle. They can also monitor when an autonomous system is under stress, or in an error prone situation. It is not unlike the intrusion detection problem in security, where one tries to infer potentially dangerous behaviour from the complex system state and knowledge of threats. The Defense Advanced Research Projects Agency (DARPA) Assured Autonomy programme for example, extends the safety monitor concept to include a dynamic assurance case, as monitors can be seen as a form of run-time certification that shifts the certification or assurance challenge from the design and development part of the lifecycle to operation [31].

Safety monitors are one example of defence in depth. Defence in depth is fundamental to achieving safety and resilience through detection, tolerance, recovery and disaster management and through having appropriate system architectures. Defence in depth and diversity are discussed in [40].

The need for defence in depth is in part driven by the performance of AI/ML based sensor systems. Although the current performance of AI/ML sensors is much improved in the past decade, it is often not sufficient. This improvement is illustrated in Figure 17 [33] where a metric, the mean average precision (mAP), is shown against time using standardised image sets as a ground truth comparison (see Appendix A.2). One technical issue for such sensors is how to relate the metrics used in benchmarks to operational measures of failure per demand. Some example plausible claims for current technologies are shown in Table 3, and in Section 9 we explore this in more detail.

| The traffic light detection system correctly identifies 95% of traffic lights in Vitoria with confidence 60%[6]. |
|---|
| The pedestrian tracking system identifies 80% of pedestrians which are visible for at least one second. |
| The addition of a GPS guard reduces false positive traffic light detections by 80%. |

**Table 3: Plausible claims for AI/ML sensors**

---

[6] The study did not provide in depth details of the environmental conditions in which the tests were carried out.

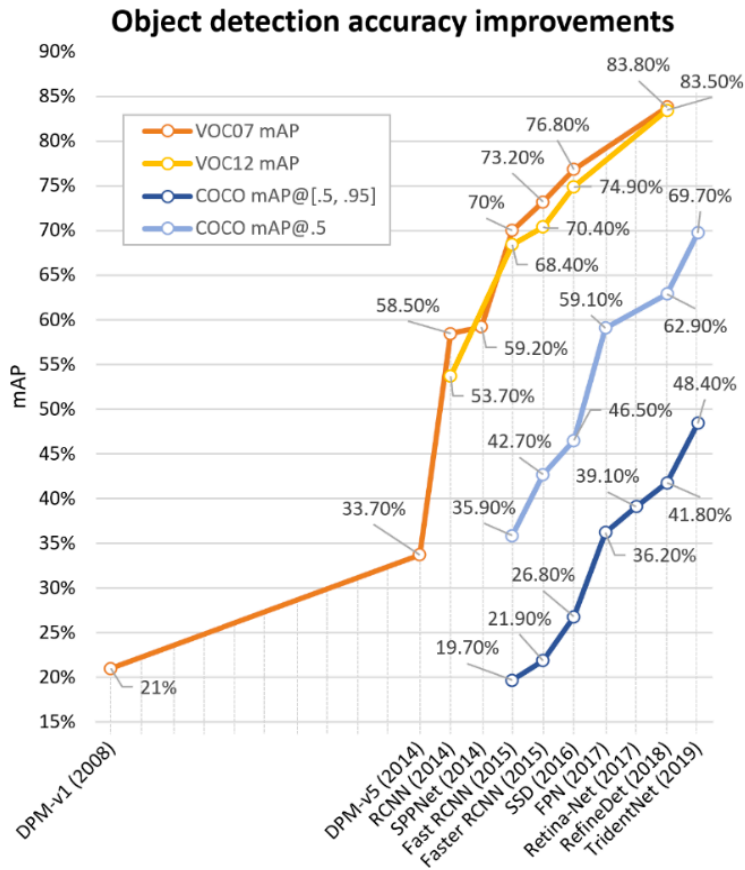

Figure 17: Object detection accuracy improvements [33]

In summary, current AI/ML based sensors are likely to need other architectural measures such as different forms of diversity to improve them to acceptable levels. Our approach aims to support the assurance of an architecture that limits reliance on sub-components of the system that need to be highly trusted (e.g., ML algorithms). In Figure 18 we have adapted the safety monitor architecture of [28] to include both a safety monitor and a complex function monitor, implemented for an AI/ML based system (note that we use the term AI/ML rather than the term Learning Enabled Component (LEC) used in [28]).

Assurance of the AI/ML based sensor presents challenges because of the very nature of the technology and the complex tool chains used to develop it (we discuss this further in Section 8). As with conventional software components, evidence can come from dynamic analysis, static analysis and formal verification.

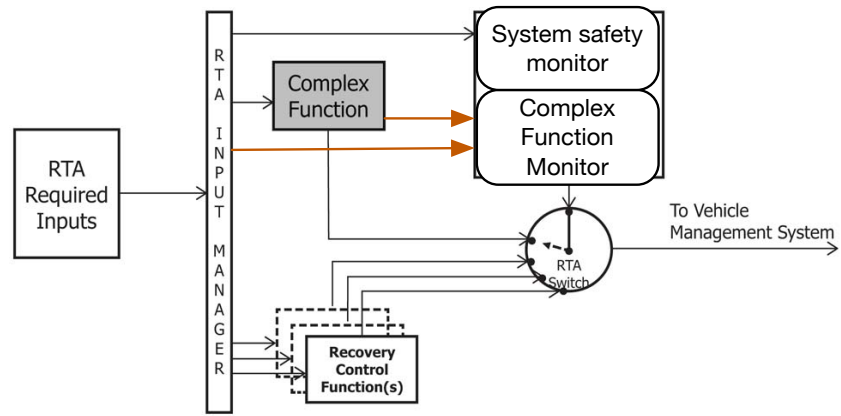

**Figure 18: Safety monitor architecture**

The recovery functions must address a number of design challenges. A key design challenge for an AI/ML monitor is illustrated in Figure 19.

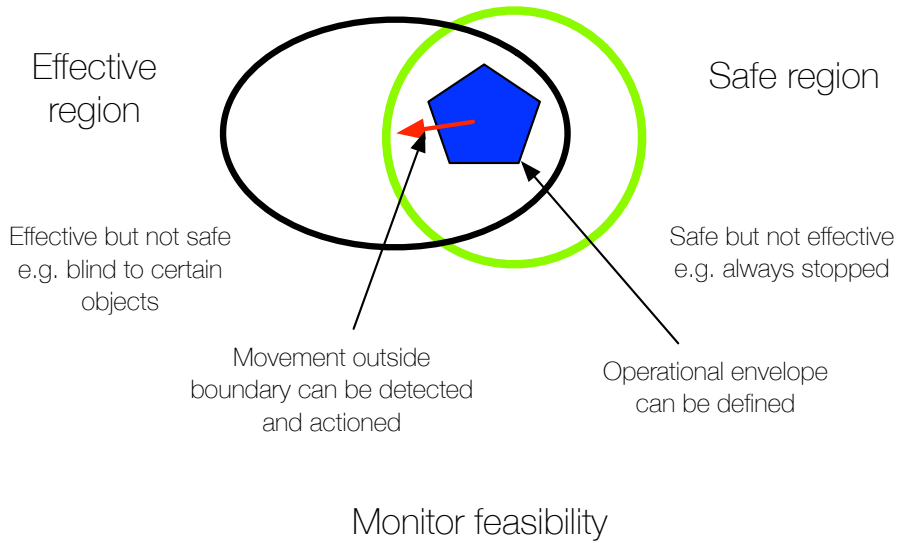

**Figure 19: Monitor feasibility**

During the design of a monitor architecture, a number of questions must be addressed:

• is a monitor feasible – can a region, the operational envelope in Figure 19, be identified that is both safe and effective, and can this be described in terms that can be assured more readily and to greater extent than the AI/ML itself?
• can transgression of the monitor region be detected and actioned upon by a recovery control function?

A number of approaches can then be taken, which we characterise as:

• *Environment Monitors:* these monitor an ML system's input space where there are known performance issues – e.g., bad weather, flying upside down, etc.

- *Health Monitors:* these monitor the ML system's internal state and identify states that might be "stressed" or indicative of a problem, (e.g., monitoring activation patterns, resource utilisation, simple tasks for which diverse measurement).
- *Behaviour Monitors:* these monitor the ML system's outputs and inputs to see if they violate bounds on specified behaviours or invariants.

Finally, the recovery strategies are very application specific. They may involve:

- Use of other sensors in the case that the ML system has degraded performance, but will still allow for safe behaviour (e.g., moving to a minimum risk position or reduced performance while recovery is planned).
- If all sensors fail, have an "eyes closed" safety strategy. For example, when solving for reachable sets in which the autonomous vehicle and other vehicles maintain safe separation and safety despite unobservability, a simple approach would be to slow down and stop.

In practice, the architectures could be far more complex than shown in Figure 19. There is also the inverted architecture where AI/ML is used to learn the difficult tasks (e.g. how to conduct difficult and dangerous manoeuvres in an aircraft) and has the authority to take over from the more conventional operation: this might be as some sort of super-operator. In this case the monitors could become nested with an AI/ML monitor of a conventional monitor of an AI/ML based sensor.

### 7.2.1.2  Limitations on the monitor/sensor approach

In the previous section we discussed the feasibility of defining a suitable monitor functional specification, and in particular a safe and effective operational envelope. A further challenge to developing a monitor specification is that in some cases, the outputs of an ML component can change significantly with only a minor change in input. Such changes are often erroneous and indicate that the output of the ML component is uncertain. However, if all outputs remain within the operational envelope, such instability in the outputs may not be detected and acted upon by the monitor.

It may also be the case that the associated performance requirements are hard to achieve, and that the approach cannot achieve the performance required for autonomous systems. A discussion of the performance of current ML sensors is provided in Section 9, which provides some indication of the limitations of monitors.

A sensor containing an ML component will typically be used in conjunction with other sensors (possibly also containing ML components) to produce a model of the environment of an AV. Where it is not possible to assure that an individual sensor/monitor achieves sufficient performance or stability, an alternative and complementary approach to the sensor/monitor architecture is to use model based reasoning that requires high confidence in the output of the resulting model, rather than in any individual sensor.

One such approach is the predictive processing proposed in [71]. The outputs provided by the sensors are compared with predicted sensor outputs based on the model of the environment. If all sensors indicate high confidence in their outputs, and these outputs only differ slightly from those expected based on the model, then we have high confidence that the model is an accurate representation of the environment.

However, if there is a large error between the prediction of the model and the output of a sensor, then this must be resolved. If the error is only present in one sensor, or one group of related sensors, e.g. all front-facing cameras, then it may be determined that this is a sensor error, and the model can be updated based on the data from other sensors. If such errors persist, then it may be due to some external cause, e.g. the cameras are being dazzled by the sun. Finally, if a large number of sensors provide a prediction error, then it is likely that the world has not developed as predicted by the model, e.g. a new object has appeared, or an object has moved in an unexpected way. In this case, the model must be updated to reflect the new data. The precise approach to handling unexpected sensor inputs will depend on the reliability and failure modes of the individual sensors, e.g. an unexpected sensor output from an ML component may be considered a sensor error, while an unexpected output from a conventional sensor with substantial operating experience may lead to refinement of the model.

Using such a predictive processing model presents a possible approach to arguing that an autonomous system's interpretation of its environment is accurate with a sufficiently high degree of confidence, without requiring a potentially infeasible level of performance from an individual sensor. If a sufficiently high level of accuracy, e.g. measured by precision and recall, can be achieved, then these can support a higher level of reliability in the model. For example, if five independent sensors each have precision and recall of 0.9, then the "majority verdict" has precision and recall of 0.991.

Demonstrating the reliability of the model in this way is similar to increasing performance using ensembles. In particular, we require arguments and evidence that the failure of sensors is independent or, as is more likely, some bounds on the dependence between them. There could be many causes of systemic failures, such as incomplete training data, or sensors having reduced performance in the rain. The impact of such systemic failures on the performance of the model will depend on the precise nature of the failure. Reduced performance in rain will affect the accuracy of the sensor data, however the errors are unlikely to be consistent, and these errors are likely to be identified. On the other hand, if an object is not seen in the training data, then the sensors may consistently fail to identify it, leading to an incorrect model. To support a claim for independence of the sensors, we would require evidence identifying any potential sources of systemic errors, and arguments as to why they do not occur.

However, the predictive processing model also allows for additional reasoning about the model itself, which can increase our confidence further. One such method is to keep track of any uncertainties in the model, which can then be resolved by further observations. For example, if an object is identified but the sensors cannot determine whether the object is a bicycle or a pedestrian, this uncertainty can be maintained within the model (or as multiple models) until the object can be identified from further observations or a safe action must be performed (e.g. brake). Assurance of such multiple models would likely take the form of showing that the true state of the world is represented within the space of possible models with sufficient confidence, and that the actions planned and taken by an AV are acceptably safe in all possible models.

The model would also be able to track objects which are partially or totally obscured. This includes both previously detected objects which have since become hidden, and potential objects in areas which are not visible to the sensors, e.g. if a parked vehicle is blocking the view of the road. Doing so would be necessary to ensure the accuracy of the model. Some work on detecting and representing the properties of objects and their relative positions in an image, and answering queries regarding these, has been performed in [72]. Using a combination of YOLO and answer-set programming (ASP), this algorithm was able to answer 93.7% of queries correctly. The vast majority of the errors were caused by incorrect object detection with YOLO, and we assume that the small number of errors arising from incorrect parsing of the natural language queried could be eliminated. It is not clear how well this performance will translate to the objects of interest to AVs, but it is plausible that ASP could be used to develop systems that can reliably reason about the relative positions of objects in a model.

The challenges and limitations of a monitor will vary according to the sensor, and the types of error not detected by the monitor must be understood. Even where it is not feasible to provide sufficient assurance in an individual sensor/monitor, the sensor/monitor architecture may be used in conjunction with predictive processing or other model based reasoning to provide sufficient confidence in the system as a whole. The implementation of such reasoning will depend heavily on the design and application of the system, and may itself include ML components, e.g. for the predicting the future positions of objects.

## 7.3   Outline of structure and reasoning

We have chosen an example in which a sensor containing AI/ML is part of a monitor architecture subsystem. As noted previously, in theory, the monitor could contain a number of guards which themselves could include AI/ML. Further, AI/ML could also be used with a planner and in the fusion of multiple sensor components.

However, in this example we limit the AI/ML to just the sensor component; as if it was also used in the guards, it would add to the complexity of the guards and may make the guards' behaviour unverifiable. It also limits some of the traditional system engineering techniques that can be used to build confidence and provide assurance in the guards. We will require more confidence and a higher degree of assurance in the guards compared to the AI/ML sensor, so avoiding AI/ML in the guards themselves is a reasonable design choice.

Our high level CAE template structure of the architecture is shown in Figure 20.

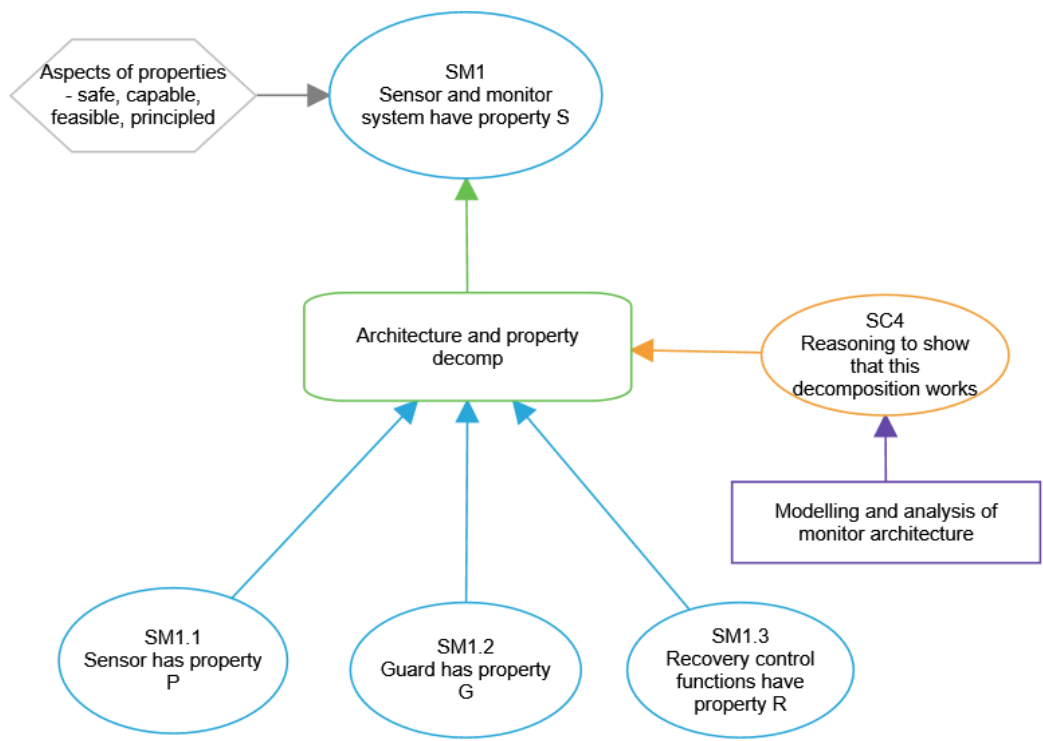

**Figure 20: Monitor - high level claim structure**

As with the other templates, there is nothing remarkable in the *structure* of the case but in the reasoning needed to support the argument, located in the side claim (SC4). The reasoning is important since it justifies that the system property is satisfied by the combined behaviour of the guard, sensor and recovery functions.

The sensor part of the case is based on a generic approach to component cases, and will be discussed in detail in a separate template (see Section 8). The generic component case highlights the need to consider how the sensor case can enable and impact other parts of the case, such as the monitor architecture. Some of these interactions will be dealt with at a component level and some at a system level/architecture level (such as future behaviour of the sensor, see Section 10).

The guards (SM1.2) must be able to detect possible transgressions to outside the permitted operational envelope (viability regions), and the monitor system must implement an effective recovery action (SM1.3) at this time to bring the system back into the permitted region. A balance is needed, as if the recovery action threshold is set too low, the recovery action may be active too often reducing confidence in the stability of the system. This could have other negative impact such as constraining the ML too much, making it less effective.

An analysis is needed of the architectural and functional separation between the sensor and that of the guard, so that the subsystem is not susceptible to Common Cause Failure (CCF). For example, if the sensor and guard rely on a similar input data stream. Some of this argument is captured in the recovery control claim and the main subcomponent side claim, but also within the architecture of the guard that we describe next.

Figure 21 provides a partial expansion of our template, decomposing the guard claim (SM1.2) into our three different types: environment (EG1), health (HG1) and behaviour (BG1) and we now discuss these in turn. When using this template sub-arguments for their feasibility, implementation and requirements will have to be addressed individually. For example, the confidence measure for the system being stressed is likely to be more complex than using a photocell sensor to detect that you are not flying a drone in bad lighting conditions.

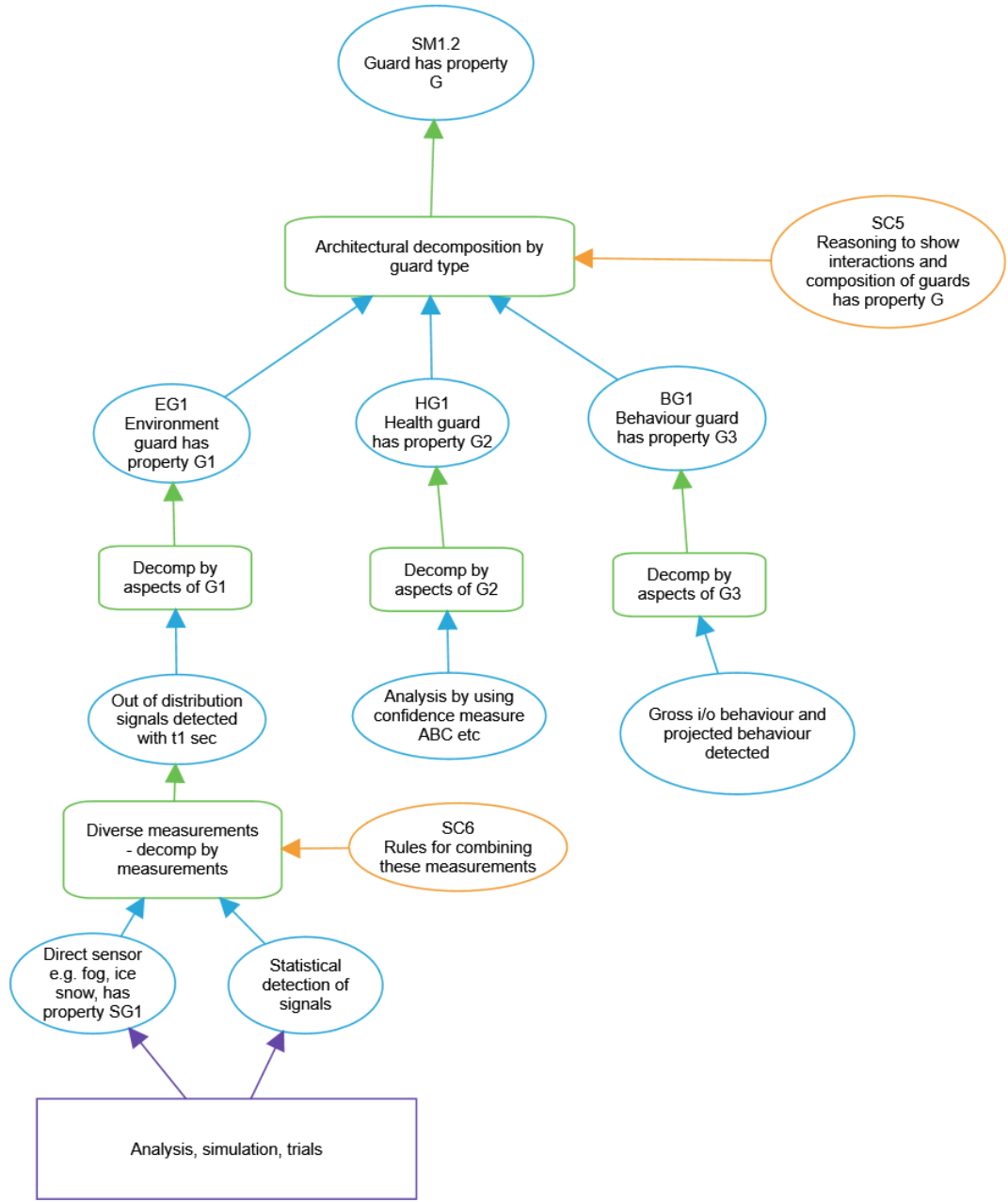

Figure 21: Guard CAE fragment

### 7.3.1   Environmental guard

We have presented an example of using the environmental guard where claim EG1 is expanded to look for out of distribution signals in Figure 21 to detect

- gross differences in the environment
- subtle differences in the environment that require analysis from multiple inputs to determine

In our example expansion we assume there would be a time constraint on the guards, requiring response within a certain time frame in order to be up to date and useful. This would require composition of diverse measurement types within that time frame. Each of these can be assessed individually if rules for composition are justified (SC6).

We are expecting gross differences in the environment to be measurable directly from sensors. Some of the guards' checks could be performed at start up and then periodically throughout use[7].

Monitors detecting subtle environmental differences or stresses in the model are potentially more challenging than those monitoring gross differences. Confidence measures for ML models are an active area of research. A recent study comparing confidence measures applicable to the problem of determining depth from stereo images identified over 50 different confidence measures [42].

With such a wide variety of confidence measures available, the outputs of which may not be directly comparable, any confidence measures implemented as environmental or health guards must be well understood to determine what assurance they can provide. For example, a measure of confidence based on the similarity of the input to training data may not provide suitable monitoring of the health of the system, and vice versa. We consider two examples of confidence measures here to demonstrate this variety.

One possible monitoring mechanism would be to build a confidence measure into the ML algorithm. Bayesian methods can be applied to many ML models, providing an output as a probability distribution. However, the computational resources required for both training and active inference mean that this is currently infeasible for all but the smallest ML models.

Confidence can be learned explicitly in neural networks by requiring that the model to predict both an output and a confidence value $c$. When training, rather than computing the loss function of a given prediction $P$ for a ground truth $T$, the prediction is first adjusted according to the confidence $c$, i.e., the loss function is computed for $cP + (1-c)T$. This shift of the prediction towards the ground truth proportional to the confidence encourages low confidence when the prediction is likely to be inaccurate. A term is also added to the loss function to provide a penalty for such low confidence scores.

Confidence learned in this fashion can be used as an effective environment monitor, detecting when the system is operating outside of trained situations [43]. However, the confidence value $c$ is not a probability, and so testing would be required to compute a suitable threshold defining a safe region. Any change in the model will require recalibration of this threshold; this may be impractical for some applications.

The assurance case claims would need to reflect the safety performance which is actually provided by the technology, as well as the guard design and implementation.

---

[7] An example could be an autonomous vehicle that used a different ML model at night or during foggy conditions. The model could be selected at start up based on the initial environmental conditions and if needed switched over when an ensemble architecture is used.

### 7.3.2   Health guard

Computing confidence as part of the ML prediction would be unlikely to provide sufficient diversity for a monitor on its own. An additional guard would therefore be required to monitor the state of the ML model itself (claim HG1).

Attribution-based confidence [44] is an alternative approach to monitoring, providing a health guard which does not require any changes or additions to the ML. Instead, confidence in a prediction is measured by the stability of the prediction under small changes to the input data (perturbation). If small changes to the input produce a significant change in the output of an ML model, this is an indication that the system is easily stressed and this reduces confidence in the resilience of the output of the model.

With the potential for a very large number of input features, adequately sampling the entire region around one input may be infeasible, particularly given the likely response time requirements of the ML component in an AV. To obtain a feasible confidence measure, attribution-based confidence reduces the number of samples needed by focusing on changes to the most important features. In this case, the importance of the $j$th feature for an input $X$ is measured by an attribution score $A(j,X)$, computed by comparing the output for $X$ to the output on a baseline input $B$.

For a given input $D$, attribution-based confidence randomly generates $n$ different inputs by modifying the $j$th feature of $D$ with probability $p(j)$, where $p(j)$ is proportional to the attribution $A(j,D)$. These inputs can then be used to provide the confidence in a predicted class, computed as the proportion of the $n$ inputs that agree with the prediction for $D$. Since the random sampling is more likely to produce inputs which modify features with a greater impact, this confidence value can be considered as a lower bound for confidence in the prediction.

Attribution-based confidence does not rely on the details of the ML, essentially treating it as a black box. It can therefore be used with a wide variety of sensors, including in systems where the ML model is regularly updated. Evidence may still be needed to validate the confidence given, and any assumption about the ML model required must be clearly identified.

In contrast to the learnt confidence above, attribution-based confidence could be used as a health guard, but is not suitable as an environment guard, since similar out-of-sample inputs could well result in the same prediction. Attribution-based confidence should therefore only be used in conjunction with other guards. More details of confidence measures are provided in Appendix B.

Again, claims based on the health guard would need to reflect any limits in the technology, and potentially convert complex confidence measures into safety performance measures. Additionally, there would need to be claims about the guard's implementation.

### 7.3.3   Behaviour guard

A behavioural guard (BG1) is not feasible for all system behaviours, particularly complex behaviours where an operational envelope cannot be easily defined. It is also possible that some system behaviours are acceptable in some circumstances but not in others. An example would be the maximum allowed speed of the vehicle in particular weather conditions. The maximum allowed speed may need to be adjusted to a lower limit in bad weather.

System input or output behaviour is one area which can be guarded. It is likely that developers will know permitted ranges of values for system variables so gross differences with "normal" values or erroneous/invalid values should be detectable by the system. The effectiveness of this guard relies on knowing what behaviour of the system is "normal" in certain circumstances, some of which can come from system requirements or simulation.

Behaviour guards may be less applicable at the lower abstraction level of AI/ML components as we tend towards monitoring the direct outputs of AI/ML algorithms, which can be less predictable and normally non-monotonic. Therefore, the operational envelope is difficult to determine and small changes in the inputs may have a large effect on the output. In this situation a full behavioural monitor may not be practical given the constraints, but by monitoring the outputs of the AI/ML algorithm it is possible to identify some issues (mostly related to some form of bias, such as data skews or model staleness [45]):

- incorrect training data distribution – behaviour is skewed to a particular output which does not match the real world application
- irrelevant features in model – some features may be irrelevant or not prominent in the real world.
- data interdependencies – often AI/ML algorithms are one piece of a "data pipeline" the full data pipeline may have not been used during training, and therefore, there could be unknown data interdependencies between other components
- model no longer fit for purpose – shifts/new trends in the environment, changes in behaviour of other parties such as a change in adversarial tactics

## 7.4   Recovery actions

In Figure 20, we have a high level claim about recovery actions (SM1.3). There are many different recovery actions possible and some will be specific to the application of the system, or even the system itself. Recovery actions may occur on different levels depending on the severity of the issue and the speed in which the actions are required. We have highlighted some possible examples below:

- course correction – temporarily overriding the control algorithm of the vehicle
- degraded operation – e.g. reducing speed and capability
- halting current operation – temporarily entering a recoverable pre-defined safe state
- aborting – returning to starting point or entering a non-recoverable safe failure state

In our monitor example, the recovery actions are implemented using conventional software, as we can use traditional system engineering techniques to build confidence and provide assurance. Since traditional systems engineering techniques can be used to show that the recovery functions have property "R", we have not developed this part of the case further.

## 7.5   Sources of evidence

For the monitor guards, traditional sources of evidence should be available, such as design documentation, requirements specifications, testing plans and reports. These will support claims that the guards have been designed and successfully implemented.

Evidence for the effectiveness of the guards will depend on the type of guard. Different strategies will be required as they are protecting against different types of anomalies. In Figure 21, we have decomposed the argument for the environmental guard for out of sample detection. As discussed, there are two types of out of sample protections we are attempting to detect:

- gross differences in the environment
- subtle differences that require multiple inputs to determine

Evidence from testing (such as type testing or component testing on a test bed) is likely to be easier for gross differences in the environment, where the viability region is well-defined (black and white) and the conventional analysis of sensor inputs can be used.

If a guard is deployed based on statistical analysis of data, such as a behaviour or health guard monitoring a complex function/algorithm of the system, evidence sourced from testing should be available, however this should also be supported by additional sources, such as performance in trials and simulation for stronger overall assurance.

For a guard based on the detection of subtle differences away from a baseline, such as a behaviour or health guard, there are some confidence measures that are mathematically grounded with formal proofs of the confidence they provide for predictions. However, these proofs will make certain assumptions, and we would expect evidence to verify that these assumptions hold. In most cases, evidence for confidence measures could be obtained through testing and simulation. Ensuring adequate coverage in these tests will probably require diverse, and possibly more extensive, data than used for validation of the ML model.

The separate pattern being developed for an ML based sensor in Section 8 discusses evidence sources for the ML based sensor.

## 7.6 Defeaters for the monitor architecture

We expect defeaters to be present at multiple levels of the pattern from the individual guard types to the monitor architecture design itself. There will be defeaters associated with the traditional systems and software engineering approaches and others specific to AI/ML technologies. Some examples of defeaters are documented in the table below, which were derived from the Defeater Workshops we held [20].

| Description | Part of monitor pattern | Possible mitigations |
|---|---|---|
| Implementation of intended design is not correct. Possible exacerbated by use of libraries, COTS. | All (sensor, guard, recovery action). | Increasing requirements for verification and validation evidence. |
| Guard confidence measures not validated sufficiently | Guard. | Perform independent review and analysis to ensure validity. |
| Performance requirements on guards not met (e.g. sometimes takes too long to complete checks, recover). | Guard. | Reduce functional complexity of guards. Alerts when system is struggling to meet the performance goals. |
| Security and safety interaction not considered. | All. | Adopt security informed safety approach (see PAS 11281). |
| Overzealous guards lead to unsafe behaviour in wider system. Or lack of sensitive guards lead to unsafe behaviours. | Guard/recovery action. | Overarching system level analysis of guard requirements. Analysis of trade-offs between false positives and false negatives. Analysis and simulation of wider system as part of early design validation activities. |

| Description | Part of monitor pattern | Possible mitigations |
|---|---|---|
| Operating out of permitted operational envelope not detectable/detected. | Guard/recovery action. | Well-defined operating requirements, testing. Operational restrictions. Make an explicit part of case to detect out of envelope (see Section 7.2.1.1). |
| AI/ML guard functional behaviour not fully verifiable. | Guard. | Restrict design to verifiable ML algorithms in guards. Use reliability rather correctness arguments. |
| AI/ML guard functional behaviour too complex in practice. | Guard. | Simplify guards and place restrictions on operation. |
| Not enough of diversity/independence in sensor and guard. Common cause issues, (e.g. due to external common systems GPS or due to sensors finding similar situations difficult). | Architecture level. | Functional diversity – use different type of input data provides some defence. Architectural diversity – different computer system for guards. Justify a level of dependence and use a confidence evaluation that takes this into account. |
| Architecture sensitive to complex failures, e.g. dataflow between sensor and guard vulnerable to Byzantine failures. | Architecture level. | Adopt appropriate explicit fault models, validate these and engineer architecture accordingly. |
| Evidence of false negatives undermines confidence in guard | Guard. | Improve guard or perform additional testing to increase knowledge of performance limitations. |
| Testing does not fully cover viability region | Guard | Perform additional testing. |
| Environmental model does not have an adequate level of fidelity to test the environmental guards | Guard | Improve environmental model Perform independent review and analysis of the model |

In addition, there were issues that are out of scope of this pattern but are important:

| Description | Part of monitor pattern | Possible mitigations |
| --- | --- | --- |
| Adversarial/threat model not defined or appropriate. | Guard/recovery action. | Security informed safety not within the scope of this guidance. |
| Human aspects of taking ownership when guard/monitor fail or human as one of the guards not addressed. | - | Human aspects important but outside scope of this pattern. |
| Monitor becomes out of date as sensors adapt, learnt or degrade or as operating environment changes. | - | See Section 10.2.3 for adaptation and changes to the operating environment. |
| Issues of completeness of requirements, how they deal with change. | - | See Section 5 for requirements issues for ML components. |

**Table 4: Monitor architecture defeaters examples**

The defeaters identified illustrate the range of possible mitigations:

- increase amount of evidence - greater quantities and more rigorous
- reduce design complexity - for alerts, algorithms, etc.
- constrain the claims made, e.g. using a greater precision in the requirements
- changing or detailing the arguments being made (reliability vs correctness)

### 7.6.1 Incorporating defeaters into the monitors template - justifying the composition of the guards

In each application of the template there will be a need to systematically identify and assess defeaters (a detailed process for this is an ongoing research issue) and decide how to incorporate into the CAE structure. The side claim in Figure 21 justifying the composition of the guards is substantiated by incorporating evidence from modelling and analysis.

Justifying this evidence incorporation could lead to the detailed reasoning that uses the confidence pattern to record doubts. This is shown in Figure 22, where the challenge to the case has resulted in more detail in the transition between the model results and the claim (addressing doubt about model validity) as well as the use of the confidence/doubt pattern to record other defeaters concerning the use of confidence measures and model fidelity.

In addition, defeaters on confidence measures could evolve around whether the case has sufficiently addressed the possibility of adversarial inputs designed to throw the confidence value of output out of the allowed range and causing the system to respond or enter a degraded state of operation intentionally. This could represent a denial of service attack on the systems availability.

Confidence is dealt with qualitatively and informally in the example: extensions might consider greater technical treatment.

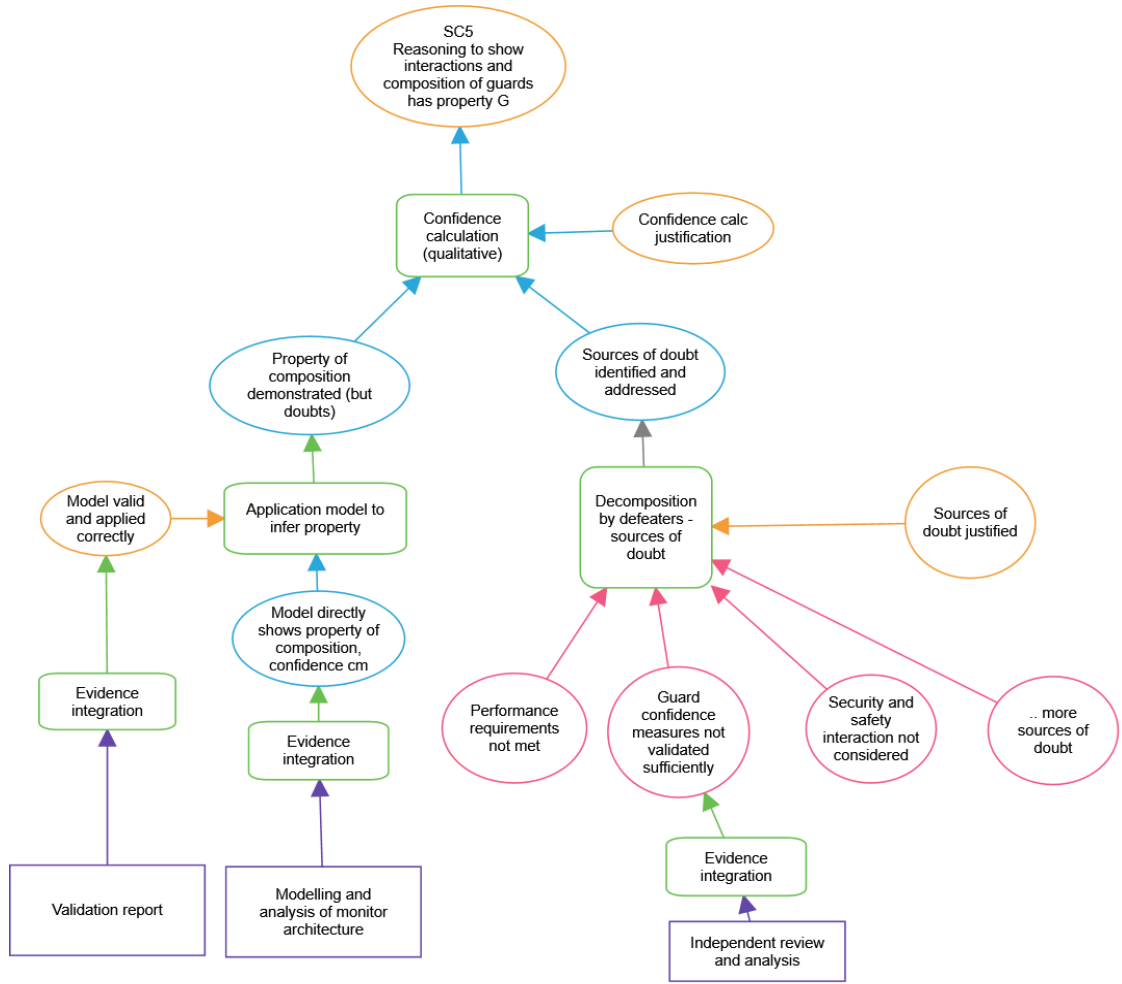

**Figure 22: Expanding the side claim justifying the composition of the guards**

Figure 22 identifies some possible evidence to support the claims, as does Table 4 (for reasons of space the full expansion is not shown). The integration of this evidence might be challenged and even more detailed reasoning required in the CAE template or a subcase developed to detail the argument.

Figure 22 also shows how we can readily identify the side claims and defeaters without evidence and decide whether they become assertions and assumptions in the wider case or whether we seek to justify them.

In expanding the argument, we can address the role of trials and simulation of the system; therefore, evidence from this testing that is used to support the guards will need to have possible defeaters addressed, since there are some subtle nuances in the setup of trials and simulations.

However, some majority are not specific to the guards themselves but to the trials and simulations in general such as difficulty in setting up test oracles. This would need to be addressed in the assurance case (see section 9).

### 7.6.2   Impact on the monitor CAE template

The evolution of the CAE to incorporate the mini-patterns for the guards and defeaters is shown in Figure 23 (this connects with the pattern in Figure 22).

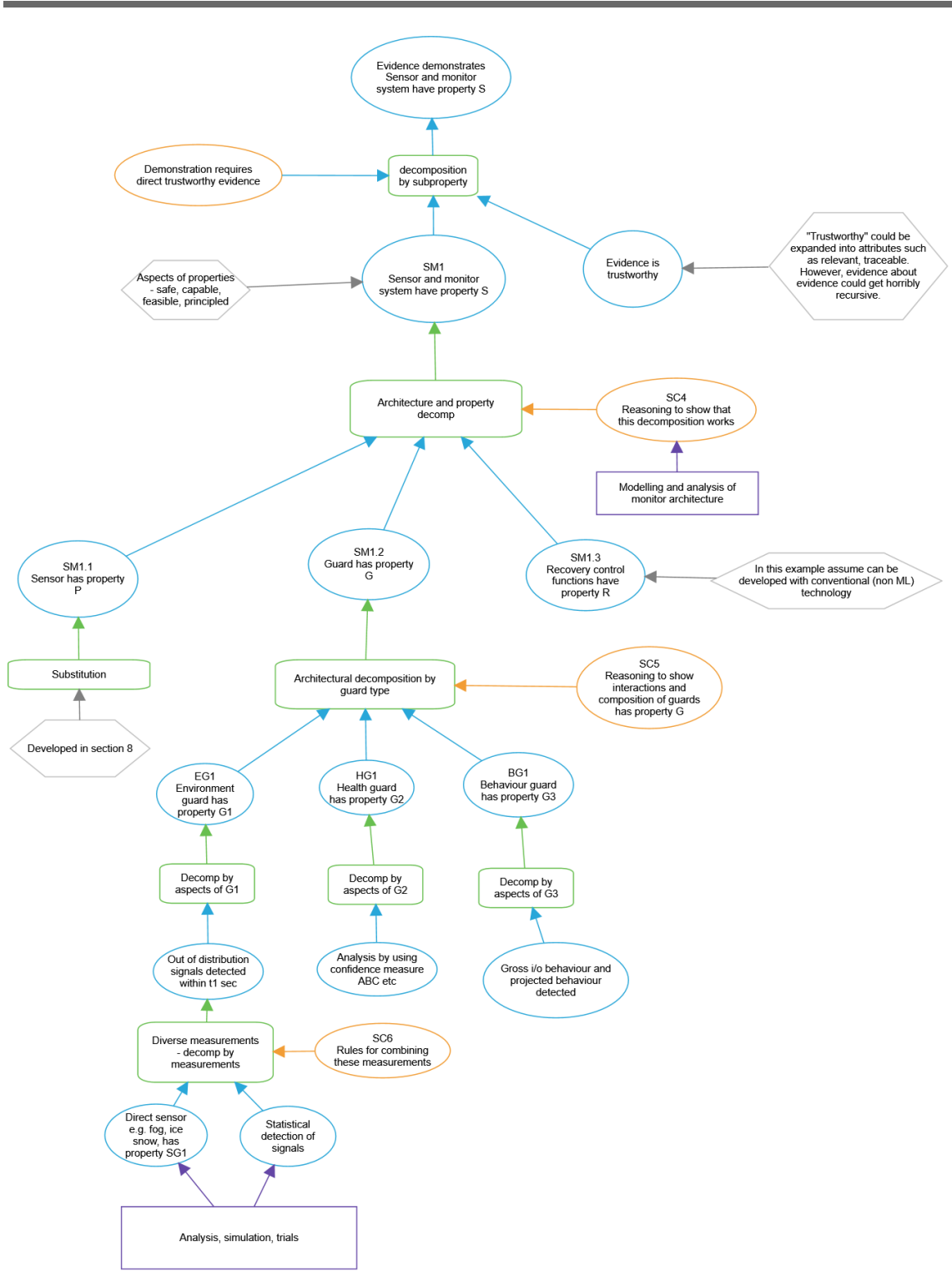

**Figure 23: Evolution of template**

## 7.7 Deploying the monitor template

In this section, we have provided an example of deploying a monitor based architecture to a system with a sensor based component containing AI/ML. The monitor acts as a complex function monitor, checking the functions of the sensor are working correctly, using guards.

In the template we have described three different types of guards (environment, health and behaviour) and how they can be deployed within the context of our example system. The template could be extended to many different subsystems where their functionality or behaviour is difficult to verify by traditional means and use a monitor/guard architecture.

Some of the key challenges for deploying this template are:

- the dependence on upstream for requirements and other patterns for the sensor part
- the overall feasibility of successfully justifying a monitor architecture including
  - definition of permitted operational region and one that is both safe and secure
  - complex system behaviour, such as AI/ML
  - additional design, validation and verification of guards
  - additional performance pressures on subsystems with guards

# 8 ML based component – sensor

## 8.1 Objectives of template and context

In this section we develop the CAE structure to address the top level claim for a ML based sensor. The safety analysis will identify a range of functional and non-functional requirements for the components of the system. These requirements will describe the set of safety properties the sensor needs to have. We capture this in the abstract claim that "Sensor has property P". This template links to the monitor architecture template detailed in Section 7 (see Figure 23) and can also be used elsewhere in the case such as if there are other sensors without guards that are used or ML is used within a planner function.

In this template, we focus on the ML aspects of the justification of the sensor, as a non-ML related sensor can be addressed through traditional engineering approaches. In our example, the sensor will be performing classification of objects from a camera feed, and the ML will be trained using supervised learning from a curated data set (see Section 5.3 for a discussion on requirements for training sets). To have confidence in the sensor we have to address, inter alia, the suitability of the algorithms, and the extent and quality of the data used to train and validate the ML component. We need to address the competency of those involved and the engineering process followed, and the trust in tools and platforms so we can trust the produced evidence. We combine all of this with confidence coming from both simulated and real trials.

In keeping with our methodology, we need to identify defeaters and other issues that may undermine the safety case. In Section we describe the CAE in the template. In Section 8.3 we summarise potential defeaters. In Section 8.4 we consider the application of the template.

Furthermore, in Section 9 we consider the difficult issue of justifying the performance of the sensor, particularly its reliability. We factor the justification to show the key evidence for supporting a judgment of reliability and make a link to Appendix C.3 where we discuss this issue in detail.

## 8.2 The CAE structure for the ML based sensor component

The guard architecture assumes that there is a sensor or other ML enabled component as part of the subsystem architecture. The first part of the sensor template structure, shown in Figure 24, selects the ML delivered aspects of the sensor's functionality and performance from the other aspects e.g. environmental requirements, and failure behaviour when hardware fails.

This allows us to separate out the more conventional issues, such as application specific requirements (electrical, temperature, humidity, etc.), which can be assured in more traditional ways such as type testing or hardware bench testing.

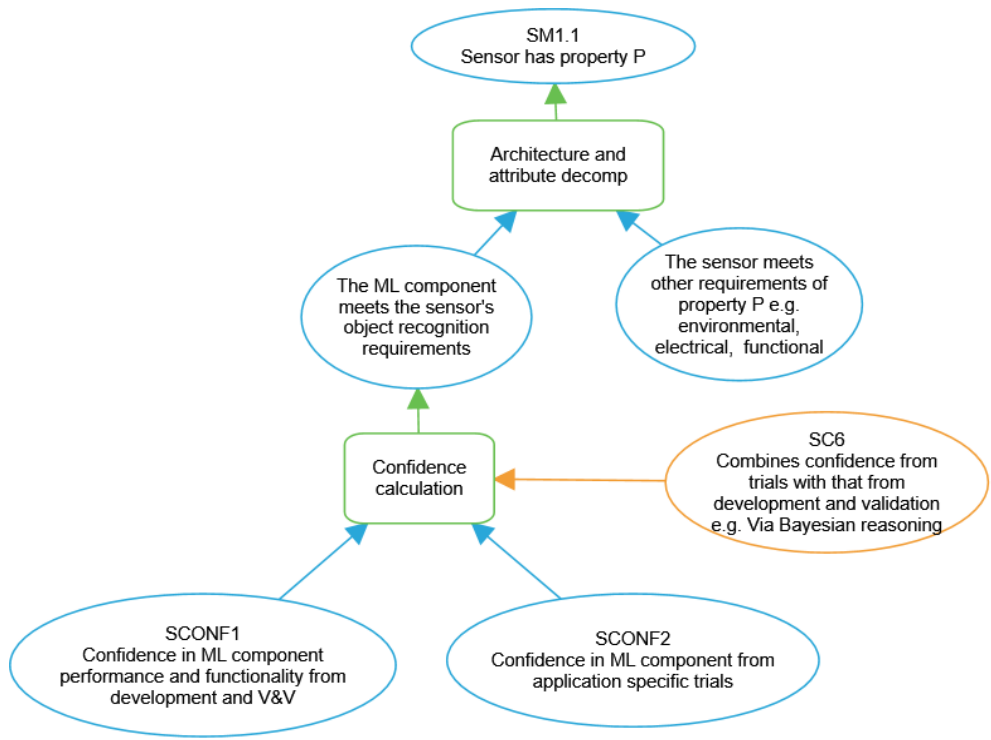

**Figure 24: Sensor case based on development and trials**

Demonstrating the ML component meets its object recognition[8] requirements (property P in SM1.1) is then argued from its observed behaviour and performance, during real and simulated trials and from the development lifecycle. The latter might provide us with a prior belief in the sensor's behaviour that is then updated/confirmed with evidence collected during the trials. The argument to support this reasoning is presented in the combined confidence side claim.

The side claim (SC6) is extremely important and technically challenging. An approach using conservative Bayesian inference methods to obtain confidence in the failure rate of AVs based on what testing has been performed and the prior confidence in the failure rate is described in Section C.3.3.

### 8.2.1    Confidence from real and simulated trials

We first develop the right-hand branch (SCONF2) of Figure 24 that deals with the confidence from application specific trials, both in real and simulated environment.

---

[8] Different properties, e.g. other forms of classification can also be argued in the same way.

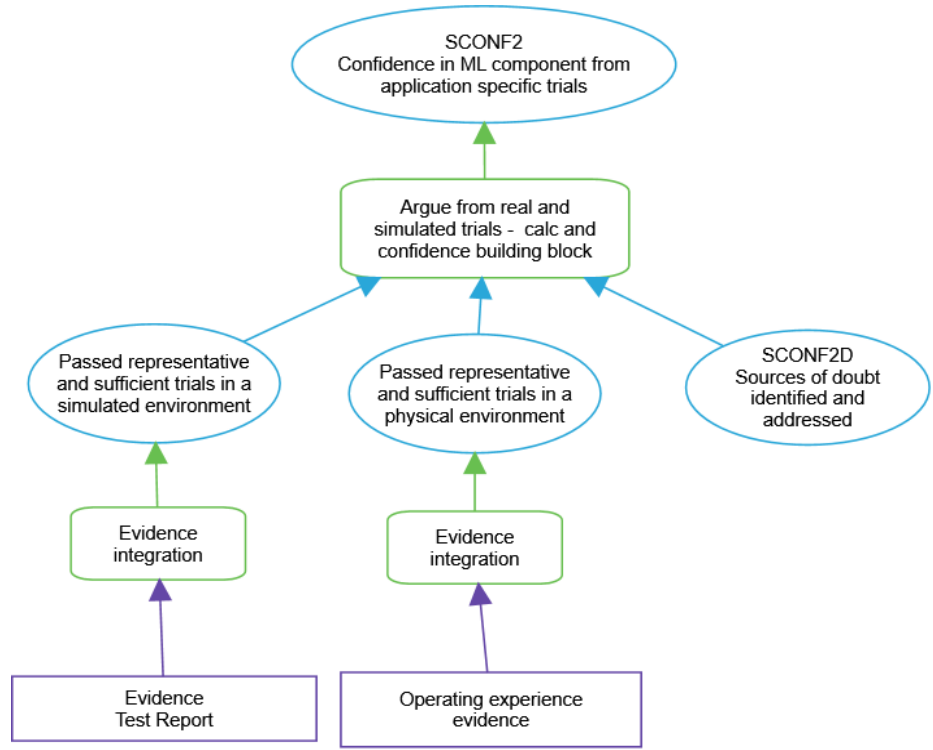

**Figure 25: Real and simulated trials**

The template could be expanded to include a hierarchy of different real/simulated tests. In [39] we summarised the pros and cons of different combinations of virtual and real-world simulation and this is repeated in Table 5. In practical terms it may be desirable to use different types of testing and simulation at different stages of ML development. This would be dependent on the risk associated with the system, as that would inform the amount of evidence required to demonstrate adequate safety.

| Environment | ML | Strengths | Weaknesses |
|---|---|---|---|
| Virtual | Virtual | Can control and model many different environment options, which may be hard to replicate in real world testing.<br><br>Can create accident sequences to test corner cases without risk of accident.<br><br>Potentially cheap and quick.<br><br>Can do early in lifecycle to assess performance.<br><br>Can monitor every aspect of performance.<br><br>Can use for reinforcement learning.<br><br>Potentially strong repeatability.<br><br>Easier to detect how/where faults occurred with monitoring. | Unrealistic input data e.g., computer generated environment[9] or modelled sensor functionality which may not match the resolution and real-time performance of a real sensor.<br><br>Extensive computer resources will be required to achieve the performance needed for adequate modelling and collecting data e.g., in terms of processing power and fast access memory.<br><br>ML may not perform this way in real life.<br><br>Hard to involve user if needed.<br><br>Potentially unrepresentative results (e.g., no feedback from bumpy surface, compromise of equipment from wet surface, temperature changes). |
| Virtual/Artificial | On target hardware (Hardware In Loop (HIL)) | Can control many different environment options.<br><br>Can create accident sequences with very limited or no risk.<br><br>Can involve end user.<br><br>Gain trust in ML hardware.<br><br>Potentially strong repeatability.<br><br>Easier to detect how/where faults occurred with monitoring. | Unrealistic input data – ML may be real but some of the input data may not be realistic e.g., if working in a room with lots of monitors.<br><br>Computing power required may be large.<br><br>Outputs may be more realistic but still constrained by environment (e.g., no actual movement or slower/faster responses).<br><br>Users may not behave as they would in real environment or may have simulation sickness [62]. |

---

[9] Consider the situation where the simulation provides conflicting and unrealistic sensor data e.g., blocky low-resolution models, moving trees and unrealistically fast pedestrians [61]. Whilst it might be useful for the ML to identify this as invalid input data, if used for training the system, care is needed not to reinforce invalid behaviour.

| Environment | ML | Strengths | Weaknesses |
|---|---|---|---|
| Real world but controlled e.g., test track | On target hardware. | Input data is real and may contain unanticipated events.<br><br>Can get useful feedback on performance with low risk to third party.<br><br>Can involve users if needed. | Less control over the environment.<br><br>Much harder to repeat results.<br><br>Harder to detect how/where faults occurred. |
| Real world trials | On target hardware. | Input data is real and may contain unanticipated events.<br><br>Can involve users if needed. | No control over environment.<br><br>Riskier to third parties depending on mitigations in place.<br><br>Hard to repeat results.<br><br>Hard to detect how/where faults occurred. |

**Table 5: Simulation variants and their strengths and weaknesses**

Table 5 informs the detail of the sources of doubt and some examples are shown in Figure 26.

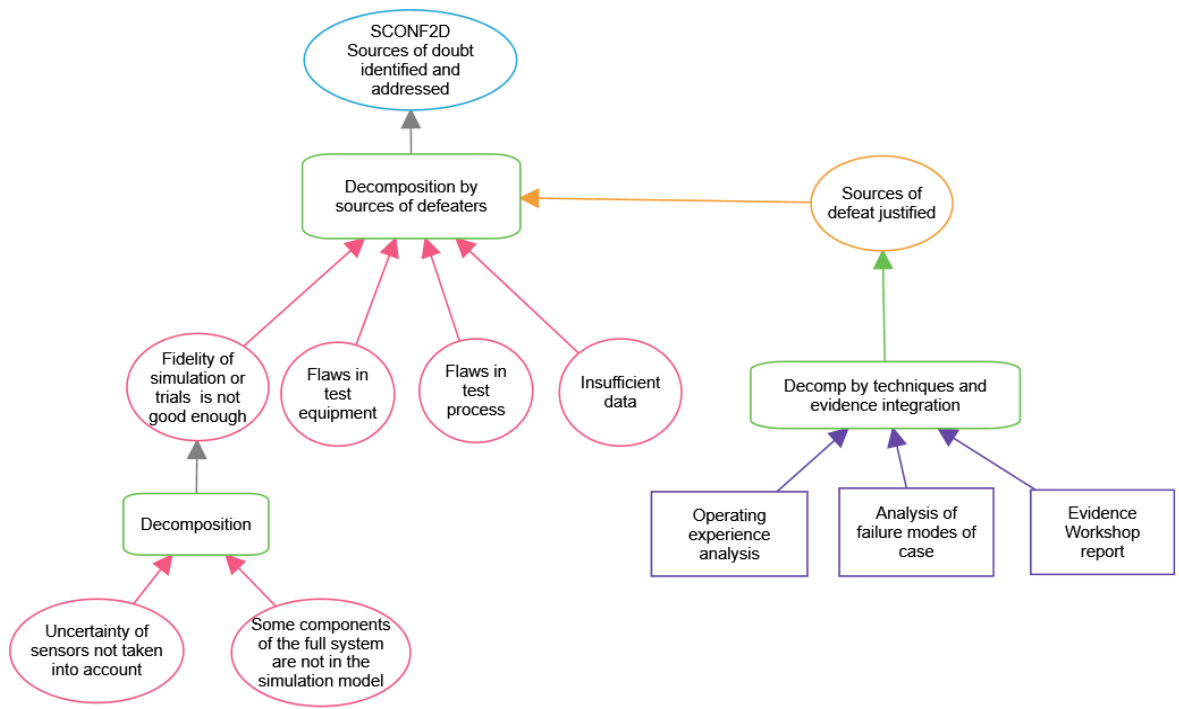

**Figure 26: Examples of doubt sources and their sources**

The side claim seeks to justify the sources of doubt drawing on operating experience, an analysis of failure modes and from workshops, such as the experimental defeaters workshop we held that is reported in [54]. The narrative supporting the side claim would explain how this evidence confirms (or otherwise) the sources of defeaters that need to be addressed.

### 8.2.2 Confidence from development lifecycle and associated V&V

We now develop the left-hand branch in Figure 24 addressing the confidence we might gain from the development lifecycle and V&V (SCONF1). The claim expansion is shown in Figure 27.

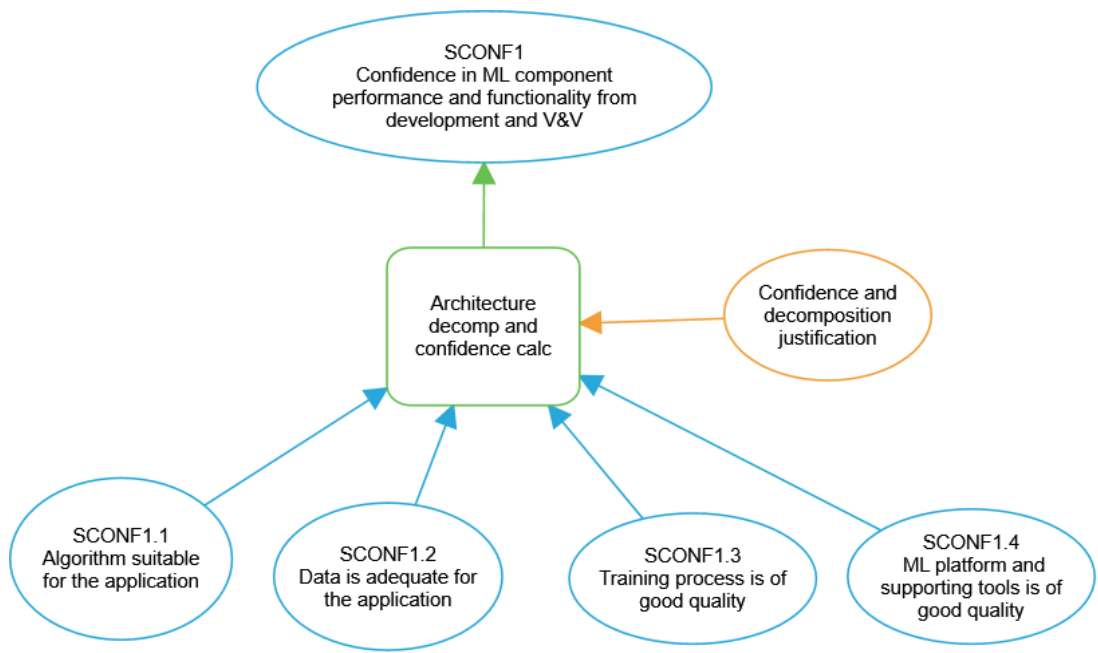

**Figure 27: Main areas to consider in justifying confidence from ML component development**

The sub-claims could be elaborated as follows. The suitability of the algorithm should address both the type of algorithm and its specific manifestation e.g. in terms of types of layers. The former could be established from journal papers about it and also from previous operating experience. The claim about the ML platform and tools will be a sub-case in its own right and is not addressed here but information can be found in [35] which has examples of traditional methods such as static analysis and formal methods being applied to ML software libraries. This leaves the two sub-claims on data adequacy and the training process which we now address.

#### 8.2.2.1 Data adequacy

Data adequacy was briefly discussed in the requirements case in Section 5.3. Here we expand the claim set for a specific application, which is decomposed into three aspects in Figure 28. One is that it assumes that some generic data set has to be curated and processed for developing the ML component for the application in mind. There is then a training process that takes the curated data and develops the algorithm to be deployed. Our usual claim about sources of doubt is also included.

During the training process, the selection and balance between the training, validation and testing data sets should be defined and justified.

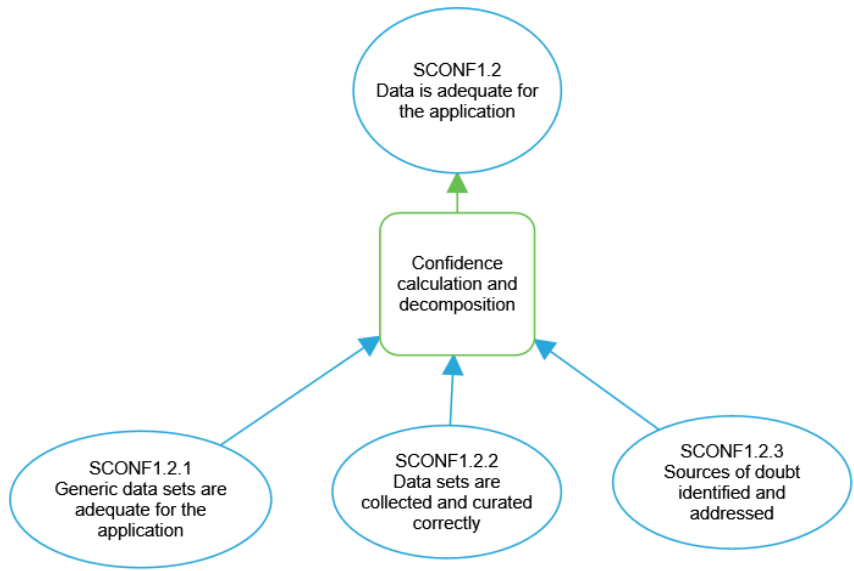

**Figure 28: Key aspects of data adequacy**

In assessing the adequacy of the generic data we should include whether

- the amount of data is sufficient
- data is adequately diverse
- data has been validated
- it contains the correct relevant feature set

The "defeaters" we need to consider include

- validation data does not match application
- biases in data sets not addressed
- datasets have been poisoned
- irrelevant features dominate algorithm

The curation needs to assess a variety of sub-claims as shown in Figure 29 along with potential sources of evidence.

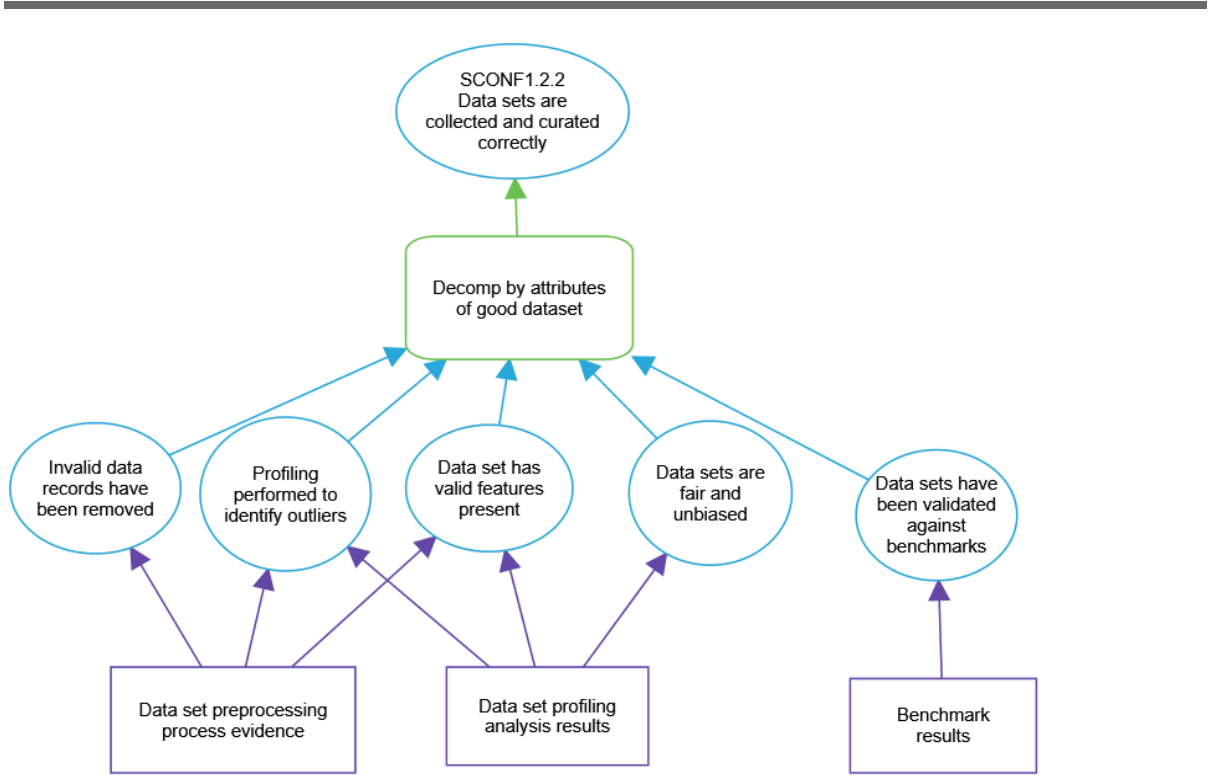

**Figure 29: Collected and curated aspects**

### 8.2.2.2   Training process

The claim about the training process is expanded by asserting that an adequate process has been defined, followed by competent staff and produced the right results. This is shown in Figure 30.

The re-training of models should also be part of the training process, as the algorithm is incrementally improved based on testing. The control of changes to the model links to the adaption and change template in Section 10.

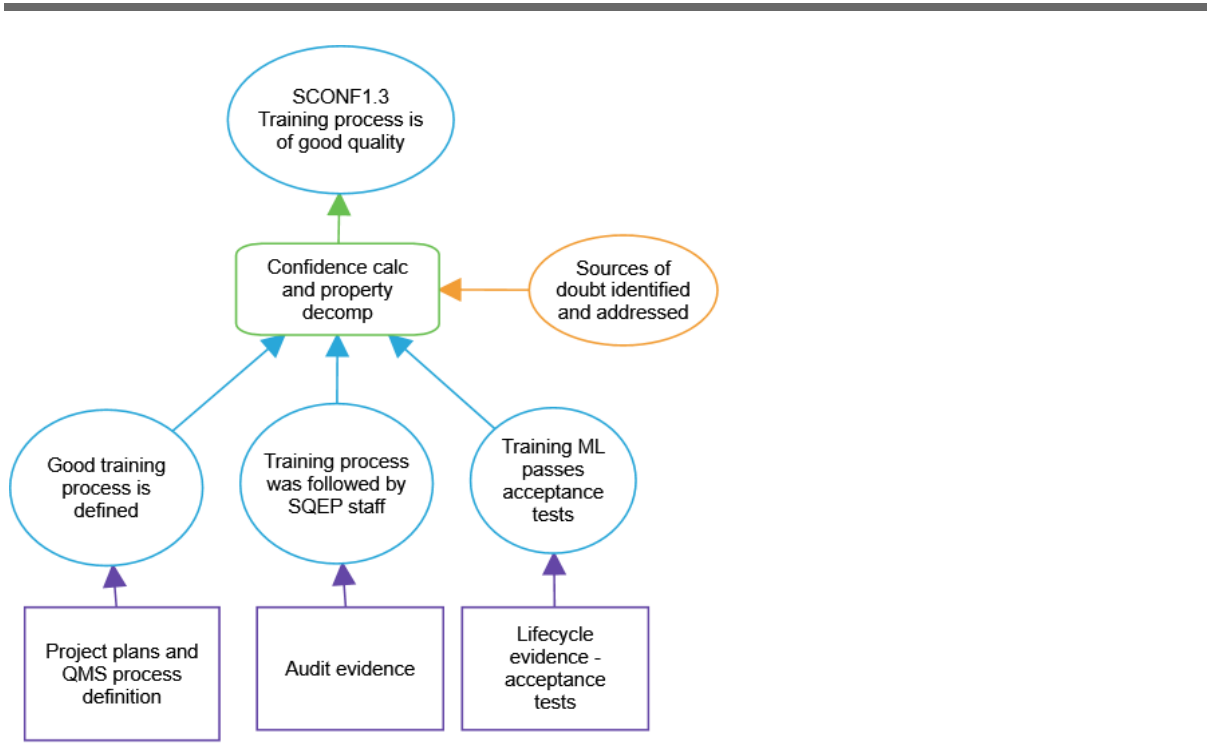

**Figure 30: Training process**

The sources of doubt would have to justify their adequate completeness and address such topics as:

• over training of the algorithm
• lack of diversity in ensembles
• wrong balance between training and validation data sets

## 8.3  Summary of defeaters

The following table summarises the defeaters that have been identified in the template elaboration.

| Description | Part of pattern | Possible mitigations |
|---|---|---|
| Flaws in test equipment. | Confidence in ML from real and simulated trials. | Validation of test approach. Audit and review of tests. |
| Flaws in test process. | Confidence in ML from real and simulated trials. | Validation of test approach. Audit and review of tests. |
| Insufficient data from trials (both physical and simulations). | Confidence in ML from real and simulated trials. | Early calculation of data requirements for safety and risk targets. Extended trials. More sophisticated use of all data sources. |

| Description | Part of pattern | Possible mitigations |
|---|---|---|
| Fidelity of trials or simulation insufficient. | Confidence in ML from real and simulated trials. | Use of hierarchy of simulations of different fidelities. Address specific issues where fidelity lacking. More focused trials or analysis of difficult aspect. |
| Trials not testing critical situations and edge cases. | Confidence in ML from real and simulated trials. | Fault injection based on previous known incidents/tail sampling methods. Analysis of test case distributions to identify difficult cases. |
| Algorithm unsuitable. | Confidence in ML from development and V&V. | Early review of algorithm. Support from literature and experience of choice made. Behaviour adequately defined. |
| Algorithm has insufficient stability. | Confidence in ML from development and V&V. | Testing with minor perturbations of inputs, such as noise injection. Use of ensembles. |
| Amount of data is insufficient. | Adequacy of data. | (see above - Insufficient data from trials). |
| Data is not adequately diverse. | Adequacy of data. | Augment data sets (whilst still ensuring it is possible to converge on a solution). |
| Data has not been validated. | Adequacy of data. | Additional validation checks. Use of diverse data. Labelling/annotation guidance. |
| Data does not contain the correct or relevant feature set or not representative of real world. | Adequacy of data. | Review at requirements stage. Augment data. Justification of data acquisition process. Comparison analysis of testing results, such as feature distributions. |
| Over training of the algorithm. | Training process of good quality. | Design V&V tests to assess this and ensure a stage in the test process to assess it. |

| Description | Part of pattern | Possible mitigations |
|---|---|---|
| Lack of diversity in ensembles. | Training process of good quality. | Increase diversity or number of ensembles (but these might be complex and of little benefit). |
| Incorrect balance between training and validation data sets. | Training process of good quality. | Ensure test process assesses this. Develop technical arguments for balance. Data partitioning guidelines. |
| Lack of validated models for inference of sensor functional reliability. | Confidence calculations from benchmarks, development and trials. | Use of more sophisticated models. Further research needed. |
| Shifting data requirements distributions over time. | Adequacy of data. Confidence calculations from benchmarks, development and trials. | Regular assessment of data sets and results. Continuous data capture and acquisition. |
| ML performance metrics are hard to compare and combine. | Confidence calculations from benchmarks, development and trials. | Further research. |
| Metrics fail to adequately measure safety. | Confidence calculations from benchmarks, development and trials. | Justification of performance metrics link to safety. Link possible metrics monitoring safety to hazard analysis. |

**Table 6: Summary of defeaters – ML based component**

## 8.4   Deploying the template

In applying the template, it would be necessary to translate the claim formulation to the particular sensor and context. In addition

- side claims would need articulating and justifying
- doubts and associated defeaters would need identifying and addressing

One of the key issues is reasoning about confidence in the performance of the sensor, and we have described how this and other aspects can be extracted from the overall case. In general, the viewpoint leads to a factorisation of "property of interest" and a set of supporting claims that need to be established.

In deploying the template, we would need to consider in more detail the difficult issue of justifying the performance of the sensor, particularly its reliability. We address this in the next section.

# 9 Performance of combined ML based sensor/monitor systems

## 9.1 Focused approach

To judge the safety of a system using an ML component, it is crucial to assess and understand the reliability aspects of its performance. In the previous sections, we have discussed the complete range of issues we need to address to make a considered judgment about both the functional and performance related requirements of a component using ML technologies. In this section, the focus is specifically on the AI/ML algorithm performance (and measures of performance); in the sense of how "good" it is – expressed as a reliability, or False Positive (FP) or False Negative (FN) rate.

We can factor the overall case so that other issues (the extent and quality of the data used to train and validate the ML component, the competency of those involved and the engineering process followed, and the trust in tools and platforms so we can trust the evidence that is produced) are all considered in a separate subcase. This allows us to focus on the key evidence we might have to consider when we argue the reliability of the ML component. This is shown in Figure 31.

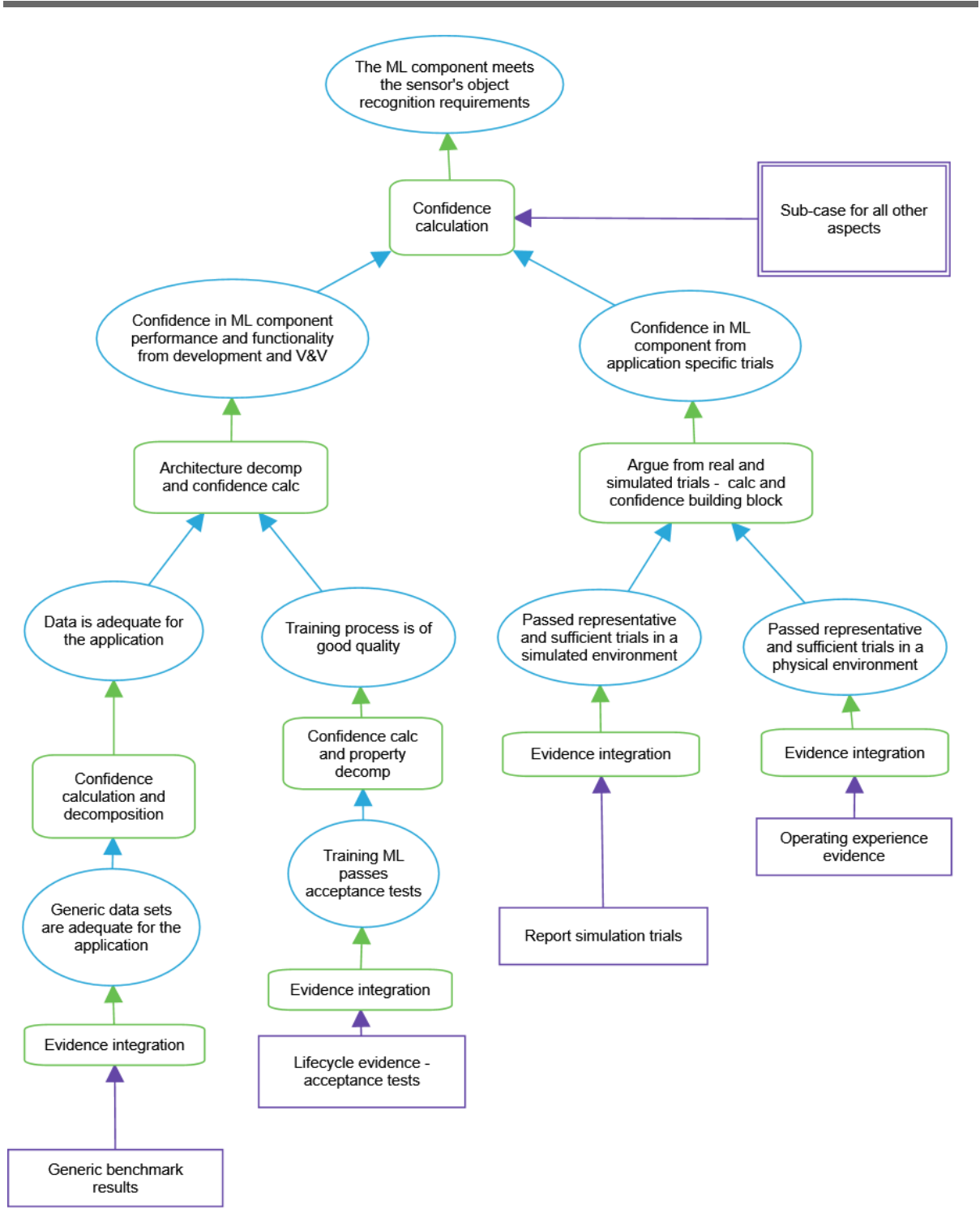

**Figure 31: Performance focus given suitable process, algorithm, data**

The arguments have retained their descriptions from the overall justification. We can see the evidence in terms of four layers corresponding to the real environment, the simulated environment, the development environment and the generic environment. This layering applies to both an ML component, and also, to the wider system that might contain it, e.g. the overall system rather than just the sensor subsystem.

What we would like is to have a hierarchy of models and calculations as shown schematically in Figure 32.

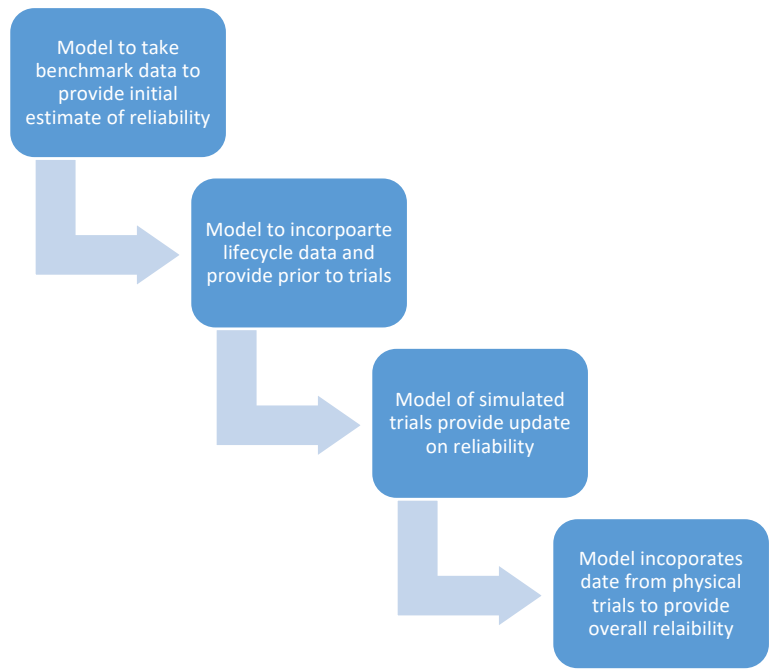

**Figure 32: Hierarchy of models**

More detail is shown in Figure 33, which brings out how we need to calculate confidence in the final reliability based on a disparate set of properties measured in different environments. It shows how claims (shown with an identifier starting with C) are parameterised on the properties they refer to (identified with a P), and their environment and associated evidence (identified by Env and E), The metrics used to describe the properties that are used at each level are hard to compare and combine. This is a key challenge and source of defeaters and is discussed next.

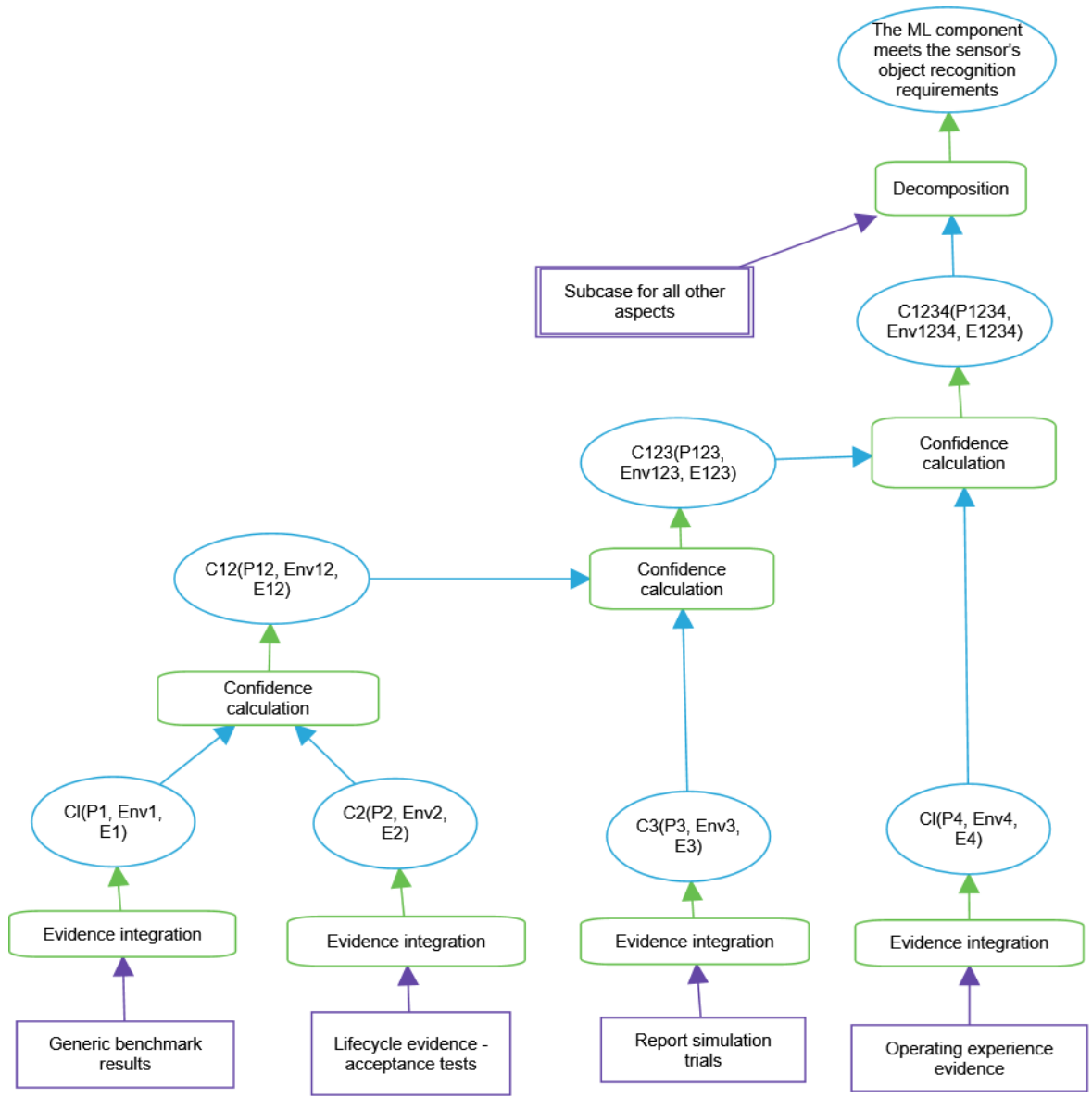

**Figure 33: Abstract reliability model**

The deployment of Figure 32 needs to be supported by specific approaches to reasoning and evidence also indicated by the "Confidence calculation" CAE Block in Figure 33. In [81] the authors present a new variant of Conservative Bayesian Inference (CBI), which uses prior knowledge while avoiding optimistic biases to assess the reliability of autonomous vehicles by applying Software Reliability Growth Models (SRGMs) to operating experience, specifically disengagement data (take-overs by human drivers). Related work [80] extends this by proposing a property-based decomposition of the safety case and assessing the system in two steps. The first step is based on assurance activities conducted at each stage of its lifecycle, e.g., formal verification on the DNN robustness. The second step boosts the confidence using field data of successful operation and a BCI approach. Some more details are discussed in Section C.2.4.

## 9.2 Reasoning and evidence for ML based systems performance

In the previous section we have shown how we propose to use a hierachy of models to justify the reliability and overall performance of the ML component. We now discuss in more detail the evidence that might be available to drive these models. In Appendix C we review

- common ML performance metrics defined for binary classifiers
- performance metrics for object detection algorithms
- experimental data measuring these performance metrics

We now relate these to three parts of the assurance case template presented earlier:

- the performance of the sensor as a component (Section 8 – examples in Table 7)
- the performance of the sensor in context of the guard architecture (Section 7 – examples in Table 8)
- the performance of the overall systems (Sections 5.2.2 and section 7 – examples in Table 9)

For each of these parts, we summarise the type of evidence that might be available, how it might be used, and give examples of what claims might be feasible with the evidence that is available. For the example claims these are rather speculative but illustrate the disjoints between the metrics needed in a case and the typical measures provided, as well as the state of the body of evidence.

| Evidence | Example | Role in case | Example claim |
|---|---|---|---|
| Performance on generic data sets | Test results demonstrating that YOLOv3 trained and tested on the COCO data set can make predictions for at least 20 frames per second with a mAP[10] of at least 0.5. | Demonstrating that pre-developed ML software is of good quality.<br><br>Demonstrating that the ML meets requirements on response time and resource usage, which are usually independent of training data. It is likely that this would also be supported by testing on the trained system as well.<br><br>It is unlikely that evidence of performance on generic data sets could support claims about the quality of a sensor's predictions in the specific application in question, since the generic data sets are unlikely to be representative of that environment.<br><br>Provides evidence to the future branch that the sensor can be retrained if the requirements change, e.g. new type of vehicle, different environment. | YOLOv3 can classify and detect relevant objects at a rate of at least 20 images per second when trained on the COCO dataset, on representative hardware. |
| Performance on specific tasks | When trained and tested on traffic light detection data, YOLOv3 achieved a recall of at least 0.87, with a precision of at least 0.85. | Supporting a claim about the performance of a sensor in a specific application. | YOLOv3 correctly identifies traffic lights in 87% of images containing traffic lights.<br><br>Traffic lights in the data set are representative of those in the operating environment.<br><br>Re-training / fine tuning of the generic YOLOv3 model has not had a negative effect on its performance. |

**Table 7: Evidence on performance of sensor as a component in an assurance case**

---

[10] mean Average Precision – a measure comparing false and true positive and negative results with ground truth.

| Evidence | Example | Role in case | Example claim |
|---|---|---|---|
| Temporal redundancy | The "Person of Interest" tracker tracked 41% of pedestrians and lost 19% of pedestrians over 20 consecutive frames.<br><br>The traffic light detection system detected all red lights in the test data within 1.6 seconds at a distance of at least 80 metres. | If the sensor output is processed further to produce a model of the world, then the frequency with which each vehicle/pedestrian is detected can support claims about the accuracy of the model.<br><br>Evidence regarding temporal redundancy is particularly relevant in detecting static objects such as traffic lights or a stop sign, which need not be detected every frame, but must be detected within a suitably short timeframe.<br><br>The sensor must also be resilient against single event upsets (if not detected or if falsely detected) to ensure the stability of its outputs. | The pedestrian tracking system identifies 80% of pedestrians which are visible for at least one second[11].<br><br>All red traffic lights are detected from a distance greater than the stopping distance of the vehicle. |
| Additional information (e.g. GPS) | The traffic light detection system correctly identified all traffic lights in the test using predictions from YOLOv3, GPS data and a map of traffic light locations.<br><br>Keeping maps up-to-date used for navigation and locations of static objects of interest (traffic lights, stop signs, junctions) needs to be made in the system is safe in the future branch. | Information such as GPS location can be combined with object detection algorithms to provide better performance for a sensor. A performance claim can be made for this combined system.<br><br>Additional information such as GPS location could also be used as a guard by, e.g. setting a maximum speed if a traffic light is not detected when expected, or geofencing the area in which the AV can operate autonomously. | The addition of a GPS guard reduces false positive traffic light detections by 80%.<br><br>The traffic light detection system correctly identifies 95% of traffic lights in Vitoria with confidence 60%[12].<br><br>The AV only operates autonomously within the city of Vitoria.<br><br>GPS information and navigation maps are adequately up-to-date. |

---

[11] Note that the impact of pedestrians not being visible for at least one second is highly dependent on the speed of which the vehicle is travelling.

[12] The study did not provide in depth details of the environmental conditions in which the tests were carried out.

| Evidence | Example | Role in case | Example claim |
|----------|---------|--------------|---------------|
| Diversity | Different object detection algorithms using different camera inputs both have recall of 0.85.<br><br>Analysis of test results demonstrating the independence of sensor failures. | Diverse inputs, e.g. different cameras, LIDAR or radar data, can be combined to reduce the probability of missing an object, or to increase the confidence in the presence of an object.<br><br>Diversity could also come from using different algorithms on the same inputs, e.g. ensembles. More care must be taken to show independence between algorithms in this case. | The probability of a vehicle being missed by both cameras is 2%[13].<br><br>The majority of algorithms in the ensemble detect a traffic light with probability 99%. |
| Additional reasoning | Test reports showing a recall of 0.87 for R-CNN when trained and tested on input data including predicted locations for objects.<br><br>Specification for algorithm generating model, showing that it compares sensor outputs with its predictions and updates the model accordingly. | Information about the "big picture" model may be used as additional input and reasoning for individual sensors, improving the performance of the overall sensor.<br><br>Evidence regarding the model generated may be used to support claims about sensor fusion.<br><br>Reasoning about expected changes in the model may support claims of fault tolerance. | The object detection algorithm identifies 87% of vehicles when provided with their expected locations.<br><br>The model of the environment correctly identifies and locates 99.5% of objects it has been trained to track. |

Table 8: Evidence on performance of sensor in guard architecture in assurance case

| Evidence | Example | Role in case | Example claim |
|----------|---------|--------------|---------------|
| Accident and disengagement data | Disengagement and accident reports demonstrating that there were 110 disengagements in 1.4 million miles driven. | Accident and disengagement data could be used to support a claim regarding number of accidents the AV will be involved in. | The number of crashes involving the AV averages at most 89 crashes per million miles driven with confidence 95%.[14] |

---

[13] This probability of the vehicle being missed will depend on the conditions, such as weather, time of day or even the specific speed of the vehicles. It should be clearly argued what the circumstances/assumptions are and if possible use a worst case figure.

[14] The equivalent figure for conventional vehicles in the USA is approximately 2 crashes per million miles driven.

| Evidence | Example | Role in case | Example claim |
|----------|---------|--------------|---------------|
| On-road testing | Report documenting each sensor failure observed during one million miles of driving in target urban environment.<br><br>Report documenting key metrics of the performance of the sensor. | Increasing confidence in the performance of the sensors.<br><br>Validating test and simulation results.<br><br>Monitoring the reliability growth over time and performance of sensors. | The object detection algorithm provided data inconsistent with other sensors on average twice per mile.<br><br>Sensor failures have been observed to be independent with 90% confidence. |

**Table 9: Evidence on performance of overall system in assurance case**

The claims in the tables above are examples of the claims that we might be able to make using evidence that is currently available. These tables can be used to assess the feasibility of and challenge inherent in the claims that we might be able to make for the case.

## 10  Adaptation and deployment template

### 10.1  Objectives of template and context

The overall assurance case argues that the autonomous system is safe initially and will continue to be safe in the future. There are many changes, in which the autonomous vehicles will have to adapt to in order to maintain its safety in the future. In order to achieve this, we are taking the Open Systems Dependability (OSD) perspective based on IEC standard [55] and developing previous work in the TIGARS project [56][57][58]. The model covers changes to the system (the vehicle itself), the supporting infrastructure directly associated with the vehicle(s) and the wider ecosystem and environment.

Adelard developed a code of practice and a publicly available specification, *PAS 11281: Connected automotive ecosystems Impact of security on safety*, which gives recommendations for managing security risks that might lead to a compromise of safety in a connected automotive ecosystem [59]. We focused on the application specific to autonomous vehicles, as all levels of vehicle automation and autonomy are in the scope of the document. The aspects of developing and maintaining a safe, secure and effective vehicle over its lifetime has been brought into this template.

Our assumed system overview was outlined in Section 3.4 and shown in Figure 7.

### 10.2  Outline of structure and reasoning

The top level claims for the adaption and change template are shown in Figure 34. The objective of this part of the assurance case is to demonstrate that the deployed autonomous system/vehicle will continue to be safe and effective in the future. A balanced approach to assessing the risk and potential impact of updating the system should be taken, weighing the risk of updating verses not updating. There may be a greater risk in not updating the system, such as if a critical flaw was identified or a change in operating environment needed to be addressed.

The future refers to different perspectives in time after the initial deployment of the autonomous system. One of which relates to changes that have already occurred with the assurance case being a "living document" updated, reviewed and re-issued after each design change. The other relates to possible changes that may happen in the future that the system must be able to address, such as obsolesce, regulatory and legislation changes, or vulnerabilities.

In our example template, we are expecting some form of high level In-Service Safety Management System to be in place; for example based on POSMS SMP13 [18] for defence systems. We would expect regular safety reviews to take place and metrics important to safety to be monitored throughout the in-service phase of the lifecycle.

The change infrastructure should also consider changes required for the end of life and disposal of the system as the future time scale is not infinite and at some point the system would need to be retired.

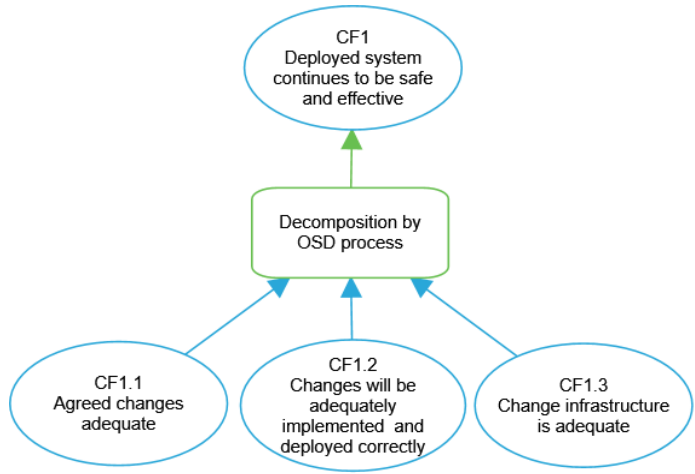

**Figure 34: top level claims based on OSD model**

The top level claim is decomposed based on the OSD model [55] into the following areas:

- the changes that have been agreed between stakeholders are adequate
- the changes will be implemented and deployed correctly
- the change infrastructure is adequate

## 10.2.1 Agreed changes are adequate

The agreed changes rely heavily on the correctness of requirements of the proposed changes and the changes being suitable for all stakeholders. Figure 35 shows the structure of this section of the template. This branch should have a similar structure to that of the requirements template Section 5. Some of the issues for requirements of autonomous systems apply here as well, such as adequately specifying the requirements of AI/ML components and traceability of requirements; guidance for addressing these areas of concern can be found in Section 5.2.2. Therefore, if the difficulties with requirements between autonomous and none autonomous components in the system is too great then the assurance case argumentation should split at this point and separate sub arguments for each type of component should be made (Figure 38 shows an example of this methodology being used).

An impact assessment on the system requirements and potential hazards caused by changes should be undertaken to determine the full extent of the changes required. For example, changes to subcomponents may impact the wider system, e.g. modifying an ML component may require additional protections to be implemented in guards.

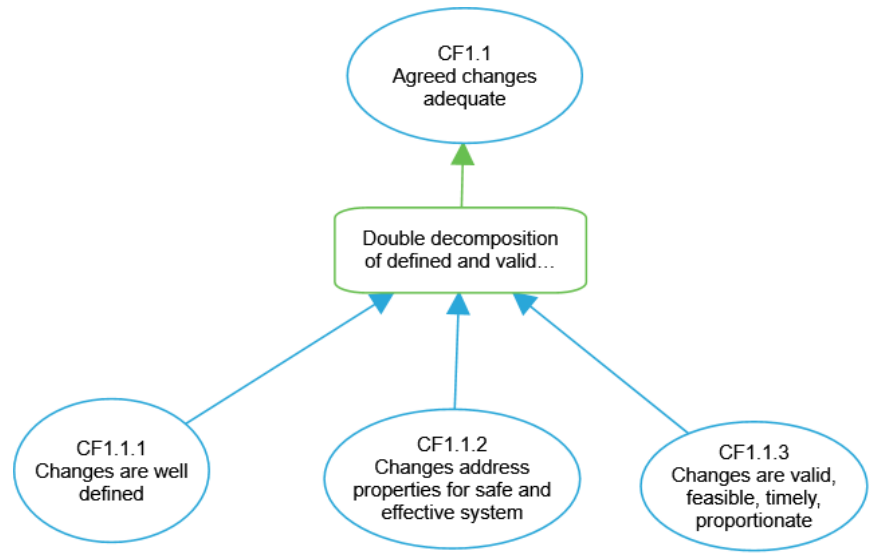

**Figure 35: Agreed changes are adequate**

### 10.2.1.1 Types of evidence

Similar evidence for the initial system requirements claims can be used to support these claims (see Section 5.2.2.1). However, there should be particular focus on the impact of changes to the system.

- impact and risk assessment of proposed changes
- design and change planning, including reverification plan and test specifications
- processes and procedures (change control, periodic review, change identification, etc.)

## 10.2.2 Changes are implemented adequately and deployed correctly

This part of the template focuses on performing the changes to the system and being able to effectively deploy them. Processes and procedures should be in place to allow the safe modification of the autonomous system. The processes and procedures for performing system changes should be similar to those for the development of the initial system in other areas of the assurance case. Evidence and defeaters will overlap between these two sections of the case.

Deployment of changes can be a difficult area if a fleet of autonomous systems is in operation and a phased approach may be required. This could mean that multiple versions of the same autonomous system could be in operation at any one time. Therefore, processes are needed to ensure that updates are deployed safely and interactions between the different versions are safe.

Furthermore, it may be difficult to determine the full impact of changes especially for ML components, and therefore, a phased roll out approach could be useful to further assess the impact of changes, with initial deployment of the changes to systems in a limited capacity. This would give confidence in the changes made and the deployment system.

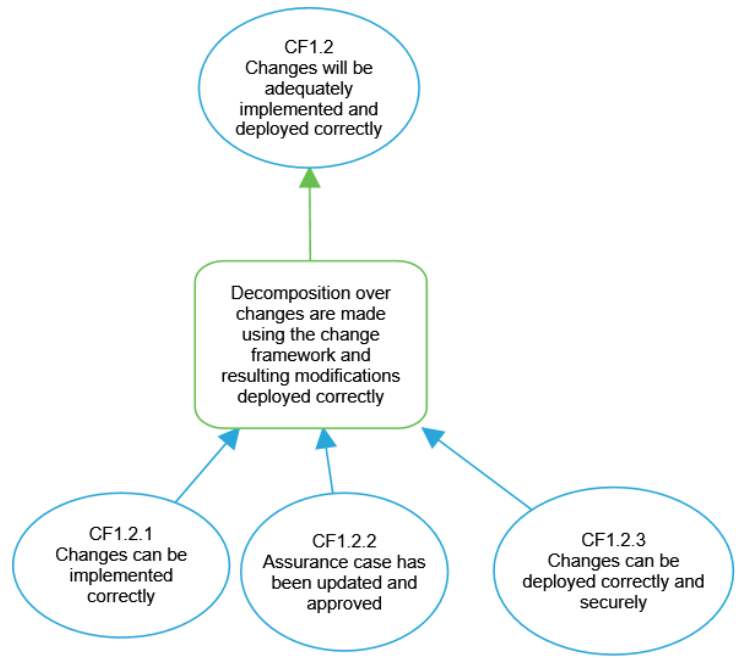

**Figure 36: Changes will be implemented adequately and deployed correctly**

At the end of the change process, the assurance case needs to be updated and approved to account for the changes made. A Safety Management System (SMS), which had sign off as part of its procedure, could be in place to help control the process of updating the assurance case. Evidence documents need to be updated if they were impacted by changes or where new evidence is available. The processes of updating and approving the assurance case could link into the supporting processes on governance and accountability (see Section 10.2.3).

### 10.2.2.1 Changes implemented correctly

The implementation of changes to the system is dependent on the type of change required. In the template, we focused on changes to the subcomponents of the autonomous system and split the assurance argument between changes to AI/ML components and more conventional components without AI/ML. This is shown in Figure 37. We assume conventional system lifecycle processes and procedures are adequate to assure changes to non–AI/ML components.

It is almost certain that additional assurance and verification activities would be required to ensure the correctness of changes to AI/ML components and that no unintended behaviour or functionality has been introduced. If the algorithm has been re-trained or modified, some regression testing would be expected to show that the latest version of the algorithm has the same level of performance as the old version.

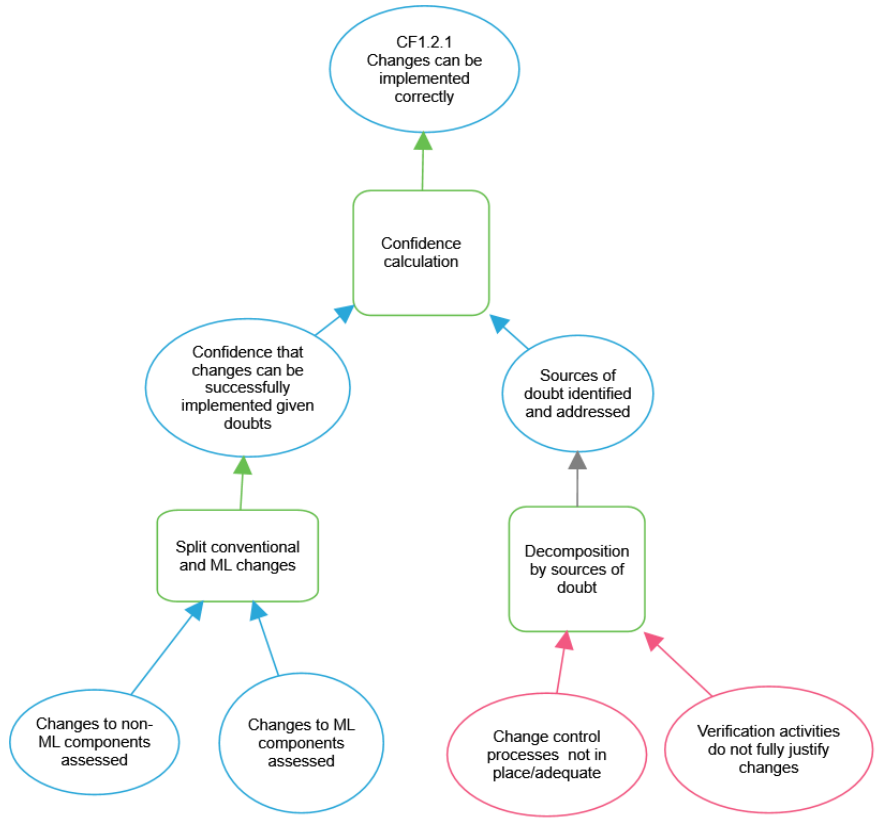

**Figure 37: Changes can be implemented correctly**

We would expect the defeaters listed in the right side of Figure 37 to be addressed as the case is developed, as evidence for change control and verification activities is produced. Some of this evidence may be repurposed from the initial development branch of the case.

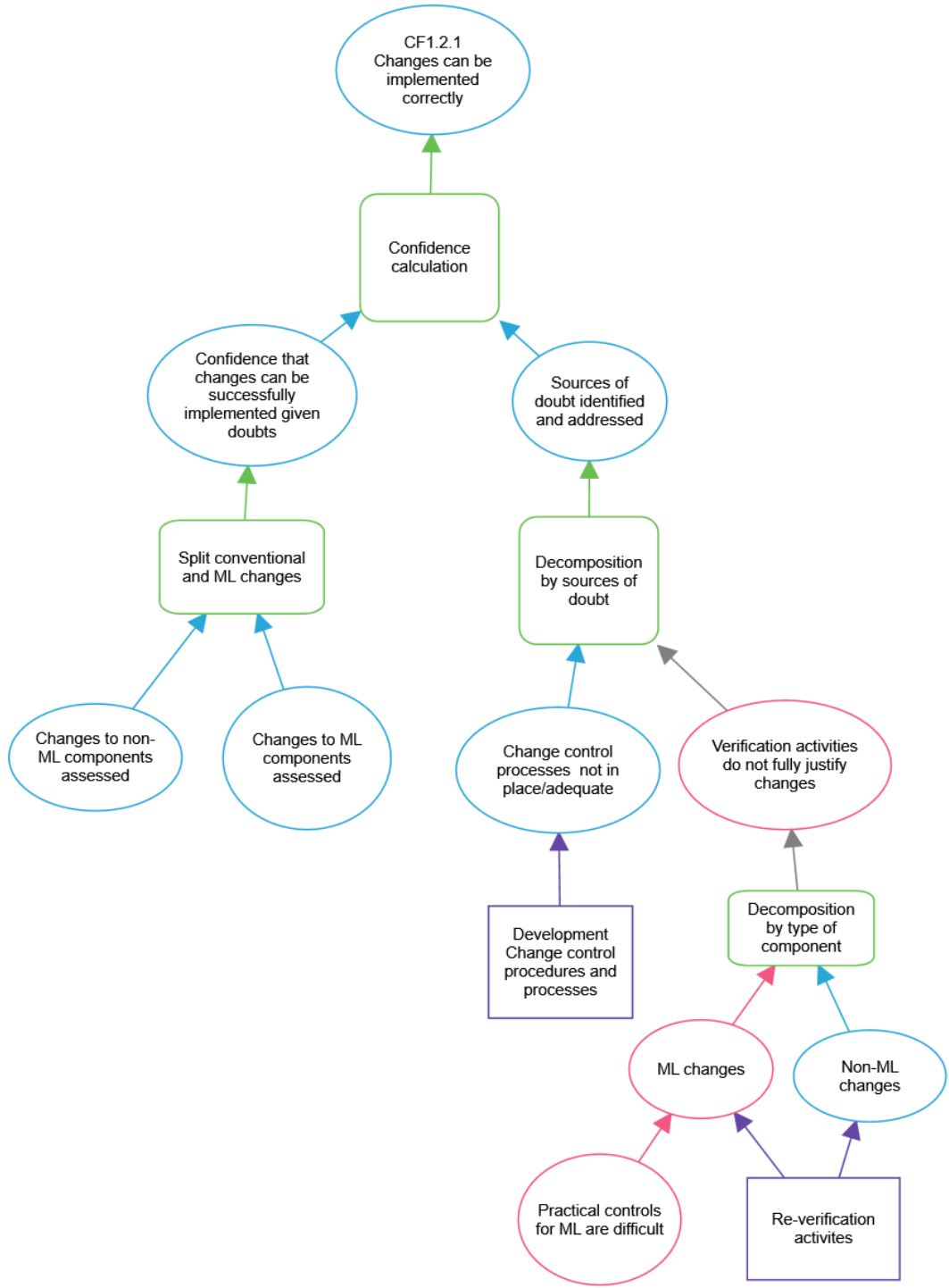

Figure 38: Evolution of defeaters for changes implemented correctly

The evidence for the re-verification may not be enough for the defeater to be rebutted and turned into a claim supporting the confidence calculation; therefore the claim is shown remaining as an undercutting defeater in Figure 38. As the case evolves the defeater that is no longer relevant ("the Non-ML changes" is now no longer a source of defeat) could be hidden but kept as a record of the case evolution) The most appropriate notation for capturing defeaters and defeaters of defeaters is work in progress and one of the novel aspects of these templates.

### 10.2.2.2 Sources of evidence

Similar evidence for the development, implementation, and verification and validation for the initial system can be used to support the claims in this part of the template (see Section 5.3.1.3). Note that different evidence is likely to be required for ML and non-ML component changes.

Potential required evidence for verifying the implementation of changes should be identified early on in the change process; this can allow for the potential of modifying the change infrastructure to accommodate any difficult to obtain evidence before changes to the system are performed and avoid retrofitting evidence post change. This is also useful for updating the assurance case; a review should be carried out during the change process to identify evidence that will require updating.

Evidence should focus on correct implementation of changes and reverification activities, such as simulation, testing and trials. Evidence for an effective plan for the deployment of the new changes should also be provided. Possible evidence sources include:

- plans for implementation of changes and deployment
- reverification and validation planning, such as testing and trials and building a complete release
- simulation and prototyping testing results
- processes and procedures (change implementation, regression testing, deployment, etc.)
- assurance case review report, update and sign off

### 10.2.3 Change infrastructure

The change infrastructure part of the template ensures that the policies are in place to accommodate changes to the system in a controlled manner, and also that the change infrastructure itself is able to accommodate changes. We have divided the change infrastructure to changes to the system and changes to the wider ecosystem infrastructure.

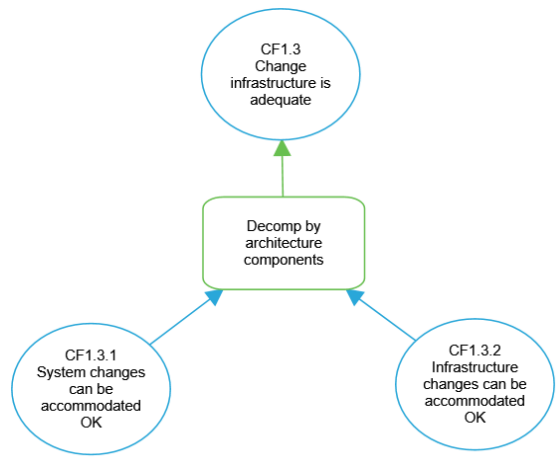

**Figure 39: Change infrastructure is adequate**

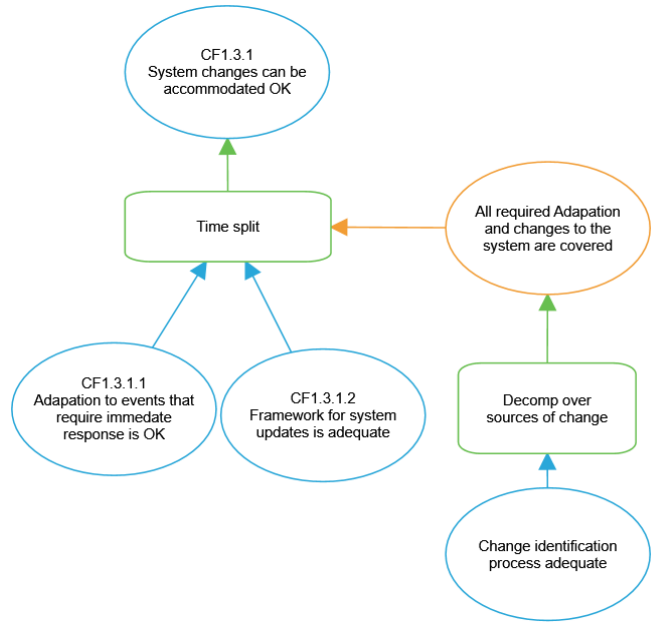

Figure 40: System changes can be accommodated

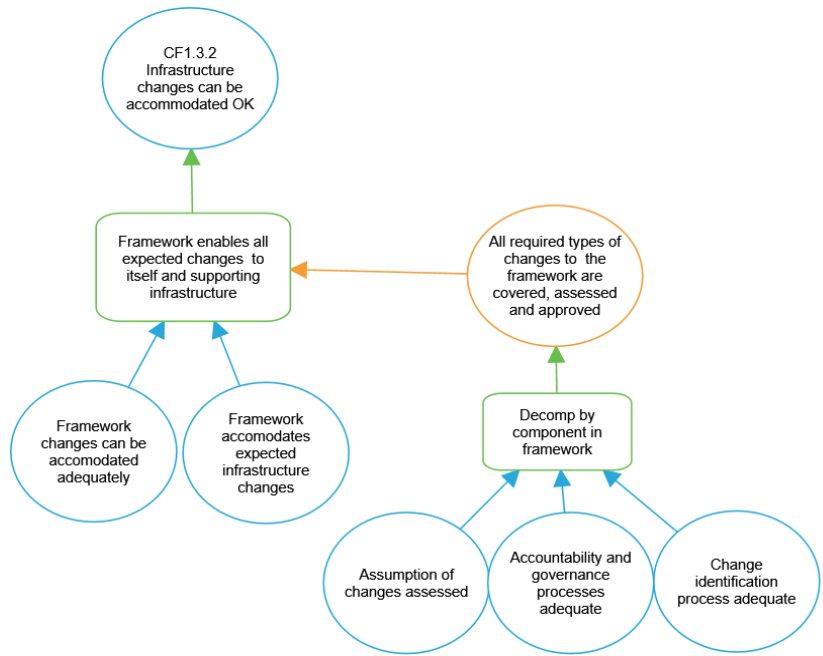

Figure 41: Infrastructure changes can be accommodated

The autonomous system will need to be able to adapt to immediate issues, but also medium and long term change cycles. Some of these changes may be planned while others may not be (such as unpredictable component failures). Therefore, the change infrastructure framework must be able to accommodate different feedback cycles. In the template, we incorporate this by performing a time split on CF1.3.1 for different changes to the autonomous vehicle system. Some types of changes that need to be considered for the autonomous system include:

- immediate response
  - component failures
  - malicious attacks
  - unexpected behaviour from other actors in the environment
- medium change
  - development lifecycle changes
  - security updates
- longer change
  - maintenance and operational changes
  - regulation and legislative changes

Furthermore, the change infrastructure must accommodate changes to the supporting infrastructure and wider ecosystem. This change cycle is likely to be much slower, as the supporting infrastructure affects multiple autonomous systems, and therefore, consensus will involve multiple parties and stakeholders. It is likely changes in this area will be driven from modifications to best practice, standards, regulation and legislation.

Some of the wider principles would be more difficult for AI/ML components, for instance accountability and governance would be more difficult particularly if the system is taking decisions with light oversight. A split in the assurance argument may be needed here if different governance systems are in place for the AI/ML components. Operating Centres or equivalent are required by POSMS [60] to maintain a procedure for identifying and accessing the relevant safety and environmental legislative and other requirements that are applicable to their projects.

Finally, the framework itself will need to be adaptable, as new forms of change may need to be accounted for or old practices become obsolete. This can start to become a recursive argument, with changes to all parts of the change infrastructure needing to be accommodated. Therefore, we have limited the claims to only modifications to the system change framework and supporting infrastructure framework.

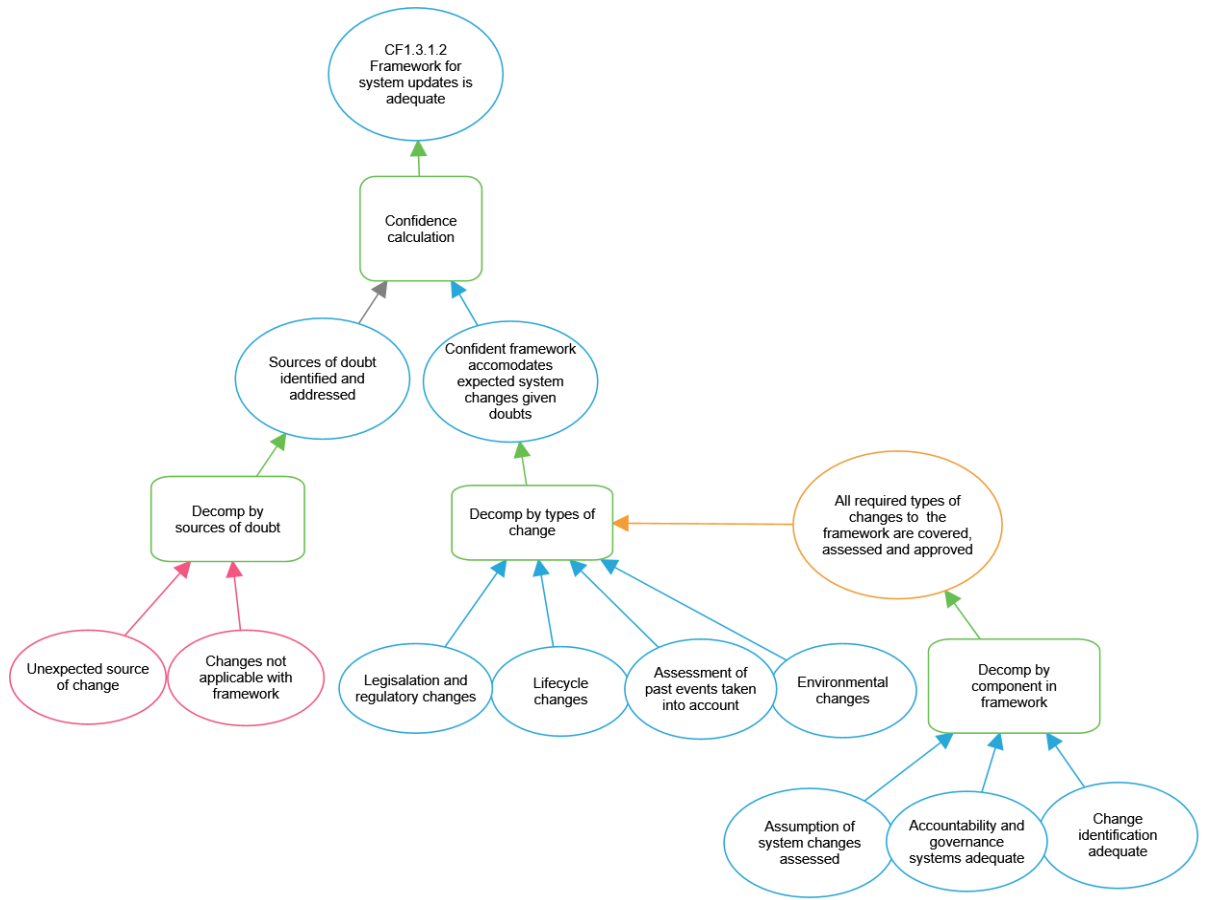

**Figure 42: Framework for updates to system is adequate**

Figure 42 shows the fragment for the framework for updates of the system. The types of change we have included:

- events based changes and learning from experience
  - incident and accident records/post event analysis, e.g. security breaches or failure incidents
  - changes to the operating environment, congested and competing factors, physical, weather and climate changes
  - legislative and regulatory changes, new standards or best practice
  - in-service audits, non-compliance reporting
- lifecycle changes
  - system improvements - better software, AI/ML model performance or hardware components
  - ongoing maintenance and obsolescence changes
  - end of life and disposal

This is not an exhaustive list and we expect domain experts to adapt and tailor the categorisation of changes applicable to the system using their own domain knowledge. Different regulations may be in place for the monitoring and reporting of significant occurrences and failures or post-event analysis information required. This process is important and should be carried out by SQEP, as it also provides evidence for the change identification process and counter evidence to address the incompleteness defeater related to unexpected sources of change (see Figure 42).

Potential changes should actively be sought out through periodic reviews and internal auditing. An example would be reviewing hazard logs and incident reports to see if assumptions made about the frequency of hazards still stand.

The exercise is important to discover sources of new potential changes or review current practices based on new information. It is possible that new techniques have been developed or technology has become available. Cyber security often evolves at a fast pace as new vulnerabilities are continuously discovered, these should be reviewed for potential impact on the system and supporting infrastructure and patched if possible while maintaining safety, or mitigated. Some areas of security monitoring may be a continuous process rather than a periodic event, such as threat intelligence gathering.

The lifecycle changes may need to make a distinction between AI/ML components and that of non-AI/ML components. Additional life cycle activities may be required to fully assess the impact of a change to AI/ML components. This can be included into the template by decomposing the types of change claims between ML and non-ML component changes.

Figure 43 shows how some of the defeaters of the framework for system changes may be addressed. The defeaters shown in the diagram are related to unexpected changes occurring which have not been taken into account and the framework for system changes not accommodating types of changes. Some suggested evidence is provided in the figure, with Section 10.2.3.1 providing more details on possible sources of evidence for this CAE fragment.

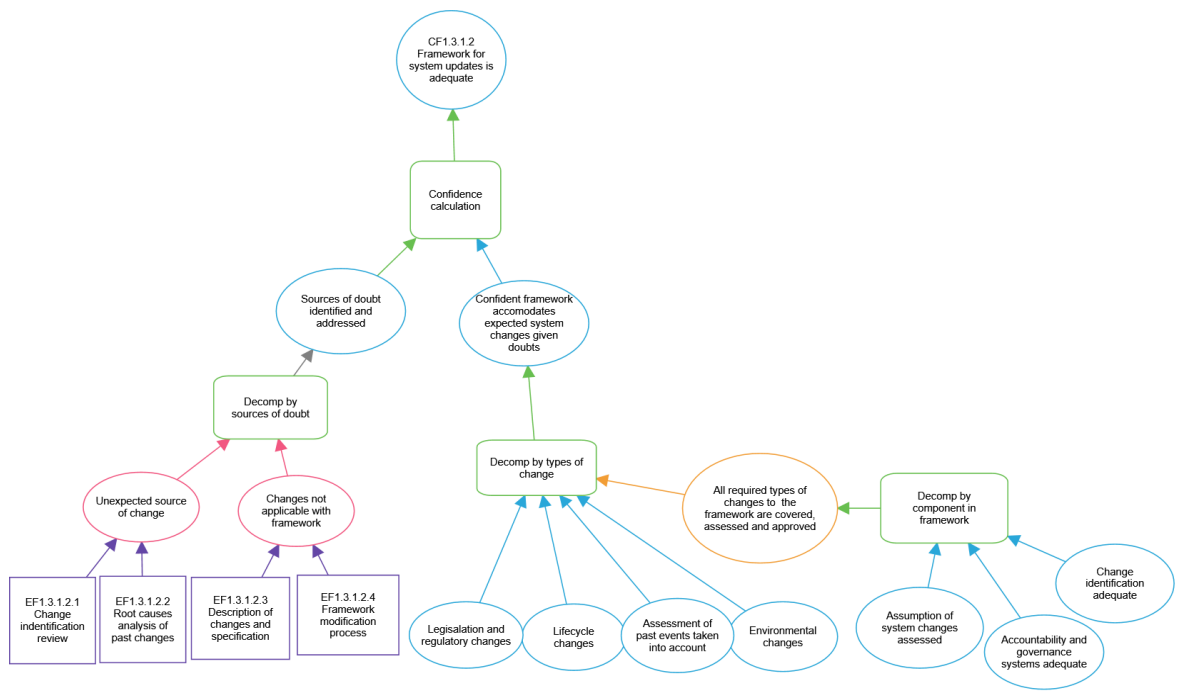

**Figure 43: Framework for updates to system is adequate defeaters addressed**

### 10.2.3.1  Sources of evidence

The majority of the evidence for this section of the template relates to the change framework procedures and processes, including supporting activities. Evidence can be sourced from an SMS if one is in place, which could include sign offs of various stages of the processes to update the system. However, processes and procedures are likely to be different for each organisation, so the potential evidence should be adapted to the particular organisation.

Outputs from the periodic reviews to identify new changes can be used as a source of evidence (EF1.3.1.2.1). This may not necessarily be formal documents or analysis reports; minutes from regular meetings (for example, regular system safety panels) may also be useful evidence. Experience from previous changes could also provide useful information (EF1.3.1.2.2)

Another potential source of evidence can be based on compliance with standards and industry best practice frameworks for change (EF1.3.1.2.4); though for the autonomous vehicle domain some of these standards have not reached maturity and in some areas there is no industry best practice or consensus as the field is evolving at a rapid pace.

Our evidence list includes:

- lifecycle processes and procedures
    - o  change control
    - o  quality management and continuous improvement
    - o  governance and accountability
    - o  best practice and standards compliance
    - o  safety and environmental targets based on legislation
- reviews to discover and identify required changes
    - o  impact and risk assessment
    - o  post-incident analysis review
    - o  security - threat intelligence, vulnerability review, etc.

## 10.3  Defeaters

Many of the defeaters to requirements and safe implementation of the initial system could also apply to adaptations and changes (see Sections 5.4 and 7.6). These sections should be reviewed with respect to changes to see if any of the defeaters are relevant.

| Description | Part of pattern | Possible mitigations |
|---|---|---|
| Traceability of changes to requirements may be difficult for ML components. | Agreed changes adequate - Well-formed (CF1.1). V&V. | Compensate for lack of detailed traceability with more testing, including impact analysis test after changes and regression testing to ensure performance has not diminished. Use of agile development process revising requirements. |

| Description | Part of pattern | Possible mitigations |
|---|---|---|
| Difficult to define test cases for simulation for testing of modified ML. | Changes implemented correctly. (CF1.2) V&V. | Review initial test specification with impact analysis to identify test candidates. Use of agile development process revising requirements and testing. Perform additional testing to obtain greater confidence. |
| General existing issues with systems engineering amplified by use of ML. | All. | Use of existing quality criteria. Consider use of expanded quality criteria. |
| Accident investigation should be supported. | Change framework (CF1.3). | Ensure appropriate and feasible level of logging and explainability is included for any post incident analysis. |
| Maintenance of requirements, versions, compatibilities, components long-term. | All. | Use of agile development process revising requirements. Identify obsolesce and component changes early to be proactive rather than reactive. |
| Limited best practice and industry consensus on approaches. | Change framework (CF1.3). | Use of broad range of SQEP. Use available good practice even if not well established with justifications for deviations. Ensure robust risk and safety analysis. Deploy in limited capacity to gain confidence. |
| Inability to certify and accept updated system. | Changes implemented correctly (CF1.2). Hazard analysis template. | It may be unclear how to accept risk or to certify the system. To mitigate we would need requirements which are clear to a potentially non-technical audience. There may be a greater risk in not updating the system, such as if a critical flaw was identified, so both cases should be sufficiently investigated and credible impacts determined. |
| Inability to have practicable controls for changes to ML. | Agreed changes adequate (CF1.1) – Valid. Changes implemented correctly (CF1.2). | Deploy in limited capacity to gain confidence following changes. Use a staged roll out deployment model. |

| Description | Part of pattern | Possible mitigations |
|---|---|---|
| Long term issues such as sensor calibration and degradation of components. | Change framework – System changes (CF1.3.1). | Use of monitor architect can mitigate problems once performance of ML sensor is too low. Proactive maintenance and ongoing in-situ testing. |
| Changes too complex to be specified adequately. | Agreed changes adequate (CF1.1) – well-defined. | Impact and risk analysis, increased testing to understand full impact of changes. Deploy in limited capacity to gain confidence following changes. |
| Change/adaptation too frequent cannot be captured in framework. | Change framework (CF1.3). | Deploy in monitor architecture to continually monitor ongoing performance. Identify allowed/acceptable regions of change for parameters. |

Table 10: Defeaters for adaptation and change

## 10.4 Deployment

In this section, we have presented the adaption and change template in order to assure that the autonomous system continues to safe, secure and effective future after initial deployment.

Some of the key challenging areas to address to ensure the successful deployment of this template are

- Ensuring that credible sources of change are identified and reviewed over the life time of the system, and that you are reacting to change in a proactive meaner as possible.
- Proposed changes are reviewed for their impact and possible risk. The changes should be fully understood, well-defined (scoped) and detailed enough for SQEP to make informed decisions. Preferably, changes should be linked to system requirements.
- The change process should follow a robust change infrastructure framework, which should follow the core principles of the OSD model (consensus, traceability, accountability, etc.).

Parts of the justification and arguments of this template can be referenced from the "system is OK initially" branch, in particular the requirements of changes and implementation of changes. These patterns should follow a similar structure to their counterparts in the initial time branch.

## 11 Discussion and next steps

The aim of this project is to provide safety assurance argument templates to support the deployment and operation of autonomous systems which include ML components. This report presents example safety argument templates covering the development of safety requirements, hazard analysis, a safety monitor architecture for an autonomous system, an ML based sensor and a template for adaptation of the system over time. It also presents generic templates for "argument defeaters" and evidence confidence that can be used to strengthen, review, and adapt the templates as necessary.

As we discussed in the introduction to this report, there are a number of approaches to "templates". We take the view that in this area we need to provide generic structures as scaffolding for the user so they can develop and communicate their understanding. To enable this, there needs to be sufficient discussion of the issues and explanations of how the template can be adapted. Therefore, we support the argument structure with a discussion of the issues involved and consider the reasoning and identification of relevant and proportionate evidence.

For a complex system in a complex environment to be judged safe there are a plethora of issues that need to be addressed and the templates we have developed provide a structure for identifying, prioritising and understanding these. While this detail is essential for making the safety case we should also not lose sight of high level questions for autonomous systems:

- do we understand and have we described what it is we are developing and assuring?
- do we understand how big a challenge it is to demonstrate the effectiveness of ML based systems?

These templates should help address both these questions. The evidence of the effectiveness of ML based systems is needed to shape our judgment of feasibility, to define what should be collected during the project and to form the basis for the safety case. This evidence is challenging to define, as there are many candidate and incomparable metrics and the associated data and benchmarks are often hard to assess in detail. However, we have reviewed the current state of the art in metrics and evidence and this is summarised to provide a basis for using the templates.

This stage of the project is complete, however we will continue to work with others to provide feedback and to support an evaluation of the approach as well as to explore further the ideas for defeaters and more detailed guidance.

Future considerations for templates would include human interactions and verification of machine learning platforms. The latter can be supported using largely traditional approaches as discussed in [35].

## 12  Conditions of Supply

This work is published in accordance with Contract no DSTLX-1000140535, and under the R-cloud Framework Agreement v3.0 terms and conditions. The document is made available as a resource for the community providing all use is adequately acknowledged and cited. This document provides a snapshot of work in progress. We welcome feedback and interest in this work: please contact the authors or admin@adelard.com.

## 13  Glossary

| ABC | Attribution-Based Confidence |
|------|------------------------------|
| AI | Artificial Intelligence |
| ALARP | As Low As Reasonably Practicable |
| ASP | Answer-Set Programming |
| AUC | Area Under Curve |
| AV | Autonomous Vehicle |
| CAE | Claims Argument Evidence |
| CCF | Common Cause Failure |
| CIFAR | Canadian Institute For Advanced Research |
| CNN | Convolutional Neural Network |
| COCO | Common Objects in Context |
| DARPA | Defense Advanced Research Projects Agency |
| DE | Deep Ensemble |
| EAR | Easy Approach to Requirements |
| FN | False Negative |
| FNR | False Negative Rate |
| FP | False Positive |
| FPR | False Positive Rate |
| GPS | Global Positioning System |
| GTSRB | German Traffic Sign Recognition Benchmark |
| Hazops | Hazard analysis and operability study |
| HIL | Hardware In Loop |
| HMI | Human Machine Interface |
| ICP | Inductive Conformal Prediction |
| IoU | Intersection over Union |
| LC | Learned Confidence |
| LEC | Learning Enabled Component |
| LIDAR | Light Detecting and Ranging |
| MAA | Military Aviation Authority |
| mAP | mean Average Precision |
| ML | Machine Learning |
| NN | Neural Network |
| ODD | Operational Design Domain |
| OSD | Open Systems Dependability |
| PAS | Publicly Available Specification |
| POSMS | Project Oriented Safety Management System |
| QMS | Quality Management System |

| RAR | Remaining Accuracy Rate |
|---|---|
| RER | Remaining Error Rate |
| ROC | Receiver Operating Characteristic |
| RSS | Responsibility-Sensitive Safety |
| RTA | Run Time Assurance |
| SASWG | Safety of Autonomous Systems Working Group |
| SFAIRP | So Far As Is Reasonably Practicable |
| SMS | Safety Management System |
| SQEP | Suitably Qualified and Experienced Personnel |
| SRGM | Software Reliability Growth Model |
| STPA | Systems Theoretic Process Analysis |
| TN | True Negative |
| TP | True Positive |
| TSS | The System Shall |
| V&V | Verification and Validation |
| YOLO | You Only Look Once |

## Appendix A
## Example requirements

### A.1    System requirements

For the purposes of this document we have produced the following set of typical high level example safety requirements with criteria for acceptance, which will need to be decomposed into low level design requirements where possible. This is to illustrate the types of requirements and the difficulties which may be had in decomposing these for an actual system.

Some of these requirements can be demonstrated through the implementation of a combination of functional design and operational procedures. Some will be demonstrated by a qualitative approach of arguing legislative compliance. Some will require detailed hazard analysis to demonstrate risks identified, assessed and reduced SFAIRP.

| ID | System requirement | Criteria for acceptance |
|---|---|---|
| 1. | The System Shall (TSS) conform to health and safety legislation and other required standards e.g. Def Stan 00-056 [3] | All non-compliances justified and demonstrated not to impact on safety of system S. |
| 2. | A safety and environmental case shall be provided for the system | All non-compliances justified and demonstrated not to impact on safety of system S. |
| 3. | TSS protect against data errors which could cause unnecessary risk | Review of data errors, where they might occur and their impact. |
| 4. | TSS provide information in sufficient time to deploy response | Design evidence demonstrating time properties with adequate evidence of validity of timing requirements. |
| 5. | TSS reduce the risk of injury or death to involved persons | Demonstrate risk is tolerable and ALARP. |
| 6. | TSS limit the risk of injury or death for non-involved persons | Demonstrate risk is tolerable and ALARP. |
| 7. | TSS meet principles for ML and autonomous systems | Demonstrate no bias, fairness in outcome, protection of data privacy and transparency based on established good practice or guidance[15]. Non-compliances should be justified based on context of use and trade-offs such as operational capability, safety impact and cost. |
| 8. | TSS meet all legal frameworks. | Demonstrate compliance and assess impact on other features. |

**Table 11: Example system requirements**

---

[15] We have not specified good practice or guidance, see [32] for examples.

## A.2    COCO dataset

This section includes some example images from the Common Objects in Context (COCO) dataset[16] used for many image classification machine learning examples. The training and validation data sets form the requirements data set for a classifier.

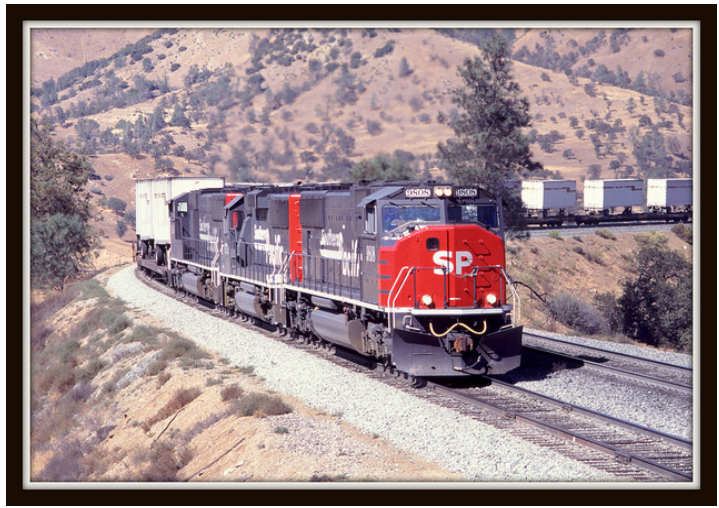

**Figure 44: Training image example – annotations are stored in an indexed file**

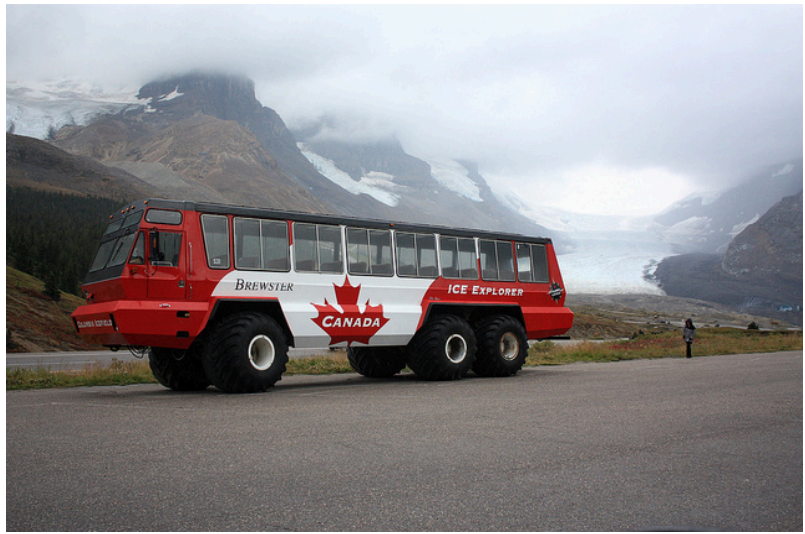

**Figure 45: Validation image example (no annotation)**

---

[16] https://cocodataset.org/

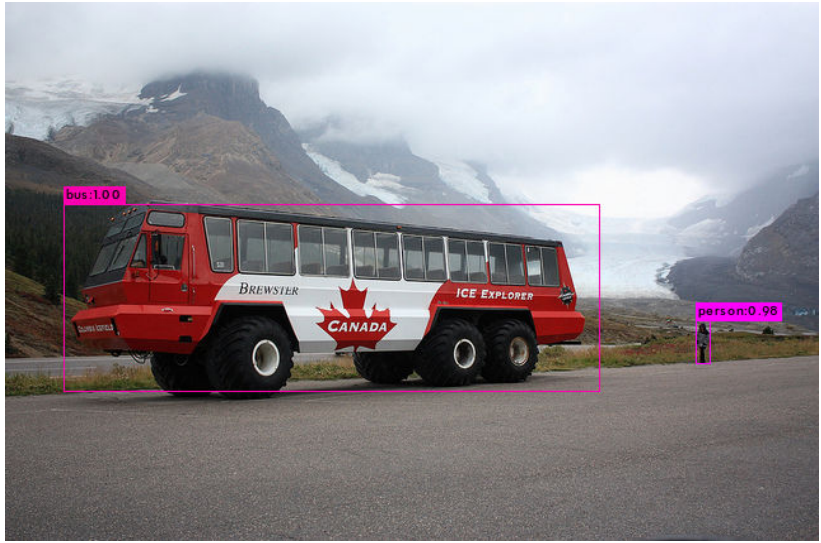

Figure 46: Validation image processed by YOLO

## Appendix B
## Confidence measures for ML

In this appendix, we describe in more detail several confidence measures to show the variety of measures available. In particular, these confidence measures below demonstrate the different uncertainties detected by each measure, and the incomparable meanings of the confidence computed by different confidence measures. This highlights the need to understand fully any such measure before implementing it in a ML system.

### B.1    Conformal Prediction

Conformal prediction [48] is an algorithm for producing predictions from a training set based on the argument that the more similar a prediction is to data we have seen before, the more confidence we should have in this prediction. However, rather than producing a single output, conformal prediction is given a confidence level $C$, and produces a set of possible outputs such that we have confidence $C$ that the true answer is within this set. In some contexts, this may be sufficient, e.g. an autonomous vehicle could still plan a path with confidence $C$ if nothing in this prediction set conflicted with that path.

We first require a measure of how similar two inputs are. Given an unordered set $D$ of examples, consisting of pairs of input data and classification/result, and a particular example $x$ from $D$, we must define a nonconformity measure $A_D(x) = A(D\backslash\{x\},x)$ on the elements of $D$. This nonconformity measure is a representation of the extent to which $x$ conforms to the rest of $D$. The choice of nonconformity measure is application dependent, but can be very simple. For example, in the case where the input attributes and the result are numerical, a typical choice of nonconformity measure would be to define $A_D(x)$ as the shortest distance from $x$ to another point of $D$, i.e. $A_D(x) = min_{y \neq x}(|y-x|)$. Nonconformity measures have also been defined based on the output of other ML techniques, e.g. random forests [51].

Conformal prediction provides an algorithm that takes the previous data, a confidence level $C$ and a new input $y$, and produces a prediction set $P$. This prediction set is computed such that, for a new input taken from the same distribution as the previous data, there is probability at least $C$ that the prediction set contains the true result. It is important to note that this algorithm is deterministic, and a confidence claim is that the algorithm produces a prediction set containing the correct result on at least a proportion $C$ of randomly chosen inputs. The confidence in any specific prediction may be higher or lower than $C$.

For a new prediction $x = (y,p)$, consisting of the input $y$ and a predicted class $p$, let $D$ contain $x$ and all previously seen inputs labelled with the correct classification. Conformal prediction calculates $A_D(z)$ for every point $z$ in $D$, and computes the value $C_p = |\{z: A_D(z) \leqslant A_D(x)\}| / |D|$, the proportion of examples in $D$ which have a lower nonconformity measure than $A_D(x)$. Informally, the pair $x = (y,p)$ would be among the $C_p$ least unusual examples in the data we have seen, and we therefore estimate that the probability of this event happening is at most $1-C_p$.

For a confidence level $C$, define the prediction set $P$ consisting of all possible predictions for which the probability $C_p < C$, i.e. the prediction set is the set of all possible predictions which would rank among the $C$ least unusual examples we have seen. Viewed from a different perspective, if the true class $q$ is not in the prediction set, then $(y,q)$ has a higher nonconformity measure than $1-C_q < 1-C$ of the previous data, and so an event of probability less than $1-C$ has occurred. Observe that as we increase $C$, the size of the prediction set will increase. This is what we would expect – to be more confident our set includes the correct prediction, the set must include more possible classes.

The use of a suitable nonconformity measure allows conformal prediction to be relatively effective at detecting individual out-of-distribution values. An input which is very different from trained data will likely have high nonconformity measure for any prediction, i.e. $C_p$ will be very close to 1 for all $p$, and so the prediction set will likely be empty. However, an important assumption made in the confidence calculations is that the inputs are drawn from the same distribution, so several similar inputs outside of this (whether malicious or not) could lead to inaccurate measures of confidence since these inputs will conform to each other. It may be possible to mitigate this by careful choice of nonconformity measure, or by being more selective with which new data, if any, to add to the set $D$. Conformal prediction is also potentially vulnerable to an adversarial attack via careful manipulation of the input attributes.

The confidence value in conformal prediction is the probability that the algorithm will produce a set containing the correct output, rather than confidence in any specific prediction, i.e. it is confidence in the *algorithm* rather than in the prediction. As a result, the algorithm can produce prediction sets which are apparently nonsensical, such as the empty set, or which are of limited use to the application, such as the set of all possible classes.

Conformal prediction can be adapted to return a single prediction. To do so, we consider all possible predictions, and select the prediction $p$ where $C_p$ is lowest, i.e. the prediction that most conforms with the previous data. Computing confidence in this prediction is not straightforward. The confidence in this prediction is typically given as the largest value of $C$ for which the prediction set contains only $p$. However, since conformal prediction provides a probability of correctness for an algorithm, rather than for an individual prediction, this value $C$ does not represent the probability that $p$ is the correct prediction. For example, in the case of an out-of-distribution input, $A_D(y,p)$ will be high for all predictions $p$, and hence $C_p$ will be very close to 1 for all $p$. As a result, using conformal prediction in this sense will return a prediction with very high confidence when given an out-of-distribution input.

To alleviate this, a prediction is also given a credibility, defined to be $1-C_q$, where $q$ is the prediction where $C_q$ is the second lowest such value, i.e. $q$ is the second most likely prediction. In the case of an out-of-distribution input, this would lead to a prediction with very high confidence, but extremely low credibility. Any guard for an ML sensor would likely require a minimum level of confidence and credibility. A minimum threshold for credibility could be a somewhat effective guard detecting out-of-distribution inputs, but further work would be needed on combining these confidence and credibility scores to provide confidence in the output of a sensor.

## B.2    Inductive Conformal Prediction

The computational resources required by conformal prediction grow quadratically with the size of $D$. To obtain a feasible confidence measure, it may be necessary to limit the size of $D$. Inductive conformal prediction (ICP) [49] is designed to provide such a computationally feasible implementation of a conformal predictor.

Given a training set, ICP removes from it a relatively small calibration set $D$. We learn an ML model $M$ based on the remainder of the training set. We define a nonconformity measure $A$ representing the distance between the output of the predictor and the true value, i.e. for $y = (x,p)$ with input $x$ and output $p$, $A(y) = d(M(x),p)$ for some metric $d$. This could be the proportion of trees in a random forest which predict $p$, or the distance between the output of a neural network and the unit vector corresponding to $p$.

This nonconformity measure $A(y)$ is computed for every data point $y$ in $D$. $A(y)$ depends only on $y$ and so does not need to be recalculated when a new prediction is made. Given a new input $x$, we proceed as for conformal prediction by calculating the nonconformity measure $A(x,p)$ for each possible prediction $p$. This nonconformity score is compared with the nonconformity scores of the examples in $D$, giving a prediction set for a fixed confidence level, or giving a single prediction with some level of confidence and credibility, computed in the same way as for conformal prediction.

Since the nonconformity measure of each prediction depends only on the output of the learned model, we do not need to recalculate these values for the calibration set every time we make a new prediction. As a result, computing the confidence in a prediction requires little further computation beyond running the model on a new input. However, it must be ensured that the calibration set is drawn from the same distribution as the training set.

## B.3 Attribution-based confidence

Attribution-based confidence (ABC) [44] is an alternative approach to computing the confidence in a ML prediction. Confidence in an ML model is lower when the prediction is in a noisy region, or close to a boundary between two classes, and hence small perturbations to the input can have a relatively large effect on the outcome. Attribution-based confidence detects such situations by randomly generating a weighted sample of nearby inputs and judging whether the prediction is stable in the region of the current input.

As described in [44], ABC applies to classification problems and assumes that the ML model is a neural network. However, the features required of the predictor are that the output is in the form of real numbers $F_i$ for each class $i$, and that this output changes continuously, i.e. an arbitrarily small change in input values results in an arbitrarily small change in the $F_i$. These are usually both satisfied in neural networks.

ABC begins by defining a baseline input $b$, which could be an input consisting of all zeroes, an element of the training set, or any other fixed point. For simplicity, assume the baseline is the zero vector and the values of the $F_i$ have been rescaled to be zero on this baseline. We can then define attribution functions $A_j(x)$ for an input $x$, measuring the contribution of the attribute $x_j$ to the overall output $F$, where $F$ is the highest value among the $F_i$, i.e. the value corresponding to the prediction made. Game theory and explainable AI offer a variety of methods for calculating precise values for these attribution functions. The key property required of the functions $A_j$ is that the sum of $A_j(x)$ across all attributes $j$ is equal to $F$.

In generating examples to compare our prediction with, we wish to explore the space of inputs around $x$. However, since $x$ may consist of many attributes, many of which have little impact on the classification, it is not feasible to sample the neighbourhood of $x$ widely enough to be sure of accurately measuring how stable our prediction is. We therefore wish to focus on varying the attributes with the greatest effect on the classification. To do so, define the probabilities $P(x_j) = |A_j(x)/x_j|/\sum|A_k(x)/x_k|$ to be proportional to $|A_j(x)/x_j|$, i.e. the attributes which bring about the greatest rate of change in $F$ have the highest probabilities. We generate sample inputs from $x$ by replacing the value $x_j$ with the baseline value $b_j$ with probability $P(x_j)$. The confidence in the prediction made is then the proportion of these randomly generated inputs for which the predicted class is the same as the predicted class for $x$.

Assuming that the attribution functions $A_j(x)$ accurately represent the contributions each attribute made to the outcome, which would need to be justified, the confidence value returned is a conservative estimate for the result that would follow if each variable were modified with equal probability.

Attribution-based confidence provides a means for detecting when an ML system is stressed or unsure about a prediction without relying on internal features of the ML component. Treating the ML as a black box in this way ensures that confidence values produced correspond to probabilities, and hence are comparable between multiple ML models. Moreover, unlike some other measures of confidence, ABC requires no additional testing or calibration when the ML is updated or retrained.

There is some evidence that ABC is also able to detect out-of-distribution inputs [44], however it is not clear whether this would be the case in every application. In particular, one could question whether it would be as effective if the out-of-distribution input were in a stable region, or produced the same predicted class as the baseline. It is likely that any monitor architecture would require a separate environmental guard as well as an ABC-based guard.

## B.4    Learning confidence

Conformal prediction and ABC both provide potential confidence measures in an ML model which do not require the use of any additional ML components, allowing for simpler and more easily verifiable guards. Several approaches have also been proposed which make use of the existing ML algorithm to compute a confidence measure alongside any other predictions. While these are unlikely to build sufficient confidence on their own, in part due to the potential for common cause failure, they can still contribute to building confidence in the system.

### B.4.1    Learning confidence for out-of-distribution detection

A relatively straightforward approach to adding confidence learning to a neural network is detailed in [43]. This approach requires the neural network to output a confidence value in addition to the probability distribution it already produces. When calculating the loss function for learning, we take a sum of the model's prediction and the true answer, weighted according to the confidence. This ensures that wrong answers with low confidence are penalised less than wrong answers with high confidence. To encourage predictions with confidence as high as possible, the loss function contains an additional term penalising low confidence.

Suppose that given an input $x$, a neural network gives the output $p$, when the expected answer is $y$. Training a neural network consists of optimising weights to minimise the loss function $L_t(p,y)$, which is greater the further $p$ is from $y$. We augment the neural network to produce a confidence value $c$ in the range $[0,1]$. Given the outputs $p$ and $c$, we define the confidence-adjusted output $p' = cp + (1-c)y$, moving the model's output closer to the true answer proportional to how confident the model is in its answer. The loss from the accuracy of the answer is then $L_t(p',y)$, ensuring the model is penalised less for wrong answers when it is less confident in these answers.

It is clear $L_t(p',y)$ is minimised when $c = 0$, since in this case $p' = y$ and $L_t(y,y) = 0$. In order to prevent the model from simply producing random answers with no confidence, we include a loss function $L_c(c)$ penalising low confidence. The overall loss function for this new neural network is therefore $L(p,c,y) = L_t(p',y + kL_c(c)$, where $k$ is a parameter we adjust to strike the desired balance between confidence and accuracy.

The described implementation in [43] includes a couple of additional features to bring about a model with a useful notion of confidence. The first is to include a budget $b$, which is used during training as a target value for the total loss due to confidence $kL_c(c)$. If the total loss from confidence is above $b$, then the value of $k$ is increased to encourage predictions made with more confidence. Conversely, $k$ is decreased when the loss from confidence is below $b$, as the network is being too confident in its predictions. The second feature, to avoid the network converging towards meaningless answers with very low confidence near the boundary of classes, is to apply the confidence correction in only half of the instances when calculating the loss function. This ensures that optimising the loss function still requires a reasonable prediction in these boundary cases, even if the confidence is low.

The experimental data in [43] provides evidence that the confidence learned in this way is effective when used as a threshold to distinguish out-of-distribution examples. Further testing found the confidence threshold minimising detection error in misclassified examples to be very similar to the threshold best for detecting out-of-distribution examples. This suggests that a suitable threshold could be determined without requiring a large out-of-distribution dataset.

There are some potential challenges in using this as a confidence measure when implemented in an ML algorithm. The budget was introduced to prevent confidence from tending towards a constant 0 or 1. However, it may be difficult to justify a requirement that there must be a fixed amount of uncertainty in the predictions. In extreme cases, such as randomly generated noise or very clearly separated data, we would expect the confidence to be very low or very high respectively for all examples in the distribution. Confidence may still be appropriate for detecting out of distribution values, but provides no information likely each prediction within the distribution is to be accurate.

Despite being a value assigned between 0 and 1, the confidence does not represent the probability that the prediction is correct, as one might naturally assume. Indeed, it does not represent a probability at all. Since the confidence value can be rescaled by modifying the budget, the parameter $k$ or the loss function $L_c(c)$, it is not possible to use this confidence measure to compare the relative confidence from two different models. Any guard implemented using this measure of confidence will require extensive testing to determine and justify any confidence thresholds used. This would need to be repeated for any change in the model.

## B.4.2    Confidence in YOLO and object detection

The goal of YOLO [50] and other object detection and classification algorithms is to process input images, and highlight areas of the image containing certain objects and provide a classification of the type of object. The output is in the form of one or more bounding boxes, each of which is labelled with the type of object it contains. The algorithms are also required to produce a confidence value for each bounding box. Unlike some of the confidence values described above, this is not computed after making a prediction, but is computed alongside the bounding boxes and labels as part of the output of the neural network.

In the case of conformal prediction or ABC, confidence is a representation of the probability the prediction is correct. Confidence in these object detection algorithms has a more complex meaning, combining the probability the algorithm gives to the existence of an object in the box with the extent to which the bounding box given overlaps with the true bounding box [50]. Formally, the confidence is expected to be the product of the probability the predicted box $P$ contains an object at all, and $A(T \cap P)/A(T \cup P)$, the area of $P$'s intersection with the true bounding box $T$ divided by the area of their union.

Much like the method above, the computation of the confidence value is part of the trained model, and the accuracy of the confidence value is considered as part of the loss function used in training. As a result, assurance of a system containing YOLO will have many of the same challenges, such as the lack of transparency in the confidence calculation.

Furthermore, the overall meaning of the confidence value given is difficult to determine since it includes terms representing both the correctness of the detection and the accuracy of the box. A confidence value of 50% could therefore mean either "there is a 50% chance of an object here, and if there is one, this box is completely accurate" or "there is definitely an object in approximately this location, but this box is probably fairly inaccurate". In practice, it is likely to be some combination of uncertainty in both directions. Not knowing how to apportion the uncertainty between the location of an object and its existence is an issue if trying to ensure predictable behaviour.

## Appendix C
## Performance of ML based components

In this section we review performance metrics for machine learning (ML) algorithms. We consider

- performance metrics for binary classifiers
- performance metrics in object detection
- experimental performance

Many of the metrics relate to some form of "confidence" but this is not usually a confidence in the reliability of the device or the correctness of the algorithm.

We then relate these metrics to three parts of the assurance case template:

- the performance of the sensor as a component
- the performance of the sensor in context of the guard architecture
- the performance of the overall systems

For each of these we summarise the type of evidence that might be available, how it might be used and what claims might be feasible. This latter is rather speculative but illustrates the disjoints between the metrics needed in a case and the typical measures provided as well as the state of the body of evidence.

### C.1    Performance metrics for binary classifiers

We first define some common performance metrics for binary classifiers. A binary classifier is an algorithm that, given an input, determines whether that input has a given property. We consider throughout an algorithm which returns "yes" or "no" depending on whether it believes an input document to have the relevant property. Many ML problems, particularly in the area of autonomy, are more complex than binary classification, however several of these metrics can be extended to more general classification problems.

Most ML models will produce some numerical value in the course of making a prediction which can be used as a metric of how confidently the prediction is made. In the case of a binary classifier, the output of the model is usually a value between 0 and 1, with the predicted answer being "yes" above a set threshold. This threshold can be adjusted to optimise the performance of the model according to whichever metric has been chosen. In the case of more complex problems such as object classification, where multiple predictions can be made, there are a multitude of different methods for assigning a confidence value to each prediction. We can then consider only predictions with confidence above a certain threshold in order to calculate the performance of the algorithm.

The first such metric one might naturally think of would be *accuracy*, defined simply as the proportion of the inputs on which the model gave the right answer. This is simple to calculate and relatively easy to understand. However, accuracy does not take account of the distribution of the data on which the test is performed. If the model is being trained to recognise a rare phenomenon, which occurs in only 1% of cases, we can achieve 99% accuracy by simply answering "no" to all inputs. More nuanced measures, which allow for different distributions of input data, are therefore much more commonly used.

### C.1.1    Receiver Operating Characteristic

There are two possible types of error a binary classification algorithm can make: returning "yes" when the correct answer is "no", or returning "no" when the correct answer is "yes". We use the terms "false positive" and "false negative" respectively for these types of errors. The terms "true positive" and "true negative" are defined in a corresponding fashion correct responses by the algorithm.

|  | ML responds "yes" | ML responds "no" |
|---|---|---|
| Ground truth "yes" | True positive (TP) | False negative (FN) |
| Ground truth "no" | False positive (FP) | True negative (TN) |

**Table 12: Classification of classifier outputs**

The receiver operating characteristic (ROC) curve plots the true positive rate, $TPR = \frac{TP}{TP+FN}$ (also known as recall) against the false positive rate, $FPR = \frac{FP}{FP+TN}$, as shown in Figure 47. An ideal classifier would correctly identify all the positive cases, i.e. have a TPR of 1, while giving no false positives, i.e. have an FPR of 0. Any classifier which performs better than random guessing would be expected to have a greater TPR than FPR, i.e. be above the dashed line in Figure 47. There is clearly a trade-off between high TPR and low FPR, since setting a lower threshold to give more positive responses will increase both values. Moreover, TPR and FPR are both independent of the ratio of positive to negative inputs, since they calculate only the proportion of each which are correctly identified.

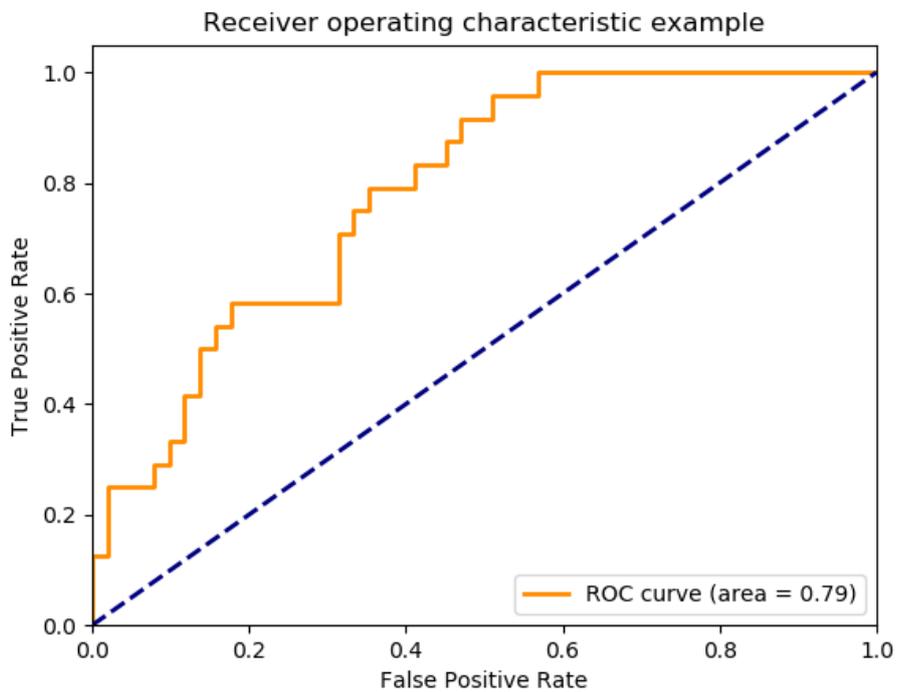

**Figure 47: An example ROC curve**

In order to encapsulate the relationship between TPR and FPR in a single metric, the area under the ROC curve (AUC) is used. The ideal classifier would achieve an AUC of 1, and random guessing, which corresponds to the dashed line in Figure 47, achieves an AUC of 0.5.

### C.1.2    Precision and recall

#### C.1.2.1    Precision

Precision is a measure of how likely it is that an input has the property when the ML answers "yes". Formally, the precision *p* is defined as the proportion of the "yes" responses which were correct, i.e.

$$p = \frac{TP}{TP+FP}$$

Precision is a useful metric in situations where the consequences of a false positive are particularly severe. Examples (albeit very oversimplified), would include facial recognition for access control, or an autonomous vehicle determining whether it is safe to turn right. In both cases, the potential consequences of a false positive are substantially more severe than a false negative.

#### C.1.2.2    Recall

Recall is a measure of "how good the ML is at identifying the property". Formally, the recall *r* is defined as the proportion of genuine "yes" instances identified by the ML, i.e. $r = \frac{TP}{TP+FN}$.

Recall is a useful metric where it is important to avoid false negatives. Examples could include an ML algorithm determining whether an MRI scan is potentially cancerous, or an autonomous system determining whether there is another vehicle in its path.

#### C.1.2.3    Precision vs recall

The measured precision and recall values of an ML model can be adjusted by modifying the threshold at which the model will return "yes". In general, precision and recall are inversely correlated – if the model returns more positive answers, then a greater proportion of the positive cases will be identified, i.e. greater recall, however a larger number of false negatives will also be identified, i.e. lower precision.

Given this correlation, when assessing the performance of an ML model, it is natural to consider the best precision achieved by an ML model for various minimum values of recall. This usually results in a precision-recall curve of the form shown in Figure 48.

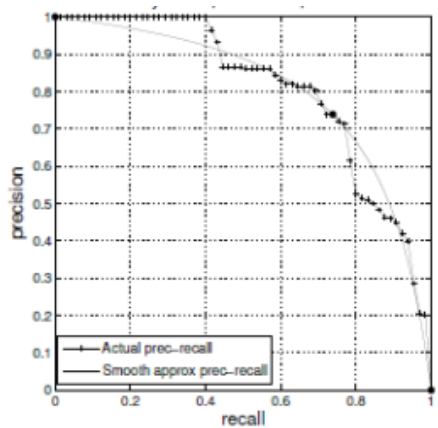

**Figure 48: Example precision recall curve [63]**

This gives rise to the *average precision* metric for measuring the performance of a binary classifier. The average precision metric is calculated by approximating the area under the precision-recall curve.

Formally, if we define precision as a function of recall, i.e. $p(r)$ is the maximum precision we can achieve with recall at least $r$, then $AP \approx \int_0^1 p(r)\, dr$. As with precision and recall, this value is between 0 and 1, with 1 being a perfect classifier. This has the advantage of summarising the overall performance of a classifier, including the trade-off between precision and recall, in a single number allowing for easier comparisons between classifiers. However, the simplicity of a single number can hide more complex details that may be relevant in some situations, such as the examples given above where precision may be considered more important than recall or vice versa. Claims in an assurance case are claims about the specific classifier with a particular threshold for making predictions. The precision and recall at this particular threshold are therefore more relevant metrics than the average precision across the entire precision-recall curve.

## C.1.2.4    R-precision

R-precision is an alternative method for measuring the performance of a binary classifier with a single metric. The different values of precision and recall used to obtain the average precision are obtained from a fixed set of test data by adjusting the threshold required to give a positive answer.

The R-precision of a classifier is obtained by measuring the precision when the threshold is set such that the number of "yes" responses by the classifier is the same as the number of "yes" responses in the test data. For example, suppose the performance of a classifier was measured using a test set of 100 inputs, of which 60 had the desired property. The threshold would be calibrated such that the classifier returned "yes" on precisely 60 of the inputs. If, of these 60 inputs, 45 were true positives, the R-precision of this classifier would be 0.75.

It has been observed that R-precision is strongly correlated with average precision. A geometric argument as to why this is the case was put forward in [63], by observing that the precision-recall curve must pass through the point *(R,R)*, where $R$ is the R-precision, since when the classifier returns the correct number of "yes" answers, the number of false positives is equal to the number of false negatives, and hence precision and recall are equal. If we also assume that the curve passes through (0,1) (by answering no to everything) and (1,0) (by answering yes to everything), then a linear approximation of the curve gives the area under the curve as $R$.

Given the similar shape of most precision-recall curves, we can use R-precision to provide an approximate precision-recall curve based only on this value, passing through (0,1), (R,R) and (1,0). The curve suggested in [63] for this model is $p(r) = \frac{1-r}{1+\alpha r}$, where $\alpha = (\frac{1}{R} - 1)^2 - 1$ to ensure the curve passes through (R,R).

Figure 49 gives examples of this curve for various values of R, with the curve $p(r, k)$ modelling the curve for $R = 0.1k$ Observe that in the case R = 0.5 (i.e. k = 5), the curve becomes a straight line, and the model is no better than random guessing. Values of R lower than 0.5 are worse than random guessing, although inverting the response given would rectify this problem, producing a classifier which outperforms random guessing.

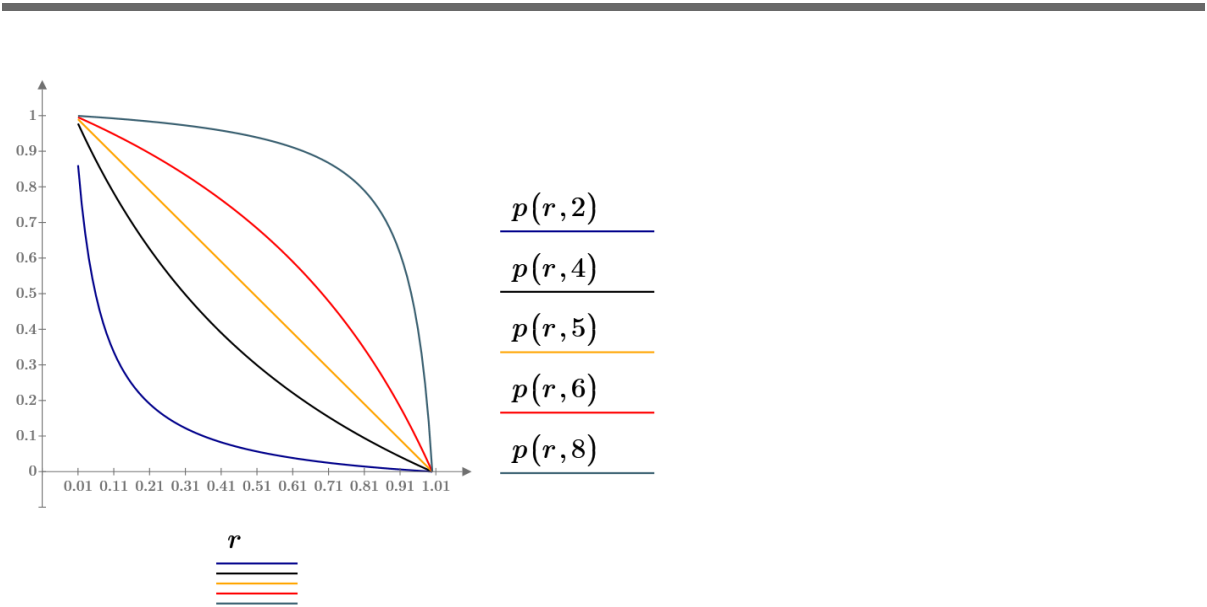

**Figure 49: Example estimated precision-recall curves for different R-precision values**

From the point of view of requirements for an ML model, we are likely to have minimum requirements on both the recall and the precision. This results in a tolerable region in the top right corner of the graph in Figure 49, a region which can only be reached by a precision-recall curve with a sufficiently high R-precision. The precise shape of this region (i.e. is it simply a box formed by absolute lower bounds on precision and recall, or do we allow some trade-offs between them) will determine whether R-precision alone would be an appropriate measure for assessing the performance of ML in an autonomous system.

### C.1.2.5    F score

An alternative method which is sometimes used in the context of machine learning to combine the precision, $p$, and the recall, $r$, into a single measure is the $F_1$ score. The $F_1$ score is the harmonic mean of the precision and the recall, i.e. $F_1 = \frac{2}{\frac{1}{p} + \frac{1}{r}} = 2\frac{p \cdot r}{p + r}$. As with many other measures, the ideal scenario is an $F_1$ score of 1, achieved when precision and recall are both 1, whereas the lowest possible value is 0, achieved when precision and recall are 0.

Unlike R-precision, which is a measure of the performance of the model and does not account for the particular threshold used, the $F_1$ score depends on the precision and recall values given at a particular threshold. However, since $F_1$ is a measure of performance, its value can be optimised over all thresholds to determine the best performance of a model. If the precision-recall curves are assumed to follow the approximation from [63] shown in Figure 49, then the maximum $F_1$ score is also equal to the R-precision; this maximum is obtained when the precision and recall are equal to the R-precision. The maximum $F_1$ score can therefore be used in a similar fashion to R-precision.

The $F_1$ score can be generalised to account for the relative importance of precision and recall in a particular application. In an application where recall is considered β times more important than precision, the performance would be measured using the $F_\beta$ score $F_\beta = (1 + \beta^2)\frac{p \cdot r}{\beta^2 p + r}$. Measuring the maximal value of $F_\beta$ would provide a measure of the performance of a classifier weighted suitably towards recall (or precision, for values of β less than 1).

## C.1.3    Metrics incorporating confidence

Various approaches have been proposed to model confidence in the prediction of an ML model, from simply comparing the relative values provided by each output of a neural network (NN) classifier, to learning the confidence as an output of the NN itself, or various techniques using sampling methods. To incorporate this confidence into a measurement of the performance of an ML algorithm, two measures were proposed in [64].

Fix a confidence threshold $t$, and consider any predictions made with confidence at least $t$ to be "certain". Predictions with confidence less than $t$ are "uncertain". The following metrics are defined:

- Remaining Error Rate (RER): the proportion of inputs that are classified incorrectly with certainty, i.e. confidence above threshold $t$.
- Remaining Accuracy Rate (RAR): the proportion of inputs that are classified correctly with certainty, i.e. confidence above threshold $t$.

RER/RAR are fairly incomparable with precision/recall, since precision and recall measure the performance of an algorithm in identifying a particular class, whereas RER and RAR consider the performance on all classes. Even in the case of a binary classification problem, RER and RAR recognise three classes since predictions can be uncertain. RAR bears some similarity to recall, in that both are measures of the proportion of inputs which are correctly classified by the ML. However, recall measures only the proportion of objects in a specific class which were correctly identified by the ML, whereas RAR is measured across all classes. Similarly, it is tempting to relate RER to precision as measures of how often the predictions are wrong, but directly comparing RER and RAR with precision and recall would be unhelpful as they measure performance on two distinct problems.

The motivation behind definition of these metrics is the assumption that a system relying on the predictions made by this model has some safe backup strategy for when it has low confidence in the outputs of the ML, e.g. an autonomous vehicle can slow down and stop. As a result, unsafe situations occur only when the ML provides a certain prediction which is incorrect. RER is therefore a measure of how frequently such unsafe predictions are made, and should be as low as possible.

However, it is possible to minimise RER by only making uncertain predictions, giving an RER of 0. Given the safe backup strategy may be relatively ineffective, as in the case of a vehicle stopping, this would result in a very safe, but very ineffective system. RAR is a measure of how often the ML provides correct outputs with certainty, allowing the system to operate safely and effectively, and is a value that should be maximised.

The values of both RER and RAR can be adjusted by altering the threshold $t$ for a certain prediction. Decreasing this threshold will result in more certain predictions, increasing both RER and RAR. This trade-off is shown in Figure 50 with experimental data for various approaches to measuring confidence in NNs trained for two datasets: the Canadian Institute For Advanced Research (CIFAR-10) dataset requiring classification of images into one of 10 classes, and the German Traffic Sign Recognition Benchmark (GTSRB) dataset requiring classification of images of German street signs. In both cases, the learned confidence (LC) and deep ensemble (DE) methods performed best.

While this does not provide a single measure with which to compare ML models and confidence calculation methods, it may still be of use in defining requirements for ML, in the sense that RER and RAR roughly correspond to how safe and how effective the ML is (the precise interpretation would depend on the context). The requirements would therefore specify maximum and minimum thresholds for RER and RAR respectively, resulting in a tolerable region in the top left of the graphs in Figure 50.

Some of the methods tested in [64] attained an RAR above 80% with an RER below 0.1% on the GTSRB data set. While these results are on a relatively limited dataset, and would still need significant improvement to be considered sufficiently safe and effective, they suggest that any requirements defined in these terms may be feasible. It is interesting to note that the results obtained for LC and DE are fairly similar on all data sets considered, despite the fact that the confidence value produced by LC is usually lower than that produced by DE (Figure 51).

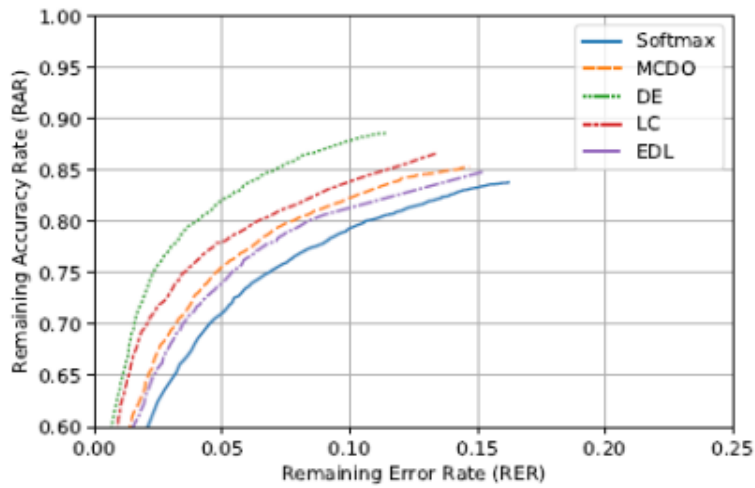

(b) CIFAR-10

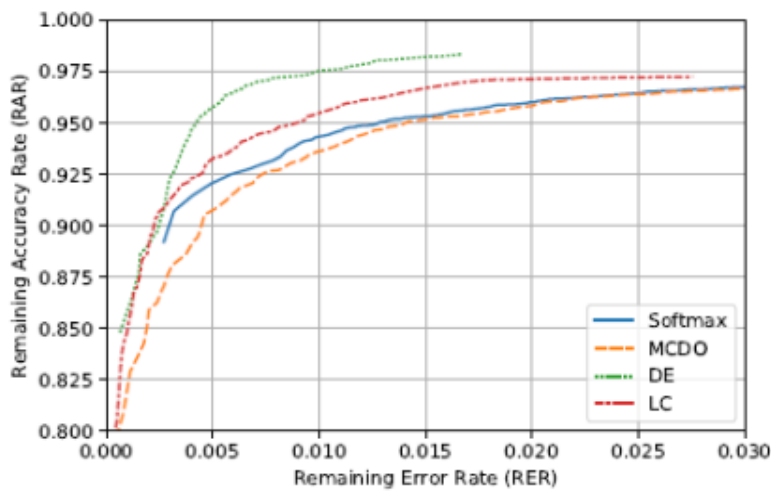

(c) GTSRB

Figure 50: Trade off between RER and RAR for various confidence measures

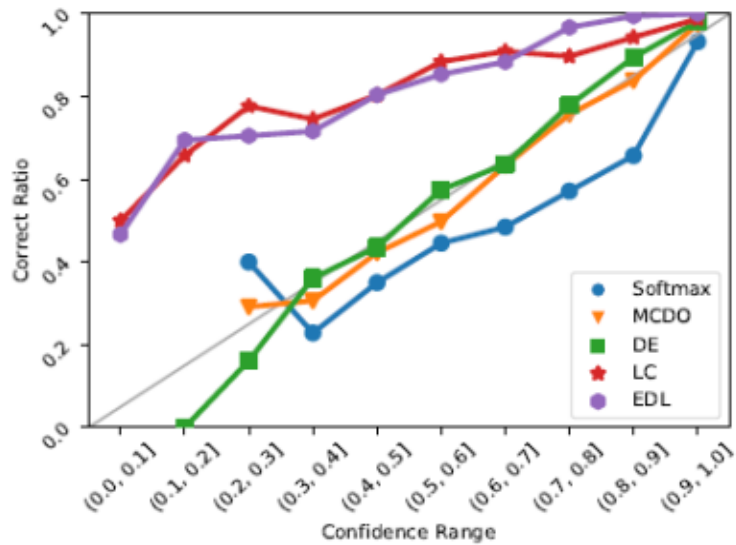

**Figure 51: Calibration of confidence values**

The reason this does not rely on the calibration of the confidence values is shown in Figure 52, which provides a cumulative count of the proportion of correct and incorrect predictions for each confidence level. Despite LC generally giving lower confidence than DE, both confidence measures produce a relatively large distinction between the correct and incorrect predictions for a suitable choice of threshold (e.g. compare the thresholds 0.7 for LC and 0.9 for DE). It is important to note that due to this lack of precise calibration of confidence levels, it is not possible to provide a confidence threshold as a requirement. Requirements would need to specify performance in terms of RER and RAR, and a suitable confidence threshold must be determined each time an ML model is trained or retrained.

Very generally, reducing the RER corresponds to giving lower confidence to incorrect predictions (i.e. moving the orange line higher) and decreasing the number of errors made, whereas increasing the RAR corresponds to giving higher confidence to correct predictions (moving the blue line lower) and increasing the effectiveness of the system. Given the details shown in in Figure 52, there would appear to be more scope to reduce confidence in incorrect predictions.

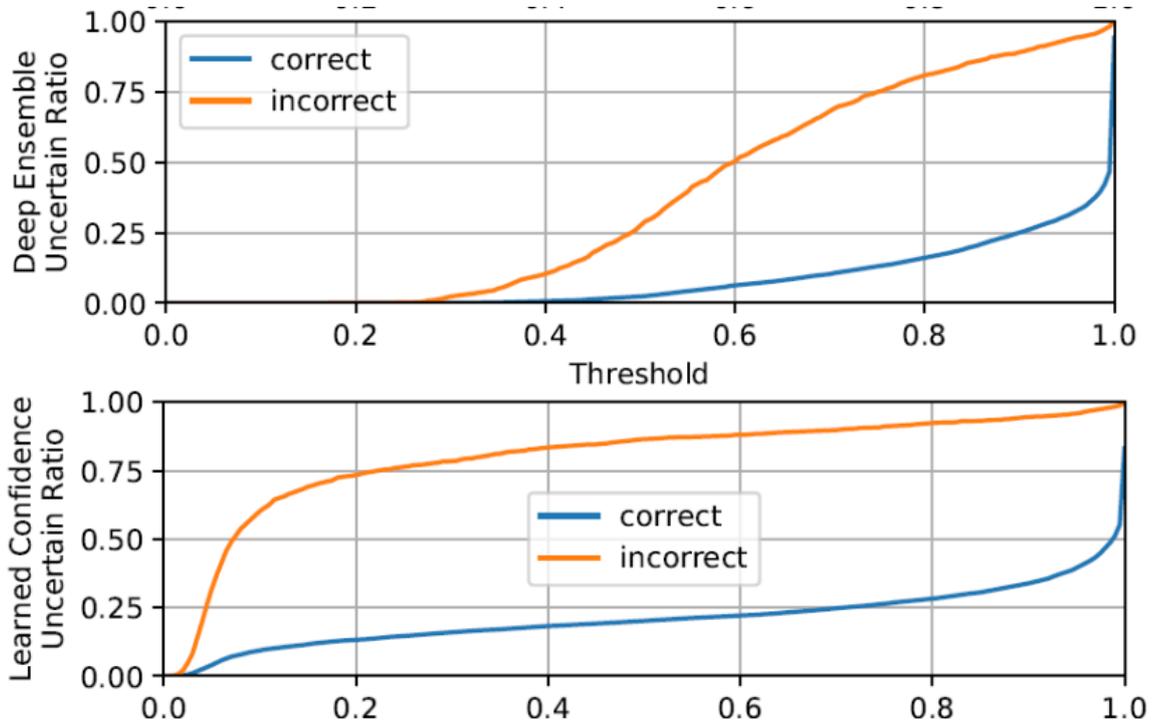

Figure 52: Proportion of uncertain predictions at given threshold

## C.2    Object detection

### C.2.1    Performance metrics in object detection

Performance measurement in object detection and classification is somewhat more complex than a binary classifier. Aside from the addition of more categories, which can be handled by the above measurements for binary classifiers, there are two further complications:

- The ML can (and is expected to) make multiple predictions on a single image
- Predictions can have varying degrees of correctness, depending on how closely the location matches the object in question

The "correctness" of a prediction is usually measured in terms of intersection over union (IoU). The IoU of a prediction is defined as the area of the intersection of the predicted region and the true region, divided by are of their union. Using the IoU does omit some information about the accuracy, e.g. predictions which are twice the actual size, and half the actual size will be equally correct, but it is the generally accepted means of determining whether a prediction is correct.

The suitability of IoU as a means of determining whether a prediction is correct may depend to a certain extent on the intended application. Figure 53 provides examples of predictions (in green) compared to the ground truth (in black), all of which have IoU of 0.5. It can be seen that errors in size are penalised less than errors in position.[17] This is not necessarily intuitive and, in certain applications, may not be desirable.

---

[17] The first prediction is 41% too large in both dimensions, whereas the second prediction is displaced by 18% in each dimension.

Indeed, it has been found to be difficult to train humans to determine the difference between IoU values between 0.3 and 0.7 [67].

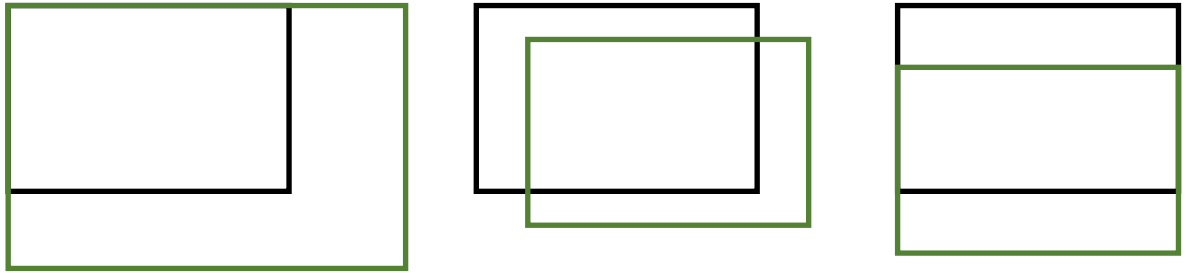

**Figure 53: Example predictions with IoU = 0.5**

All measures we discuss subsequently assume the use of IoU for determining whether a prediction is correct, but could equally apply to another (continuous) measure of the accuracy of a prediction ranging between 0 (no overlap) and 1 (perfect prediction). Some alternatives or modifications to IoU have been proposed [68].

Many object detection algorithms are tested on the COCO data set [65]. Several metrics based on average precision are defined on this dataset. The first such metrics are the average precision for a fixed IoU, i.e. predictions are considered correct if the IoU of the prediction is above some fixed value $k$, and the average precision is taken as above by altering the threshold required to make a prediction. For an IoU value $k$, this is denoted $AP^k$. In general, higher values of $k$ result in a lower average precision. The average precision $AP^{0.5}$ is a common metric that has historically been used for measuring the performance of object detection algorithms.

The overall average precision, also named "mean average precision" is the average of $AP^k$ for values of $k$ between 0.5 and 0.95 in intervals of 0.05. This average precision measure is also measured when only considering objects of a particular size – in general object detection algorithms perform better on larger objects.

Another measure used to measure performance on the COCO dataset is average recall. To measure average recall, the algorithm is permitted to make a fixed number $d$ of detections on each image. The recall is then measured for each category at a number of different IoU thresholds, and the average taken to produce the average recall $AR^d$. Given the relative consequences of a false negative and a false positive in many situations in autonomy, this metric may be more useful than average precision. However, the goal of both average precision and average recall in object detection is to allow for comparisons between algorithms, and they provided relatively limited information on the absolute performance for a particular configuration.

## C.2.2    Analysing types of error

### C.2.2.1    False positives

Average precision results on the COCO dataset can also by analysed by type of error. This is inspired by the work of Hoiem et al. in [66]. F Four different types of false positive are identified:

- Localisation error – identifying a correct object, but with IoU below the threshold (but still above 0.1). This includes duplicate detections.
- Similar object error – identifying a similar object, e.g. detecting a cow instead of a horse.

- Dissimilar object error – correctly detecting the presence of an object, but the prediction is not a similar object to the true object, e.g. identifying a dog instead of boat.
- Background error – all other errors, generally confusing aspects of the background, or objects not in the list of categories, with an object of interest, e.g. confusing a cloud for an aeroplane.

These errors are considered sequentially, so any detection which is incorrect, but is correct with the localisation error removed (i.e. setting the IoU threshold to 0.1) is considered a localisation error. If a cow is detected instead of a horse, and the IoU is 0.2, this will be considered a similar object error, despite also being an error in localisation.

To measure the impact of these different types of errors in different categories, an approach similar to R-precision was adopted. For each category in the dataset, the $N$ most confident detections were considered, where $N$ is the number of true instances of that category in the dataset. Figure 54 summarises the results obtained in [66] for two different detectors.

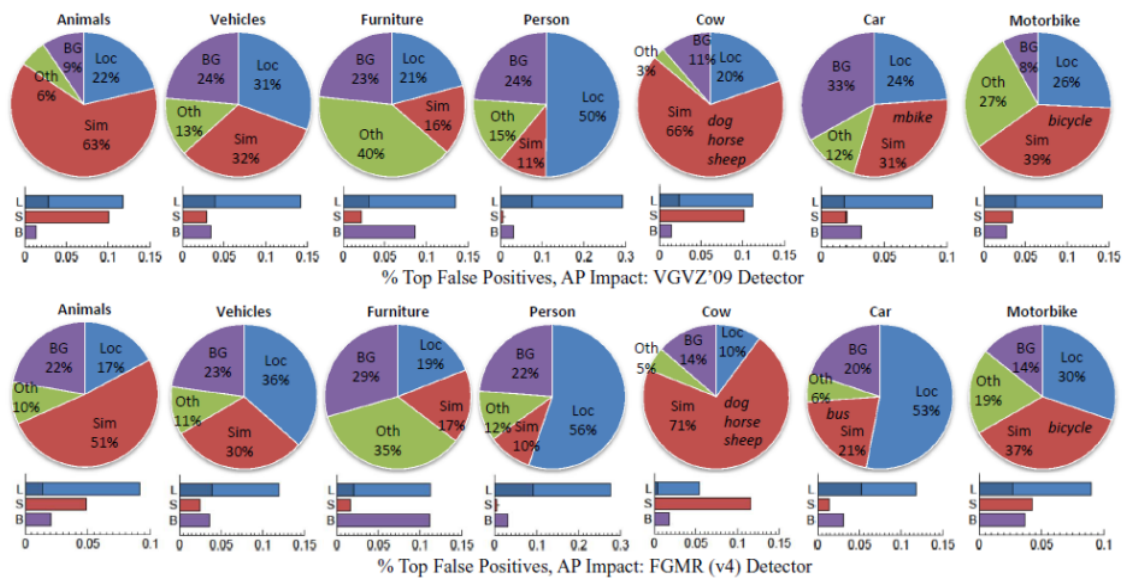

**Figure 54: Error types in object detection [66]**

The results show that the proportion of false positives resulting from confusion with the background is relatively low, usually around 20% and at most 33% in the case of a car. In the majority of cases where the object detection algorithms have a false positive, they have correctly identified the presence of an object, even if it is in the wrong place or labelled incorrectly.

The type of error also varies greatly between categories. A large proportion of errors in identifying animals were due to confusion with similar objects; given the number of four-legged animals with similar proportions among the categories, this is perhaps unsurprising. If there is frequent confusion between similar categories of object, and this confusion is considered to be acceptable, then this suggests that the categories could be reconsidered to treat these as the same object.

On the other hand, the majority of errors made when falsely detecting people were localisation errors. If evidence from the testing of object detection algorithms is sufficiently detailed, then the relative impact of these different types of errors could be taken into account when measuring the performance of an object detection algorithm. It is likely each of these error types would need to be identified independently, rather than the successive elimination of errors described in [66].

### C.2.2.2    False negatives

The relatively low proportions of false positives arising from confusion with the background could be used to argue that average recall is a more useful performance metric for object detection; since the false positives that arise from providing too many predictions are generally "not completely wrong", and false negatives are likely to have a greater impact on the safety of the system. However, the current best average recall performance documented on the COCO website provides a recall of only 0.69, even when permitted to make 100 detections per image. This is likely to be at least an order of magnitude off what would be required by a model relying solely on such a detection and classification algorithm to interpret its surroundings.

The work in [66] also attempts to provide some insight into the reasons behind false negatives. To do so, they introduce a modified version of precision in order to be able to compare precision more fairly across datasets and object categories; the definition of precision above is affected by the number of genuine positives in the dataset. This "normalised precision" fixes a value $N$, ideally close to the average number of positive instances across all categories, and defines the normalised precision $p_N(t)$ at a threshold $t$ to be

$$p_N(t) = \frac{r(t) \cdot N}{r(t) \cdot N + f(t)}$$ where $r(t)$ is the recall at the threshold $t$ and $f(t)$ is the number of false positives.

The results of this study were not particularly conclusive, although it seems clear that there are fewer false negatives for larger objects, and objects which are less occluded by other objects. This can sometimes be tempered by other patterns in the training data; for example, performance on objects such as bicycles or chairs, which are frequently partially occluded in the training data, is less variant under occlusion.

### C.2.3    Object tracking

In the context of autonomous vehicles, object detection and classification contributes to building a model of the world, allowing the AV to identify hazards and plan a safe route. Object tracking aims to identify and track the trajectories of multiple objects in a sequence of images (i.e. a video). Understanding these trajectories allows the AV to predict the future movements of tracked objects and avoid collisions.

Object detection is clearly a key element of object tracking, and most algorithms for object tracking consist of an object detection algorithm and a tracking algorithm to process the predictions made by the object detection. In contrast to object detection as considered above, objects generally do not need to be detected in every image to be tracked.

Common metrics measured to evaluate the performance of object trackers include:

- Mostly tracked (MT): objects which were tracked for more than 80% of the time they were in frame.
- Mostly lost (ML): objects which were tracked for less than 20% of the time they were in frame.
- Identity switches (IDS): instances where the same object is identified as different objects in different frames.
- Fragmentations (FM): the number of gaps in trajectories of the same object.
- False positives (FP): the number of false positive detections.
- False negatives (FN): the number of false negatives.
- Multiple object tracking accuracy (MOTA): the total number of false negatives, false positives and identity switches, divided by the total number of objects across all frames.

- Multiple object tracking precision (MOTP): the average dissimilarity between each identified object and its true position.

For understanding the performance of object detection algorithms, the false positive and false negative metrics are likely to be the most useful. The MT and ML metrics provide information on what proportion of objects were consistently identified or consistently missed, respectively. An AV need not detect an object in every single frame, and so MT and ML may be suitable metrics for assessing what proportion of objects are detected sufficiently frequently to be tracked.

Other metrics include some aspects of tracking, and so it is more difficult to associate these directly to the performance of object tracking algorithms. Nonetheless, evidence of the performance of object detection and tracking against these metrics is likely to be relevant in other aspects of the case. The consequences of inconsistent identification and tracking of objects can be severe. FM and IDS provide metrics which could be applied to the overall system for object detection and tracking (which may consist of several different sensors, including cameras, LIDAR, etc.).

## C.2.4    Additional reasoning

Even in ideal circumstances, the best object detection algorithms rarely achieve precision and recall scores above 0.9 when predictions are made on a single image. Some improvements in this performance can be made by using an ensemble of similar algorithms, or by including additional information such as GPS data in the detection but this is still significantly below the accuracy that would typically be required of a safety critical system.

In general, cameras and object detection algorithms, and other types of sensor, are used by an autonomous system to construct a model of its environment. In the case of an AV and object detection, this is a model of the physical environment, including different types of objects, their locations, motion, inferred intention, etc. Part of this model will be predictions for how the environment will develop over time, in order to plan future actions. As part of these predictions, the model can predict what data it expects to receive from the sensors. These predictions could be provided to an object detection algorithm with the aim of improving prediction accuracy by incorporating data from other sensors, in the form of predicted locations of objects.

For example, the R-CNN object detection algorithm first identifies regions of an image which are most likely to be objects, and then attempts to classify the type of object contained (or not) in each region. It is clear how such an approach could be extended to incorporate the predicted locations of objects provided by a model of the environment. When given prior knowledge on which features are expected to appear in which regions, R-CNN achieved a relatively small increase in $F_1$ score from 0.60 to 0.63 [70]. Further experiments would be needed to understand the extent to which this gain would be replicated in object detection for autonomous vehicles. However, given the frequency of localisation errors discussed in Section C.2.2, it is plausible that this could provide a significant improvement.

An alternative approach to reasoning using the predictions of the model is to require high confidence in the output of the resulting model, rather than in any individual sensor. One such approach is the predictive processing proposed in [71]. The predictions made by the sensors are compared with predicted sensor outputs based on the model of the environment. If all sensors make predictions with high confidence, and these predictions only differ slightly from those expected based on the model, then we have high confidence that the model is an accurate representation of the environment.

However, if there is a large error between the prediction of the model and the prediction of a sensor, then this must be resolved. If the error is only present in one sensor, or one group of related sensors, e.g. all front-facing cameras, then it may be determined that this is a sensor error, and the model can be updated based on the data from other sensors. If such errors persist, then it may be due to some external cause, e.g. the cameras are being dazzled by the sun. Finally, if a large number of sensors provide a prediction error, then it is likely that the world has not developed as predicted by the model, e.g. a new object has appeared, or an object has moved in an unexpected way. In this case, the model must be updated to reflect the new data.

Using such a predictive processing model presents a possible approach to arguing that an autonomous system's interpretation of its environment is accurate with a sufficiently high degree of confidence, without requiring a potentially infeasible level of performance from an individual sensor. If a sufficiently high level of accuracy, e.g. measured by precision and recall, can be achieved, then these can support a higher level of reliability in the model. For example, if 5 independent sensors each have precision and recall of 0.9, then the "majority verdict" has precision and recall of 0.991.

Demonstrating the reliability of the model in this way is similar to increasing performance using ensembles. In particular, we require evidence that the failure of sensors is independent. There could be many causes of systemic failures, such as incomplete training data, or sensors having reduced performance in the rain. The impact of such systemic failures on the performance of the model will depend on the precise nature of the failure. Reduced performance in rain will affect the accuracy of the sensor data, however the errors are unlikely to be consistent, and these errors are likely to be identified. On the other hand, if an object is not seen in the training data, then the sensors may consistently fail to identify it, leading to an incorrect model. To support a claim for independence of the sensors, we would require evidence identifying any potential sources of systemic errors, and arguments as to why they do not occur.

However, the predictive processing model also allows for additional reasoning about the model itself, which can increase our confidence further. One such method is to keep track of any uncertainties in the model, which can then be resolved by further observations. For example, if an object is identified but the sensors cannot determine whether the object is a bicycle or a pedestrian, this uncertainty can be maintained within the model (or as multiple models) until the object can be identified from further observations. Assurance of such multiple models would likely take the form of showing that the true state of the world is represented within the space of possible models with sufficient confidence, and that the actions planned and taken by an AV are safe in all possible models.

The model would also be able to track objects which are partially or totally obscured. This includes both previously detected objects which have since become hidden, and potential objects in areas which are not visible to the sensors, e.g. if a parked vehicle is blocking the view of the road. Doing so would be necessary to ensure the accuracy of the model. Some work on detecting and representing the properties of objects and their relative positions in an image, and answering queries regarding these, has been performed in [72]. Using a combination of YOLO and answer-set programming (ASP), this algorithm was able to answer 93.7% of queries correctly. The vast majority of the errors were caused by incorrect object detection with YOLO, and we assume that the small number of errors arising from incorrect parsing of the natural language queried could be eliminated. It is not clear how well this performance will translate to the objects of interest to AVs, but it is plausible that ASP could be used to develop systems that can reliably reason about the relative positions of objects in a model.

## C.3    Experimental performance

### C.3.1    Object detection

In the case of most ML algorithms trained for object detection, the performance is measured using mAP, as for the COCO dataset, or using the average precision $AP^k$ for some $k$, usually 0.5. The best performing algorithm for the COCO dataset currently recorded on *cocodataset.org* achieved an mAP of 0.588, with an $AP^{0.5}$ of 0.766.

YOLO (You Only Look Once) [74] is a state-of-the-art object detection system capable of operating in real-time, i.e. analysing tens of images per second, which has been the subject of a number of studies. One such study is documented in [75], analysing the performance of YOLO at detecting traffic lights, and their status, on the public LISA traffic light dataset captured in San Diego. The measures used in this case are based on the precision-recall curve, with a correct detection being defined as any having IoU at least 0.5.

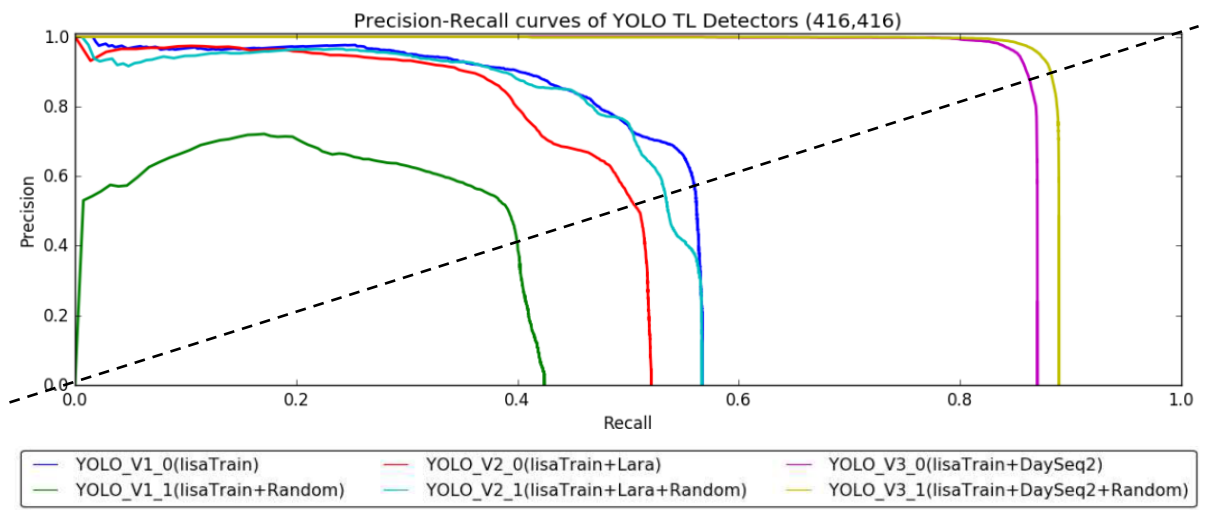

**Figure 55: YOLO performance for traffic light detection [75]**

The precision-recall curves obtained when YOLO is trained on different training sets are depicted in Figure 55. All training sets included the LISA training data. The LARA dataset is an alternative captured in Paris, and DaySeq2 is an additional test set from the LISA dataset. In cases where the training data includes the "random" parameter, the images in the training data were randomly resized between 320x320 and 608x608.

The best performance achieved in this case achieves an average precision (i.e. area under the curve) of 0.885, with a maximum recall of 0.889. This is very high for object detection algorithms, presumably due to most object detection algorithms being trained to detect a much wider variety of objects. The results of Figure 55 also serve to highlight the impact of suitable training data on the performance of ML. The benefit of additional training data very similar to the test data is clear to see from the performance of the models trained on DaySeq2, whereas the effect of the LARA data is less clear, but certainly less effective.

A further observation can be made on the relevance of R-precision to the performance of these algorithms. We can estimate the R-precision for each by its intersection with the dashed line in Figure 55, since a precision-recall curve passes through the point *(rp,rp)*. In all cases, the algorithms maintain relatively high precision for recall values below the R-precision, indeed, higher than might be predicted by the estimated precision recall curve of [63], whereas there is a sharper drop in precision at recall values higher than the R-precision. Nonetheless, R-precision relatively accurately identifies the beginning of this sharp drop in precision in the examples above, so may have use as a measure of the performance of such algorithms.

A recall of 0.889 would likely not suffice if this were the only means of detecting traffic lights in an autonomous vehicle. However, it is possible to combine the data from such object detection algorithms with other sources of data, such as location data containing the expected locations of traffic lights. This method was experimentally implemented in [76], also using YOLO for the object detection algorithm. In this case, YOLO was trained to recognise only two classes of object – green traffic lights, and red/yellow traffic lights. The predicted locations of traffic lights in the image were compared with previous data estimating the location of traffic lights relevant to the path of the vehicle, and the closest prediction, if any, was taken as the state of the expected traffic light. The training set included sequences from DTLD (Germany), LISA (USA) and IARA (Brazil).

Without the benefit of GPS information, the performance of this traffic light detection varied significantly by test set, with average precision (area under precision-recall curve) of 0.85 on a test set from DTLD, but only 0.51 on a LISA test set, and 0.55 on an IARA test set. This is indicative of the challenges faced in developing autonomous vehicles able to operate safely in a variety of environments. We would therefore expect evidence to clearly state the location of the training and test data.

The addition of the GPS information resulted in a significant improvement in the performance of the detector, correctly identifying 96% of red traffic lights which were within 100 metres of the vehicle when given the estimated location of the light (i.e. a recall of 0.96). The precision when predicting a green traffic light was 0.94 using the same threshold for predictions. This is a significant improvement on the recall and precision that may be expected from an average precision of 0.85, and an even greater improvement from the 0.55 average precision of the IARA test set, which is arguably the most similar test set to that used to test the integration of GPS information.

It is worth noting that the errors generally consist of failing to detect a traffic light, rather than mistaking green for red or vice versa. In a vehicle which knows when and where to expect relevant traffic lights, it may be possible to safely deal with this, e.g. by assuming a missing traffic light to be red. Moreover, these errors occurred far more often when the vehicle was a considerable distance from the traffic light; it could be argued that these errors are less significant than errors closer to a traffic light. An error was also noted when the ground truth was a red light, but the light was obscured by a lorry passing in front of it. The relative significance of all these different types of errors would need further consideration in requirements for any sensor; however, it suggests that specifying the performance of a sensor may require a more detailed perspective than a single measure. For example, we may have a claim about the overall performance of the traffic light detection algorithm, and a separate claim about the frequency with which a red light is mistaken for a green light.

## C.3.2    Object tracking

Multiple object tracking is an active area of research, and as a result there are a number of common datasets available to provide performance benchmarks. In general, the annotated objects to be detected are either pedestrians (e.g. MOT16 dataset) or vehicles (e.g. UA-DETRAC dataset). Datasets have been collected using both stationary and cameras mounted on vehicles.

To provide an overview of the current state of object tracking, Table 13 contains some data taken from [69] on the performance of various trackers on the MOT16 dataset according to the metrics described above. The MOT16 test set contains a total of 759 tracked objects across 5,919 frames, with a total of 182,326 individual objects.

To separate the performance of trackers from object detection, the MOT16 dataset is also published both as simple videos, and as videos annotated with the objects detected by a fixed object detection algorithm. Following [69], we refer to these as private detection and public detection respectively. The difference in performance between the two serves to highlight the influence of object detection performance on object tracking performance.

| Algorithm | Detection | MOTA | MOTP | MT | ML | FP | FN | FM | IDS |
|---|---|---|---|---|---|---|---|---|---|
| Customized Multi-Person Tracker | Public | 49.3 | 79.0 | 17.8 | 39.9 | 5,333 | 86,795 | 535 | 391 |
| Non-uniform hypergraph learning based tracker | Public | 47.5 | -- | 19.4 | 36.9 | 13,002 | 81,762 | 1,408 | 1,035 |
| Person of Interest | Private | 68.2 | 79.4 | 41.0 | 19.0 | 11,479 | 45,605 | 1,093 | 933 |
| Lifted Multicut and Person Re-identification (LMPR) | Private | 71.0 | 80.2 | 46.9 | 21.9 | 7,880 | 44,564 | 587 | 434 |

**Table 13: Performance on MOT16 benchmark dataset**

The FP and FN metrics allow us to compute precision and recall scores to provide some comparison with metrics for object detection algorithms. In the case of LMPR, the corresponding precision is 0.946, and recall 0.756. This recall is similar to the recall achieved on the COCO dataset by the best performing detectors, but with substantially higher precision. The increased precision is probably due to the more limited set of object that are being detected, as is the case for traffic lights.

It is notable that the proportion of tracks which were mostly lost is 21.9%, which is not substantially lower than the complement of the recall, i.e. the proportion of objects which were not detected. We can make similar observations for the other trackers in Table 13. This raises the possibility that those objects which are missed by a detector in an image are consistently missed in subsequent images, which casts doubt on any claimed performance gain by detecting objects across multiple images.

## C.3.3    Accidents and disengagements

Experimental data can also be gathered for the operation of the autonomous vehicle, rather than for a specific sensor. Data on accidents and disengagements of autonomous vehicles has been the subject of a number of studies. This is in part due to the relative accessibility of data – California requires all AVs tested on public roads to submit reports of each accident and disengagement. All testing is done with a human backup driver, and two types of disengagement are reported:

- automatic disengagement, where the autonomous system identifies a failure in the system or a situation where it is unsure what action to take
- manual disengagement, when the human driver believes that immediate manual control is necessary to prevent a dangerous situation

For each disengagement, a brief summary of the reason for the disengagement is also reported. A summary of such reasons in 2016 is shown in Figure 56. Human factors includes almost all manual disengagements, including for reasons such as lack of trust in the AI, or driver discomfort. System failure covers a wide range of failures, and in many cases could be attributed to a specific subsystem by the manufacturer. External conditions include events such as bad weather, heavy pedestrian traffic, poorly marked lanes, or other environmental conditions in which the AI has not been trained or is overwhelmed.

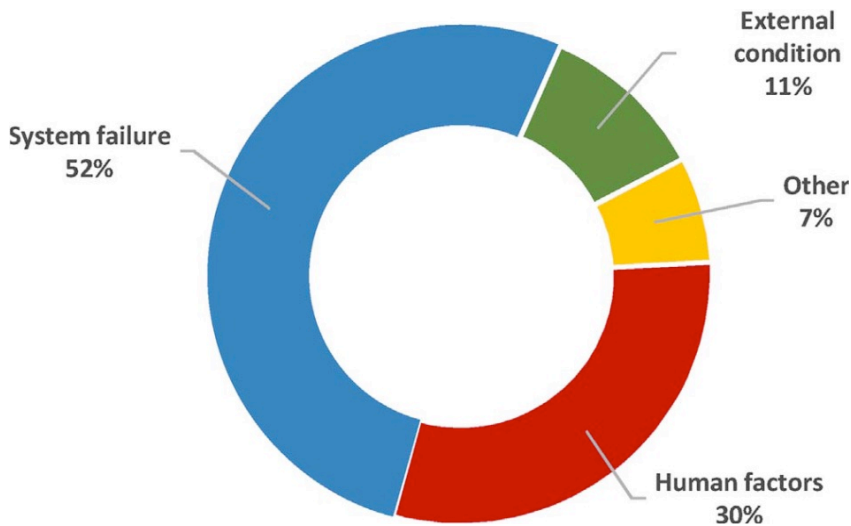

**Figure 56: Summary of reasons for AV disengagements in 2016 [77]**

There are challenges in using data from on-road testing to assess the performance of components of the system. For example, a disengagement due to incorrect route planning could be a failure in planning software, or could be a result of inaccurate data from one or more sensors. The data provided from the California DMV does not allow the identification of the underlying cause of the failure; however, this could be determined with suitable access to logs during a testing programme. Disengagement data could then be used as an additional measure of the performance of components of the AV, but disengagement data is also likely to underrepresent failures of individual sensors, since failures mitigated by other measures or undetected failures, which did not result in a crash, may not be reported.

The lowest disengagement rate achieved by a manufacturer in 2019 was 76 disengagements per million miles driven in California. This represents an improvement from 195 disengagements per million miles driven in 2016. For comparison, the rate of accidents involving cars across the United States in 2018 was estimated at 4.7 crashes per million miles driven [78]. Simulations of manual disengagements by Google/Waymo have suggested that approximately 1 in 5 manual disengagements reported would have resulted in a collision without human intervention; it is less clear how to compare automatic disengagements.

The data obtained from the California DMV is plentiful, and seems to be promising both in terms of the current rate of disengagements and the year-on-year trend. On the face of it, a requirement on the disengagement rate, which was proportionate to the crash rate of human-driven vehicles, would be feasible, if not now, then in the near future. However, this data is limited in that it only provides miles driven, and very basic information about each disengagement. There are a number of other factors which are unrecorded but could affect the rate of disengagements:

- The types of road on which the vehicles were driven is not recorded. Driving mainly on controlled access highways is likely to result in a lower rate of disengagements than driving on crowded urban streets.

- The technology in the vehicles is unlikely to be consistent from year-to-year, or even within years. Testing to gather data on a fully developed system is likely to produce a different disengagement rate from testing to assess a newly developed component.
- Different drivers may have a different tendency to initiate manual disengagements. The rate of disengagements may also reduce over time as drivers come to trust a vehicle more, even if the underlying system has not changed.
- The threshold for an automatic disengagement may change. This could be a planned part of testing, or a response to updated software or additional test data.
- Different reporting practices for disengagements may lead to data which is incomparable between manufacturers or between years. Some manufacturers choose not to report those manual disengagements which they consider to have been unnecessary.

Several of these factors have been identified both in academic analyses of this data, and by the manufacturers themselves. In building a safety case for an AV, some of these issues could be overcome or addressed with full access to testing data as part of a planned testing programme. For example, the testing programme will describe the miles to be driven in different environments, and the impact of drivers' experience on manual disengagements must be assessed and accounted for. However, obtaining sufficient evidence for a reliability claim about the autonomous driving system is unlikely to be achievable through such a testing programme alone, due to large amount of testing to achieve sufficient confidence in the reliability in all environments.

The testing of autonomous vehicles has several challenges that are not present in all autonomous systems, including operating in a wide variety of environments with which it is continuously interacting. In the case of other autonomous systems, particularly those where testing can be automated, or where we can provide a realistic simulation with high confidence, it may be feasible to perform sufficient testing of the complete system. A simpler autonomous system may also allow us to trace failures in testing to a particular component more clearly, or even directly test the performance of an individual component as part of the overall system.

An approach using conservative Bayesian inference methods to obtain confidence in the failure rate of AVs based on what testing has been performed and the prior confidence in the failure rate was described in [79]. We cannot know the precise value of the probability of failure per mile (pfm) for an autonomous vehicle, so it is modelled as a random variable $X$. Using the performance metrics described in previous sections for components of the AI, as well as any additional guards in place, we will have some prior knowledge about the distribution of $X$, denoted $F(x)$. Given additional testing which results in $k$ failures in $n$ miles, denoted $k\&n$, Bayesian inference then gives the following confidence in the failure rate:

$$Pr(X \leqslant p \mid k\&n) = \frac{\int_0^p x^k (1-x)^{n-k} \mathrm{d}F(x)}{\int_0^1 x^k (1-x)^{n-k} \mathrm{d}F(x)}$$

However, it is unrealistic to expect knowledge of the complete prior distribution $F(x)$. The information regarding the prior is likely to be limited to a confidence claim of the form "the failure rate is below $g$ with probability $\theta$", and a lower bound on the failure rate, e.g. due to the probability of mechanical failures in the vehicle. This still allows for an infinite number of priors, but we can obtain a conservative confidence bound by choosing the prior fitting these conditions which minimises the probability computed above.

Figure 57 demonstrates the effect that prior confidence in the reliability of an AV has on the miles of testing required without a failure to make a claim on the pfm $p$ with 95% confidence. The blue dotted and green dashed lines correspond to a prior belief that the pfm is below $10^{-4}$ with probability 0.1 and 0.9 respectively.

These are compared with the miles of testing required using statistical inference with no prior data. If we already have relatively high confidence in the reliability of an AV, then conservative Bayesian inference could reduce the testing required to a feasible level.

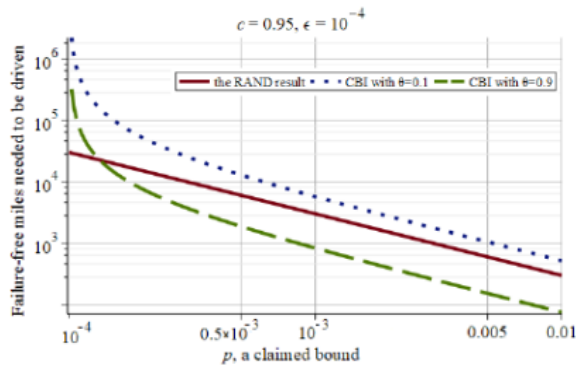

**Figure 57: Testing required to meet claimed reliability bounds [79]**

For the testing to provide evidence to support a reliability claim, the testing must be performed using the same version of the software. In the case of AVs, it is possible that this software will be updated frequently, and quantity of testing required may be prohibitive, or simply infeasible.

Software reliability growth models (SRGMs) provide a model for predicting how the frequency of software failures, in this case measured by the number of disengagements, changes over time. Such a model may be used to provide confidence in the reliability of a new version of the software. However, since SRGMs model overall trends in reliability, which need not be monotonic, predictions from an SRGM alone will not suffice to support a claim regarding pfm, and some on-road testing will be needed for each software version. In the case of ML software components, it is likely to be infeasible to use data from previous software versions to provide confidence in a reliability claim due to the lack of explainability in the ML models.